\documentclass[3p]{elsarticle}

\biboptions{sort&compress}

\usepackage{xcolor}
\usepackage{hyperref}
\usepackage{xspace}
\usepackage{amssymb}
\usepackage{amsmath}
\newcommand{\kth}{\ensuremath{K_\mathrm{th}}\xspace}

\newcommand{\kapa}{\ensuremath{\kappa_a}\xspace}
\newcommand{\kapb}{\ensuremath{\kappa_b}\xspace}
\newcommand{\kapc}{\ensuremath{\kappa_c}\xspace}
\newcommand{\kapab}{\ensuremath{\kappa_{ab}}\xspace}
\newcommand{\kapperp}{\ensuremath{\kappa_\bot}\xspace}
\newcommand{\kappara}{\ensuremath{\kappa_\Vert}\xspace}

\newcommand{\kaph}{\ensuremath{\kappa_\mathrm{ph}}\xspace}
\newcommand{\lmag}{\ensuremath{l_\mathrm{mag}}\xspace}
\newcommand{\kmag}{\ensuremath{\kappa_\mathrm{mag}}\xspace}

\newcommand{\kph}{\ensuremath{\kappa_\mathrm{ph}}\xspace}

\newcommand{\ybco}{\ensuremath{\mathrm{YBa_2Cu_3O_{7-\delta}}}\xspace}
\newcommand{\pbco}{\ensuremath{\mathrm{PrBa_2Cu_3O_{7-\delta}}}\xspace}

\newcommand{\lazn}{\ensuremath{\mathrm{La_{2}Cu_{1-z}Zn_zO_{4}}}\xspace}

\newcommand{\lacuo}{\ensuremath{\mathrm{La_2CuO_4}}\xspace}

\newcommand{\srcuocl}{\ensuremath{\mathrm{Sr_2CuO_2Cl_2}}\xspace}
\newcommand{\lacuod}{\ensuremath{\mathrm{La_2CuO_{4+\delta}}}\xspace}

\newcommand{\leucuo}{\ensuremath{\mathrm{La_{1.8}Eu_{0.2}CuO_4}}\xspace}

\newcommand{\srcuoladder}{\ensuremath{\mathrm{Sr_{14}Cu_{24}O_{41}}}\xspace}
\newcommand{\laf}{\ensuremath{\mathrm{Ca_{9}La_5Cu_{24}O_{41}}}\xspace}

\newcommand{\srca}{\ensuremath{\mathrm{Sr_{14-x}Ca_xCu_{24}O_{41}}}\xspace}
\newcommand{\srcala}{\ensuremath{\mathrm{(Sr,Ca,La)_{14}Cu_{24}O_{41}}}\xspace}
\newcommand{\cuzn}{\ensuremath{\mathrm{Sr_{14}Cu_{24-z}Zn_{z}O_{41}}}\xspace}

\newcommand{\cacuoladder}{\ensuremath{\mathrm{CaCu_{2}O_{3}}}\xspace}
\newcommand{\srcuodouble}{\ensuremath{\mathrm{SrCuO_{2}}}\xspace}
\newcommand{\srcuodoubleca}{\ensuremath{\mathrm{Sr_{1-x}Ca_xCuO_{2}}}\xspace}
\newcommand{\srcuosingle}{\ensuremath{\mathrm{Sr_2CuO_3}}\xspace}
\newcommand{\srcuosingleca}{\ensuremath{\mathrm{(Sr_{1-x}Ca_x)_2CuO_3}}\xspace}

\newcommand{\wkm}{\ensuremath{\mathrm{Wm^{-1}K^{-1}}}\xspace}

\journal{Physics Reports}

\usepackage{lipsum}
\makeatletter
\def\ps@pprintTitle{%
 \let\@oddhead\@empty
 \let\@evenhead\@empty
 \def\@oddfoot{}%
 \let\@evenfoot\@oddfoot}
\makeatother
\begin{document}

\begin{frontmatter}

\title{Heat transport of cuprate-based low-dimensional quantum magnets with strong exchange coupling}
% \tnotetext[mytitlenote]{Fully documented templates are available in the elsarticle package on \href{http://www.ctan.org/tex-archive/macros/latex/contrib/elsarticle}{CTAN}.}

%% Group authors per affiliation:
\author[mymainaddress,mysecondaryaddress]{Christian Hess}
% \address{Radarweg 29, Amsterdam}
% \fntext[myfootnote]{E-Mail: c.hess@ifw-dresden.de}

%% or include affiliations in footnotes:
% \author[mymainaddress,mysecondaryaddress]{Elsevier Inc}
% \ead[url]{www.elsevier.com}

% \author[mysecondaryaddress]{Global Customer Service\corref{mycorrespondingauthor}}
% \cortext[mycorrespondingauthor]{Corresponding author}
\ead{c.hess@ifw-dresden.de}

\address[mymainaddress]{IFW Dresden, 01069 Dresden, Germany}
\address[mysecondaryaddress]{Center for Transport and Devices, TU Dresden, 01069 Dresden, Germany}

\begin{abstract}
Transport properties provide important access to a solid's quasiparticles, such as quasiparticle density, mobility, and scattering. The transport of heat can be particularly revealing because, in principle, all types of excitations in a solid may contribute. Heat transport is well understood for phonons and electrons, but relatively little is known about heat transported by
magnetic excitations. However, during the last about two decades, the magnetic heat transport attracted increasing attention after the discovery of large and unusual signatures of it in low-dimensional quantum magnetic cuprate materials. Today it constitutes an important probe to otherwise often elusive, topological quasiparticles in a broader class of quantum magnets.
This review summarizes the 
experimental foundation of this research, i.e. the state of the art for the magnetic heat transport in the mentioned cuprate materials which host prototypical low-dimensional antiferromagnetic $S=1/2$ Heisenberg models. These comprise, in particular, the two-dimensional square lattice, and one-dimensional spin chain and two-leg ladder spin models. It is shown, how studying the heat transport provides direct access to the thermal occupation and the scattering of the already quite exotic quasiparticles of these models which range from spin-1 spin wave and triplon excitations to fractionalized spin-1/2 spinons.
Remarkable transport properties of these quasiparticles have been revealed: the spin-heat transport often is highly efficient and in some cases even ballistic, in agreement with theoretical predictions.
\end{abstract}


\end{frontmatter}

% \linenumbers
\tableofcontents
\section{Introduction}
\label{sec:intro}
The heat transport properties of a solid provide important information about the generation, mobility, scattering, and dissipation of various kinds of excitations, such as electrons, phonons and magnons. Deep fundamental knowledge in this regard exists about the heat transport by phonons and electrons \cite{Berman}, yet very little is known about heat transport by magnetic excitations. 

After the original prediction of magnetic heat transport in 1936 by Fr\"ohlich and Heitler \cite{Froehlich36}, it took almost 30 years until the first convincing experimental evidence for heat transport by classical spin waves was found in ferrimagnetic yttrium-iron-garnet (YIG) \cite{Luethi62,Douglass63,Rives69,Walton73}. 
Analogous to the case of phononic and electronic heat conduction, the analysis of this magnon heat conductivity is intriguing because it should in principle yield valuable information about the excitation and scattering of magnetic excitations (e.g. off defects, phonons, and electrons). However, most of the pioneering experiments on YIG and following experiments on other materials \cite{Gorter69,Lang77,Coenen77} focused on the mere identification of the new heat transport phenomenon. Furthermore, these early experiments typically were restricted to spin waves emerging from magnetically ordered phases at very low temperature ($T<10~\mathrm{K}$). The first signature of magnetic heat transport at higher temperatures ($T>50~\mathrm{K}$) was observed for the one-dimensional quantum antiferromagnet $\mathrm{KCuF_3}$ \cite{Hirakawa75}, remarkably \textit{above} its N\'{e}el temperature of $T_N\approx38~\mathrm{K}$ \cite{Hutchings1969}. These very intriguing first results for the heat transport of a low-dimensional \textit{quantum magnet} remained for quite some time largely unnoticed.
However, the rigorous theoretical prediction of dissipationless (so-called \textit{ballistic}) heat conduction in one-dimensional antiferromagnetic  spin-1/2 chains \cite{Zotos97} and the discovery of huge magnetic contributions in  quantum two-leg  spin-1/2 ladder compounds \cite{Kudo1999,Sologubenko00,Hess01} triggered intense research on the heat transport of low-dimensional quantum spin systems both experimentally  \cite{Kudo1999,Sologubenko00,Kudo01,Hess01,Hess02,Hess03,Hess04a,Hess05,Sologubenko00a,Sologubenko01,Kudo2003,Ando1998,Vasilev1998,Hofmann2002,Hofmann2003,Sologubenko03,Sologubenko03a,Kordonis06,Hlubek2010a,Steckel2016,Hess2007b,Hess2006,Otter2012,Montagnese2013,Ribeiro2005,Hess2007,Hlubek2010,Hlubek2012,Hlubek2011,Mohan2014}  and theoretically 
\cite{Heidrich02,Heidrich03,Orignac03,Alvarez02,Heidrich04,Gros2004,Saito2003,Saito2003a,Shimshoni03,Kluemper2002,Sakai2003,Zotos04,Karadamoglou04,Louis2006,Li2002,Li2003,Rozhkov05,Chernyshev05,Jung06,Boulat2007,Steinigeweg2016,Chernyshev2016}.

On the experimental side, more and more evidence for \textit{unconventional} heat conduction in low-dimensional quantum magnets has been observed over the years in various materials. Today, the clearest and often surprising experimental findings are known for spin systems realized in copper oxides (cuprates). In these compounds, a substantial magnetic heat conduction is often observed, despite a pronounced quantum nature of their spin systems with inherent absence of long-range magnetic order or even short range spin-spin correlations. Their actual magnetic heat conductivity can be unambiguously detected, because if present, it is usually large even at room temperature and above, often dwarfing the phononic heat conductivity of the system and thereby sometimes reaches values which are comparable to the heat conductivity of a metal. 

The content of the review at hand is as follows: In the following paragraphs of this introductory Section~\ref{sec:intro}, the essential basics of the spin models and the cuprate materials under scrutiny will be presented. Furthermore, the salient experimental signatures of magnetic heat conductivity in such systems will be presented, and the basics of analyzing the data will be introduced.  This Section~\ref{sec:intro}, i.e. the whole introductory part is based on \cite{Hess2007b}. Section ~\ref{sec:spin_planes} and Section~\ref{sec:ladders} summarize the state of the art for the heat conductivity of the $S=1/2$ two-dimensional Heisenberg antiferromagnet on a square lattice (2D-HAF) and of $S=1/2$ two-leg spin ladders, respectively. 
Finally, Section~\ref{sec:chains} addresses the heat transport and related results on the spin dynamics of $S=1/2$ Heisenberg spin chains. 
% \cite{Hess2007b,Ribeiro2005,Hlubek2010,Hlubek2012,Hlubek2011,Mohan2014}.

% \subsection{Low-dimensional quantum magnets in a nutshell}

% \section{Background}

\subsection{Spin models}
We are considering $S=1/2$ models with Heisenberg interaction
\begin{equation}
\label{Hamleit} {\cal H}=J_{i,j}\sum_{\langle i,j \rangle}{\bf S_i\cdot S_j}\,, 
\end{equation}
where the sum runs over all nearest neighbors in the system. Here we are investigating three different spin models: the 2D-HAF, the two-leg spin ladder, and spin chains. For the 2D-HAF and the spin chain $J_{i,j}=J$ and for the spin ladder $J_{i,j}=J_\Vert$ along the legs and $J_{i,j}=J_\bot$ along the rungs of the ladder.

The corresponding low-dimensional quantum spin models are characterized by very peculiar ground states which are quantum disordered in the one-dimensional chain and ladder models. The elementary excitations which emerge from this ground states bear therefore a quite exotic character. For instance, the spin-spin correlations of $S=1/2$ Heisenberg spin chains, decay algebraically with distance between the spins \cite{Kluemper1993}. The elementary excitations are nevertheless well defined. They are
fractionalized, i.e. $\Delta S=1$ spin-flip excitations of the system decay into so-called spinons which are gapless and carry a spin $S=1/2$ \cite{Faddeev1981}. On the other hand, the ground state of a two-leg ladder possesses more short-range spin-spin correlations which decay exponentially as a function
of distance \cite{Dagotto96,Dagotto99}. The elementary excitations are $S=1$ particles (usually called magnons or triplons) and are separated from the ground state by a spin gap $\Delta$ ($\Delta/k_B\approx400~\mathrm{K}$ in the case of the systems discussed here) \cite{Dagotto99}. Finally, the ground state of the 2D-HAF is a rather classical long-range ordered N\'{e}el state. However, also here due to incipient quantum disorder it only exists at temperature $T = 0$ and possesses a strongly reduced sublattice magnetization \cite{Manousakis91}. In this case the elementary excitations are well described using a spin wave framework where one should keep in mind that alternative descriptions have been discussed, too \cite{Coldea01,Sandvik2001,Ho2001}.

In the case of hole doping, all these model systems yield interesting and exotic properties. A Luttinger
liquid forms in hole-doped spin chains. Here the electronic excitations decay into collective
excitations of holes (holons) and spins (spinons). This phenomenon is often referred to as spin-charge
separation. Quite different properties have been predicted for two-leg spin ladders:
superconductivity competing with a charge ordered ground state is expected in this case
\cite{Dagotto92,Dagotto96}. Finally, hole doping has great importance in the case of the 2D-HAF, because the interaction of the doped hole with the antiferromagnetic background forms a new quasiparticle, the spin polaron, which can be understood as a hole dressed with characteristic spin fluctuations \cite{Martinez1991}. The spin polaron is thought of playing a crucial role for the emergence of exotic ground states upon doping, including high-temperature superconductivity, which is observed in such systems \cite{Chernyshev1999,Lee2006}. Note that the observation of spin-charge separation signatures upon controlled hole doping of $S=1/2$ Heisenberg chain materials has not yet been achieved experimentally, Nevertheless, signatures of it have been derived e.g. from angular resolved photo emission experiments on undoped chain compounds \cite{Koitzsch2006}. On the other hand, charge ordering and superconductivity are prominent experimental features of hole-doped spin ladder and 2D-HAF materials \cite{Vuletic2006,Lee2006}.

\begin{figure}
% \captionsetup{format=plain}
\centering
\includegraphics[width=\textwidth]{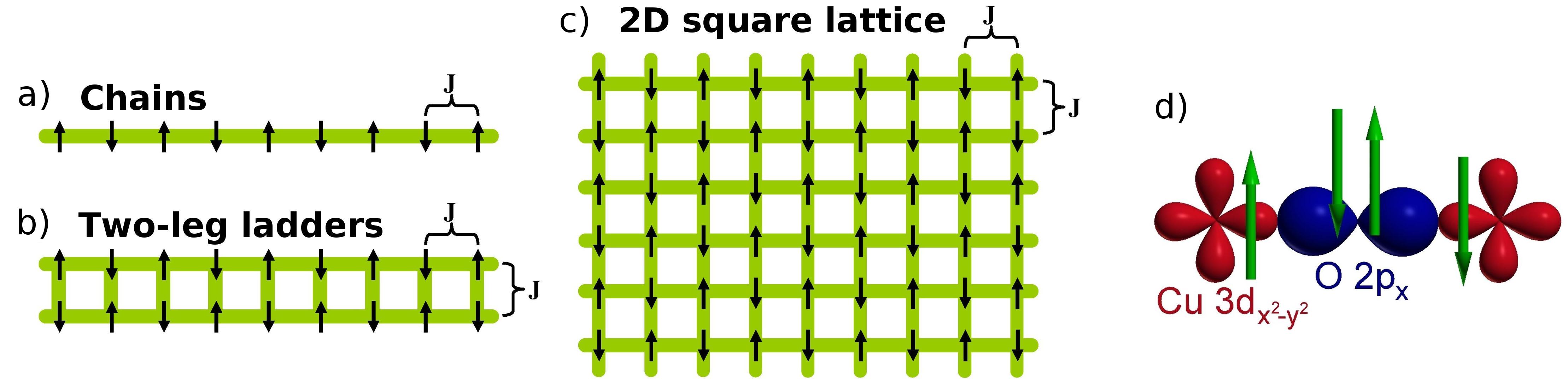}
\caption{Illustration of low-dimensional spin structures: (a) a spin chain, (b) a two-leg spin ladder, and (c) a two-dimensional square lattice. Arrows represent localized electrons with spin $S=1/2$ spin and shaded bars symbolize strong antiferromagnetic exchange between them. (d) Schematic illustration of the underlying chemical building block giving rise to the localized spins and their interaction. Only the relevant {\rm Cu}-$3d_{x^2-y^2}$ and {\rm O}-$2p_x$ orbitals are indicated. Arrows represent the spins of the electrons involved. Figure reproduced from \cite{Hess2007b}.}
\label{fig:1}       
\end{figure}

\subsection{Materials}
\label{intromat}
The materials at focus of this review are cuprate compounds which indeed host spin arrangements in the geometrical form of chains, two-leg ladders, and square lattices with a strong antiferromagnetic Heisenberg exchange ($J/k_B\approx 1500-2000~\mathrm{K}$) between nearest neighbor spins. Sketches of such spin arrangements are shown in Fig.~\ref{fig:1}a-c.
These low-dimensional arrangements of interacting spins arise from similarly low-dimensional structures composed of {\rm Cu-O-Cu} bonds, within which a dominant antiferromagnetic exchange is present if these bonds are straight, i.e. a bonding angle of $180^\circ$ is realized as depicted in Fig.~\ref{fig:1}d. 
In all the considered cuprate systems the spins have $S=1/2$, resulting from the $3d^9$ configuration of $\mathrm{Cu}^{2+}$-ions and therefore possess generally a strong quantum nature.

Good realizations of $S=1/2$ Heisenberg chains as depicted in Fig.~\ref{fig:1}a are found in the compounds \cacuoladder, \srcuodouble, and \srcuodouble, where chains of straight {\rm Cu-O-Cu} bonds and hence a strong antiferromagnetic exchange exists along one particular crystallographic direction only; the magnetic exchange perpendicular to this direction is much weaker \cite{Kiryukhin01,Motoyama1996}.
Two-leg spin ladders are realized (cf. Fig.~\ref{fig:1}b) in the $\mathrm{(Sr,Ca,La)_{14}Cu_{24}O_{41}}$ family of compounds, where parallel pairs of such {\rm Cu-O-Cu} chains (the ladder legs) are coupled to each other via bridging O-ions, producing in straight {\rm Cu-O-Cu} bonds perpendicular to the chains direction (the ladder rungs), where the
interchain coupling (or rung coupling) perpendicular to the chain direction $J_\perp$ is of a similar magnitude to the intrachain coupling (or leg coupling), i.e. $J_\perp\approx J$ \cite{Dagotto99}. Ladder structures with more legs can in principle be created by coupling more chains to the structure; eventually, this would lead to a 2D-HAF in the infinite limit. A good realization of a 2D-HAF with $S=1/2$ is given by \lacuo and other related antiferromagnetic parent compounds of high-temperature superconductors.

\begin{figure}
\centering
\includegraphics[width=\textwidth]{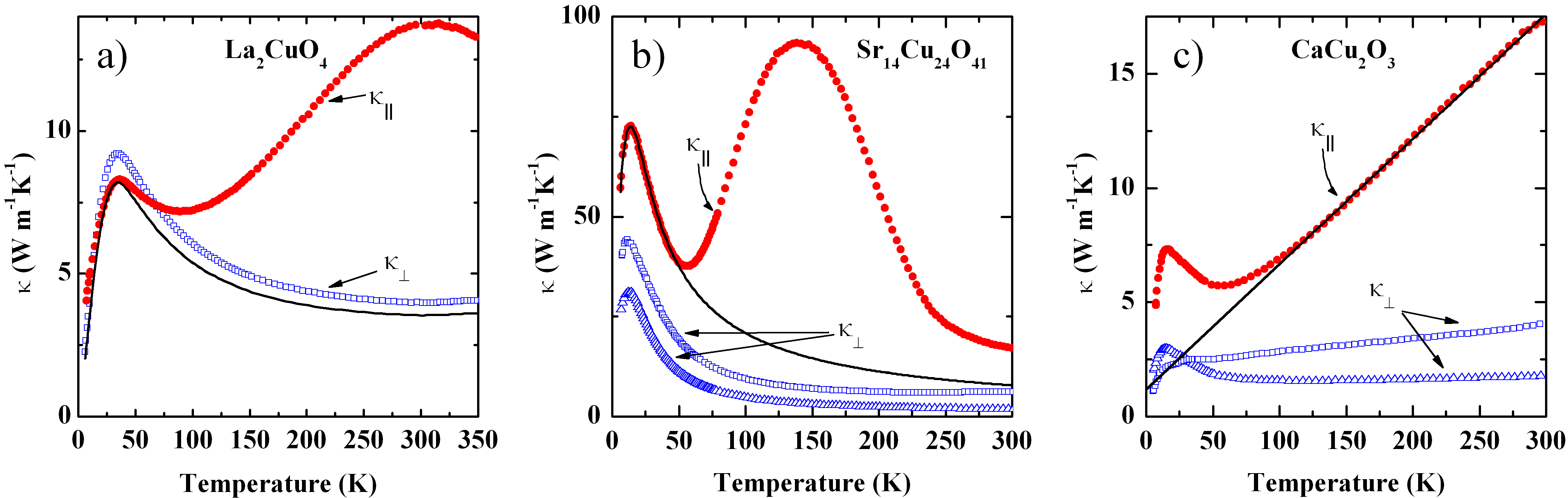}
\caption{Anisotropic thermal conductivity of various low-dimensional spin materials as a function of
temperature: a) the 2D-HAF as realized in \lacuo, b) the two-leg spin ladder material \srcuoladder,
and c) the spin chain compound \cacuoladder. Filled and open symbols represent \kappara and \kapperp of the
materials. The solid line in a) represents a linear fit to the data in the range $T>100\mathrm~K$. The axis
intercept of its extrapolation towards $T = 0$ is an approximation of \kph in the fit range. Solid lines in
b) and c) represent estimations for the phonon background. Figure adapted from \cite{Hess01,Hess03,Hess2007}.}
\label{fig:signatures}       
\end{figure}

\subsection{Experimental signatures of magnetic heat transport}
\label{sec:signatures}
The typical experimental signature of low-dimensional magnetic heat transport in a given material is a very anisotropic heat conductivity tensor of the material which reflects the dimensionality of the spin system. The origin of the anisotropy is the magnetic heat conduction of the low-dimensional magnetic structures which add to the always present phononic heat conduction. In principle, electronic heat conduction could occur as well. However, for all materials discussed here, such a contribution is irrelevant because the electrical conductivity is very low.

The experimental method which has been employed for obtaining the here discussed data is, in the largest portion of cases, the so-called standard steady-state method \cite{Berman} which is proven to be very reliable for obtaining the different components of the heat conductivity tensor. An exception concerns additional results on the two-leg spin ladder materials, where fluctuations in the magnetic signal provided the motivation for dynamic heat transport studies (Section~\ref{sec:ladders}). 

Fig.~\ref{fig:signatures} shows selected corresponding examples for experimental results of the heat conductivity $\kappa$ of \lacuo, \srcuoladder and \cacuoladder, which are good representatives of the 2D-HAF, the two-leg spin ladder, and the isotropic antiferromagnetic Heisenberg chain \cite{Hess03,Hess01,Hess2007b,Hess2007}. The data of the heat conductivity perpendicular to the low-dimensional magnetic structures  ($\kappa_\bot$), i.e. along the directions with negligible magnetic exchange interaction, the temperature dependence of $\kappa$ is characteristic of conventional \textit{phononic} heat conduction \kph \cite{Berman}. More specifically, $\kappa(T)$ has a low-temperature peak centered at temperature $T\approx20\dots30~\mathrm{K}$, with a continuously decaying high-temperature edge. An exception is found for \cacuoladder, where one component of \kapperp increases monotonically. Such an increase is well known for systems with a strongly suppressed \kph due to disorder \cite{Hess2007,Hess04}.

Completely different characteristics are found if $\kappa$ is measured parallel to the low-dimensional structures, i.e. along the directions with a large $J$ ($\kappa_\Vert$). Also in these cases, signatures of a phononic low-temperature peak are present. However, upon increasing the temperature further, \kappara is very different from \kapperp. In all three cases, \kappara strongly increases for $T\gtrsim60~\mathrm{K}$ ($T\gtrsim90~\mathrm{K}$ in the case of \lacuo) with increasing temperature, and in the case of \lacuo and \srcuoladder, a high-temperature peak at $\mathrm 310$~K and $\mathrm 140$~K, respectively, is formed, while for \cacuoladder the increase continues up to room temperature. This remarkably clear anisotropy of $\kappa$ is the clear qualitative evidence for large magnetic contributions to \kappara of these three materials, i.e. a large \kmag of the low-dimensional spin structures. 

Generally, in the shown examples, \kmag can be well extracted from the measured data for \kappara by subtracting the phononic contribution \kph, which turns out to be reasonably approximated by the experimentally obtained \kapperp. This is the essential basis for analyzing \kmag in order to yield information on the thermal generation and the scattering of the magnetic quasiparticles. 

\subsection{Modeling}\label{kinmod}
In theoretical works, the attention often is focused on the possibility of ballistic magnetic heat transport in 1D-systems: in integrable models like the XXZ Heisenberg spin chain the Hamiltonian and the thermal current operator commute, i.e. once a thermal current is established in such a system, it will never decay \cite{Zotos97}. In other words, the thermal resistance vanishes and the magnetic thermal conductivity $\kappa_\mathrm{mag}$ diverges. While such surprising properties are well established for integrable spin models \cite{Zotos97}, ballistic heat transport in non-integrable quasi 1D-systems (e.g. two-leg spin ladders) has been a subject of intense discussion \cite{Alvarez02,Zotos04,Heidrich04,Boulat2007,Steinigeweg2016}. However, in real materials scattering processes involving defects and other quasiparticles such as phonons and charge carriers must play an important role and render $\kappa_\mathrm{mag}$ finite in all cases \cite{Shimshoni03}. The analysis of $\kappa_\mathrm{mag}$ should hence provide further insight into the nature of these scattering processes and the dissipation of magnetic heat currents.

\subsubsection*{Kinetic model}
We set up a kinetic model which should be able to capture the most important features of the magnetic heat conductivity \cite{Hess2007b}. Apparently, the qualitative  temperature dependence of $\kappa_\mathrm{mag}$ often is a simple peak structure (\lacuo and \srcuoladder) or a monotonic increase (\cacuoladder) in the studied range $T=100\dots350~\mathrm K$, where the latter may be regarded as the low temperature edge of a peak. A peak structure is very common for
the temperature dependence of 
the thermal conductivity $\kappa$ of any kind of heat carrying particle, such as phonons or electrons \cite{Berman}. 
In principle, one can expect that the same kinetic considerations which successfully describe the physics for phononic and electronic heat transport, can be applied to magnetic excitations as well. 

The basic physics which determines the $T$-dependence of $\kappa$ can be inferred from the kinetic estimate
\cite{Ziman}

\begin{equation}
 \kappa=\frac{1}{d}\frac{1}{(2\pi)^d}\int c_{\bf k}v_{\bf k}l_{\bf k}d{\bf k}, \label{eq:kinetic_general}
\end{equation}
with $d$ the dimensionality of the considered system, $c_{\bf k}=\frac{d}{dT}\epsilon_{\bf k}n_{\bf k}$ the specific heat, $v_{\bf k}$ the velocity and $l_{\bf k}$ the mean free path of a mode with wave vector ${\bf k}$. $\epsilon_{\bf k}$ and $n_{\bf k}$ are the energy and the statistical occupation function of the mode $\bf k$.

Here, we are interested in magnetic excitations, i.e., a crucial parameter which determines the thermal occupation is the thermal energy $k_BT$ in relation to the magnetic exchange energy $J$.
For all systems discussed here $J/k_B\approx 1500\dots2000\mathrm{~K}$, whereas the experimental data only extend over temperatures $T < 350~\mathrm{K}$.
Thus, all considerations discussed in this review concern, from the viewpoint of the magnetic system, the situation of low temperature $k_BT\ll J$. 
The momentum of these particles naturally is confined to a small vicinity of the minima of the dispersion function. One can therefore safely ignore the momentum dependence of the mean free path, i.e., $l\approxeq l_{\bf k}$.
Furthermore, at such low $T$,
only a few magnetic modes are excited and contribute to the heat transport. If the temperature is also well below other relevant energy scales of the solid, such as the Debye temperature, scattering processes different from defect and boundary scattering which change the crystal momentum are rare.

In this situation, the low-$T$ increase of $\kappa$ is (a) characteristic of the excitation of the heat carrying particle (reflecting the $T$-dependence of the specific heat if $v_{\bf k}$ is momentum independent) and (b) proportional to the mean free path $l\approxeq l_{\bf k}$.
At higher $T$, the momentum-dependent scattering becomes increasingly important and eventually leads to a decrease of the mean free path and hence to a decrease of $\kappa$. This decrease is characteristic of the relevant scattering mechanisms and allows an advanced analysis
which potentially provides crucial information about these mechanisms.

The application of Eq.~\ref{eq:kinetic_general} for the case of 1D and 2D magnetic systems considered here leads to the general result 
\begin{equation}
 \kappa_\mathrm{mag}(T)\propto l_\mathrm{mag} f(T)~,\label{kmag_simple} 
\end{equation}
where $l_\mathrm{mag}(T)$ is a general magnetic mean free path based on the approximation $l_\mathrm{mag}\equiv l_{\bf k}$. 
As mentioned afore, 
for $k_BT\ll J$ this assumption is justified for the large-$J$ systems considered here because the heat carrying excitations exist in significant numbers only in the vicinity of the band minima, i.e. a very small fraction of the Brillouin zone. 
The function $f(T)$ depends on temperature in a manner which is characteristic of the considered spin system.
Details will be discussed in the respective sections.

\section{Spin planes}
\label{sec:spin_planes}

\subsection{Qualitative proof of magnon heat transport}
As already indicated in section \ref{sec:signatures}, the prominent signature of the magnetic heat conductivity in the 2D-HAF is a pronounced high-temperature peak in the in-plane thermal conductivity \kapab, whereas the out-of-plane thermal conductivity \kapc is of purely phononic character (see Fig.~\ref{fig:signatures}).
This strikingly anomalous heat conductivity was first observed in insulating parent compounds of cuprate high-temperature superconductors, in particular, \lacuo \cite{Morelli89,Nakamura91} as well as in \ybco and \pbco with $\delta \approx 1$ \cite{Cohn95}.  
After these pioneering experimental findings, the origin of the high-temperature anomaly in \kapab remained elusive for several years.  Nakamura et al. were the first to speculate that the high-temperature maximum could be related to heat carried by magnetic
excitations \cite{Nakamura91}. However, several attempts to explain this peak by anomalous phononic heat transport involving scattering processes
of acoustic phonons with soft optical phonons \cite{Cohn95} or magnons \cite{Morelli89} have been made.
Eventually, inspired by the observation of an exceptionally large one-dimensional \kmag in two-leg spin ladder materials \cite{Sologubenko00,Hess01} (see section \ref{sec:ladders}), several groups reinvestigated the heat conductivity of \lacuo and related compounds \cite{Sun2003,Sun2003a,Berggold2006,Hess04,Hess03,Hess03a,Hess2007a,Hofmann2003,Yan2003,Hess2007b}.
Qualitatively, all these studies came to the conclusion that a conventional explanation for the high-temperature peak in \kapab in the sense of electronic, phononic and even radiative heat transport can be excluded. Sizable electronic heat transport was readily ruled out due to the electronically insulating or only weakly conducting
properties of \lacuo \cite{Hess03,Hofmann2003} (and all other materials discussed below). The same holds for radiative heat transport \cite{Hess03} since the optical properties of \lacuo are almost isotropic in the relevant energy-range below $\mathrm h\nu\lesssim0.1$~eV \cite{Uchida91}. The possibility of anomalous phononic heat transport, however, deserves further attention here because it is difficult to exclude it based on the heat conductivity data on \lacuo without further information and accordingly has been specifically addressed in several of the mentioned studies.

\begin{figure}
\centering
\includegraphics[width=0.7\textwidth]{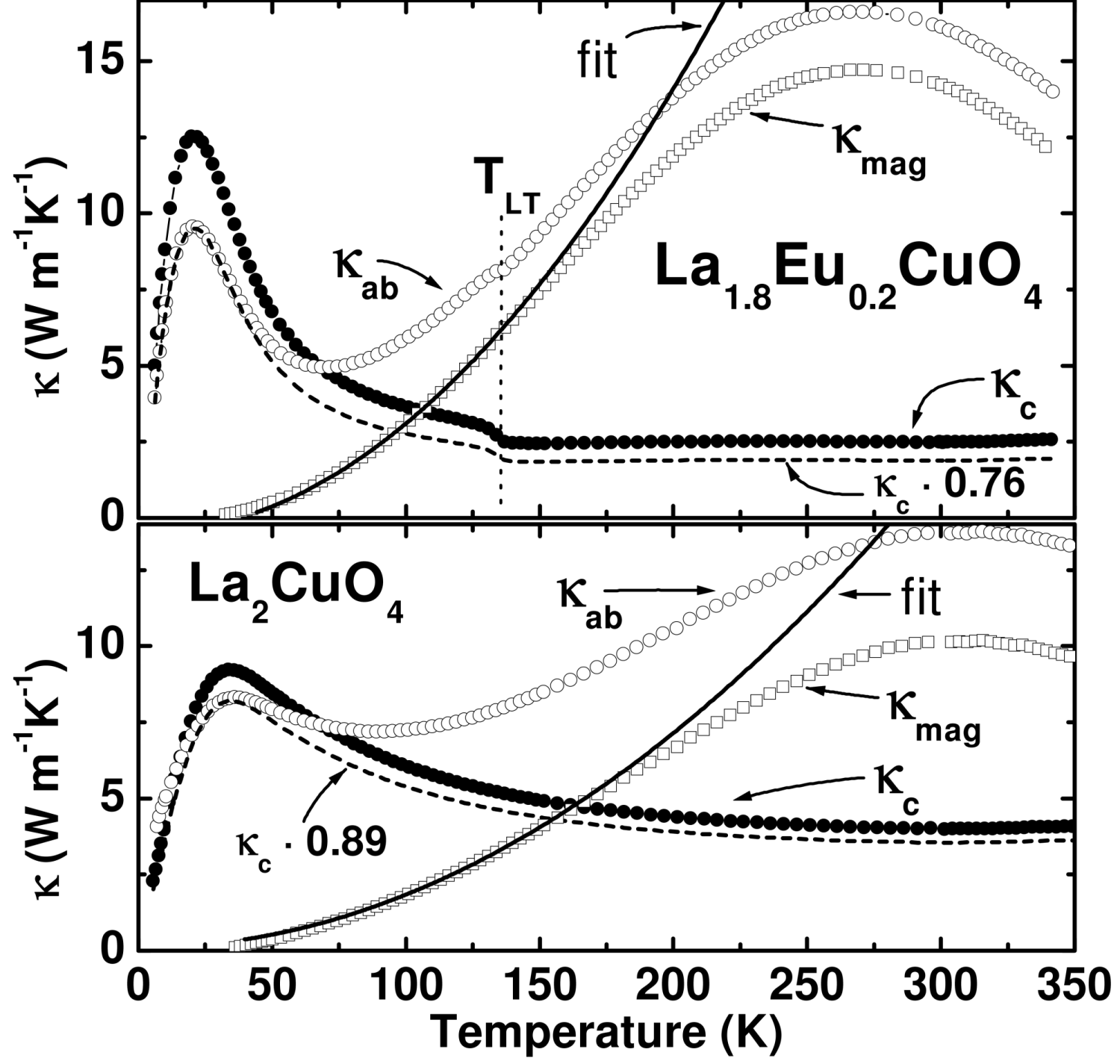}
\caption{Thermal conductivity of \lacuo (bottom) and \leucuo (top). Full circles: $\kappa_c$. Open circles: $\kappa_{ab}$.  Open
squares: $\kappa_{\mathrm{mag}}$. Solid line: fit according to Eq.~\ref{eq:fit2d}. Dashed line:
$\kappa_{ab,\mathrm {ph}}$. Reproduced from \cite{Hess03}. For similar measured data see also \cite{Nakamura91,Sun2003a,Yan2003}.}
\label{fig:leucuo}         
\end{figure}
\subsubsection*{Anomalous phonon scattering versus magnetic heat transport}
One thinkable hypothesis is that the heat conductivity of the 2D materials should be regarded as purely phononic, and thus  the high-temperature anomaly in \kapab should be regarded as not originating from an additional contribution but rather being the result of an unusual phonon damping by either magnetic excitations \cite{Morelli89} or soft phonon modes \cite{Cohn95}. Such phonon damping is known to be caused by \textit{resonant} scattering of the heat-carrying phonons off such excitations. In such a case, a double-peak structure in $\kappa(T)$ results from a strong \textit{reduction} of the heat conductivity in the temperature regime where the scattering is strongest. Well known examples where such a double peak is caused by local magnetic excitations and soft phonon modes are the heat conductivities of $\mathrm{SrCu_2(BO_3)_2}$ \cite{Hofmann2001}, $\mathrm{SrTiO_3}$ \cite{Suemune1965,Steigmeier1968}, respectively.
For \lacuo, one might consider such a scenario as being quite unlikely due to the complete absence of a double-peak in the heat conductivity perpendicular to the planes, \kapc. However, one could argue that both the magnetism and the crystal structure of the material are very anisotropic, and therefore an anisotropic phonon damping was thinkable. 
Hofmann et al. pointed out that in the 2D square-lattice cuprates, such as \lacuo, the dispersion of magnetic excitations which ranges from $\sim 0$ to $2J/k_B\sim2000~\mathrm{K}$ ($J$ is the in-plane exchange constant) is too stiff to cause phonon scattering on magnetic excitations being most pronounced in a narrow temperature interval around $100~\mathrm{K}$ \cite{Hofmann2003}. 

Damping of acoustic phonons due to soft optical phonon modes does, however, play an important role in the phonon heat conductivity of \lacuo, indeed. More specifically, soft optical phonon modes associated with the tilting of the $\mathrm{CuO_6}$ octahedra which are present in the entire so-called low-temperature orthorhombic (LTO) phase, i.e. at $T\leq T_\mathrm{HT}\approx500~\mathrm{K}$ \cite{Boeni88,Boeni1989,Birgeneau87}, have been shown to cause a significant suppression in the phononic heat conductivity of \lacuo \cite{Hess03a}. Thus, in order to unambiguously rule out soft-phonon scattering as the origin of the anomalous \kapab, the investigation of materials where the such soft phonons are not present is an obvious route.

Hess et al. therefore studied the heat conductivity of \leucuo \cite{Hess03}. In this material the presence of Eu on the La-site has only little influence on the magnetism of the $\mathrm{CuO_2}$-planes \cite{Sun2003a} but has a strong impact on the structure, because it induces a new structural phase at low temperature, the so-called low-temperature tetragonal (LTT) phase, at $T\leq T_\mathrm{LT}\approx135~\mathrm{K}$ \cite{Buchner94,Klauss00}.
Fig.~\ref{fig:leucuo} (top panel) shows $\kappa_{ab}(T)$ and $\kappa_c(T)$ of a \leucuo single crystal. Both curves strongly resemble the findings for \lacuo (lower panel), yet exhibiting obvious differences: the low-temperature peaks of \kapab and \kapc are slightly larger and more sharply shaped than in the case of \lacuo. Furthermore, a step-like anomaly
is present at $T_\mathrm{LT}\approx135~\mathrm{K}$. Above $T_\mathrm{LT}$, $\kappa_c$ remains almost constant and stays below the value for the undoped case, while $\kappa_{ab}$ also exhibits a high-temperature maximum at
$T\approx270~\mathrm{K}$ which is even larger than that of \lacuo.

The difference between $\kappa_c$ of Eu-doped and of pure \lacuo can be attributed to a
difference in phononic heat conduction: Upon doping \lacuo with Eu, enhanced scattering of
phonons reduces $\kappa_c$ for $T>T_{LT}$, where both compounds have the same structure (LTO). The
anomaly at $T_\mathrm{LT}$ (\leucuo) signals the transition to a new structural
phase for $T<T_\mathrm{LT}$ where $\kappa_{\mathrm {ph}}$ is enhanced. 
Indeed, soft phonon branches do exist in the LTO-phase of Rare Earth-doped \lacuo \cite{Keimer93}, which naturally explains a suppression of $\kappa_{\mathrm{ph}}$ of \leucuo. The change of $\kappa_{\mathrm{ph}}$ at $T_\mathrm{LT}$ then follows from the discontinuous hardening of the soft phonon branch in the LTT-phase \cite{Martinez91,Keimer93} and an
associated reduced scattering rate of acoustic phonons.\footnote{One should note that Sera et al. explain the change of $\kappa_{\mathrm{ph}}$ at $T_\mathrm{LT}$ by an observed change of the velocity of sound $v_s$ at $T_\mathrm{LT}$ \cite{Sera97}. The actual changes of $v_s$ are, however, far too
small ($\sim1\%$) \cite{Yamada94} to account for the much larger changes of $\kappa_{\mathrm{ph}}$. These changes of $v_s$ at $T_\mathrm{LT}$ should thus be regarded as a further accompanying phenomenon of the
structural phase transition.} 
The more striking observation is, however, that the high-temperature peak in \kapab of \leucuo is not affected by the transition. This unambiguously rules out that the double-peak structure in \leucuo (and thus also in \lacuo) is caused by soft-phonon scattering. Instead, the high-temperature peak has to be interpreted as stemming from another heat transport channel different from acoustic phonons which adds to the lattice thermal conductivity \kaph. 

\begin{figure}
\centering
\includegraphics[width=0.7\textwidth]{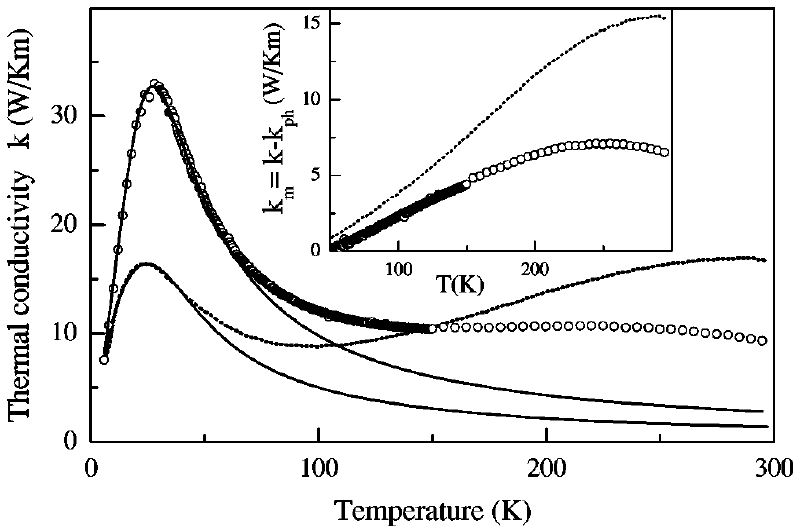}
\caption{In-plane thermal conductivity (here labeled $k(T)$) of \srcuocl (circles) and \lacuo (dotted line; data from \cite{Nakamura91}). Solid lines: fits to the phonon thermal conductivity using the Callaway model \cite{Callaway59}. Inset: The magnetic heat conductivity for \srcuocl (circles) and \lacuo (dotted line) (see text). Image taken from \cite{Hofmann2003}.}
\label{fig:SrCuOCl_Hofmann2003}       
\end{figure}

This latter conclusion is corroborated by investigations on \srcuocl \cite{Hofmann2003} and on $\mathrm{R_2CuO_4}$ ($\mathrm{R = Pr}$, Nd, Sm, Eu, Gd) \cite{Berggold2006,Jin2003}, where the soft-phonon scattering as is present in \lacuo is rigorously excluded due to different structural phases, yet practically preserving the magnetism of the $\mathrm{CuO}_2$-planes. 
On the one hand, \srcuocl is practically isostructural to \lacuo with the small difference that (apart from having $\mathrm{Sr^{2+}}$-ions on the $\mathrm{La^{2+}}$-sites) $\mathrm{Cl^{1-}}$-ions occupy the positions of the apical $\mathrm{O^{2-}}$-ions of the $\mathrm{CuO}_6$ octahedra which in \lacuo form the two-dimensional spin-1/2 planes. 
Unlike \lacuo, however, \srcuocl remains in the so-called high-temperature tetragonal (HTT) phase down to lowest temperature and thus does not exhibit any lattice instability which could give rise to soft-phonon scattering as in \lacuo \cite{Hofmann2003}. 
On the other hand, the compounds $\mathrm{R_2CuO_4}$ ($\mathrm{R = Pr, Nd, Sm, Eu, Gd}$) crystallize in a structure quite different from that of \lacuo, viz. in the so-called tetragonal $T'$-phase without any apical oxygen but with very similar $\mathrm{CuO}_2$-planes as those in \lacuo. 
For $\mathrm{R = Pr, Nd, Sm}$ the $T'$-phase is stable down to lowest temperature, whereas for $\mathrm{R = Eu, Gd}$ the structure undergoes a phase transition towards an orthorhombic phase at $170~\mathrm{K}$ and $685~\mathrm{K}$, respectively \cite{Braden1994,Vigoureux1997,Berggold2006}. 
All three heat conductivity studies \cite{Hofmann2003,Berggold2006,Jin2003} reveal clearly a high-temperature hump in the in-plane thermal conductivity with a similar, albeit with a somewhat smaller magnitude than that of \lacuo, is also present in these layered cuprates despite having very different structural properties (see Fig.~\ref{fig:SrCuOCl_Hofmann2003} for \kapab of \srcuocl in comparison of that of \lacuo  \cite{Hofmann2003,Nakamura91}) . Thus the high-temperature peak turns out as a common feature in the layered cuprates which unambiguously is to be interpreted as a two-dimensional excess thermal conductivity within the magnetic $\mathrm{CuO}_2$-planes. Note that the high-temperature peaks in the thermal conductivity of \ybco, \pbco with $\delta \approx 1$ \cite{Cohn95} should be interpreted in a similar way.

Thus, the bottom line of this section is that the high-temperature anomaly in the thermal conductivity of \lacuo and related layered insulating cuprates originates from a substantial two-dimensional excess thermal conduction which adds to the typical phononic thermal conductivity \kaph. In lack of other candidates it is straightforward to conclude that magnetic excitations (i.e. magnons) of the two-dimensional antiferromagnetic planes cause this unusual heat conduction.\footnote{Note that dispersing optical phonons can be ruled out to play an important role because their contribution to the thermal conductivity have been estimated to be about one order of magnitude smaller ($\sim1~\rm Wm^{-1}K^{-1}$) than that what is observed \cite{Hess04}.} Hence, the thermal conductivity measured parallel to the $\mathrm{CuO}_2$-planes (\kappara) can be considered as the sum of a conventional phonon heat conductivity \kaph and a magnetic contribution, i.e., 
\begin{equation}
 \kappa_\Vert=\kappa_\mathrm{ph}+\kappa_\mathrm{mag} .
\end{equation}

\subsection{Extraction of the magnetic heat conductivity}

\begin{figure}
\centering
\includegraphics[width=0.7\textwidth]{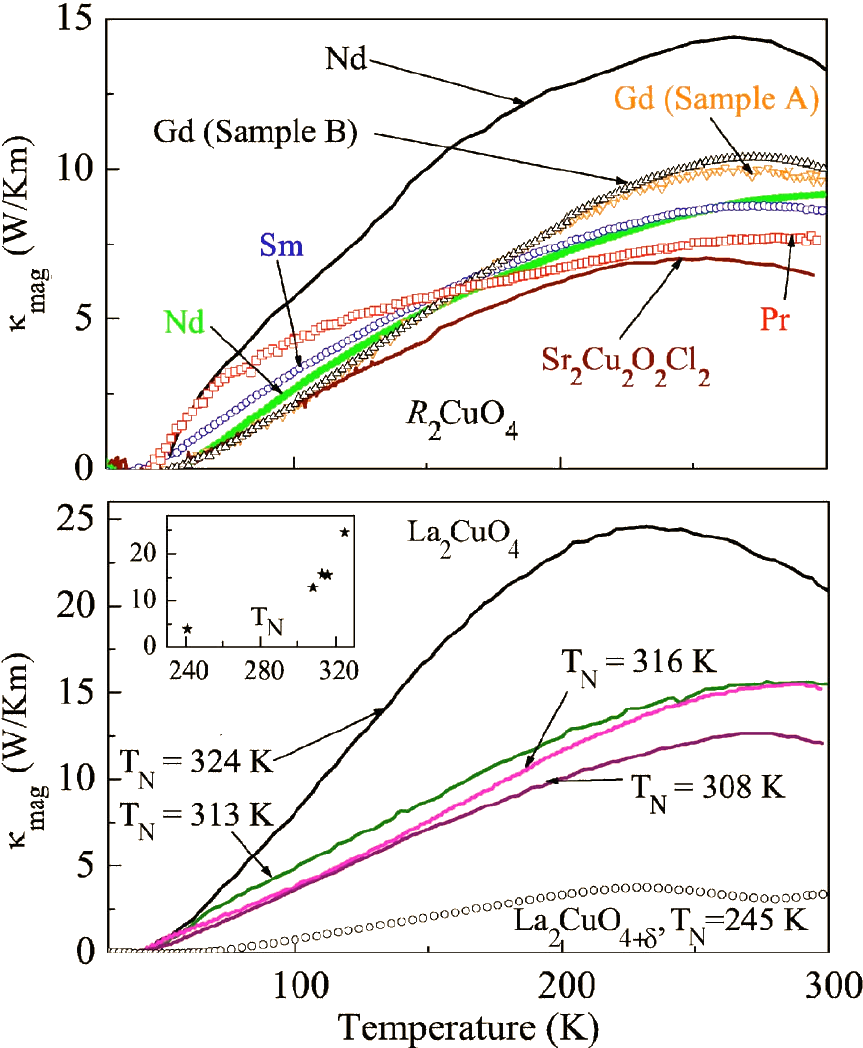}
\caption{Magnetic contributions to the in-plane thermal conductivity, calculated via $\kappa_\mathrm{mag}=\kappa_\Vert-\kappa_\mathrm{ph}$, where \kaph is
determined by a Callaway fit of the low-temperature maximum. Upper panel: Values calculated from measurements of \kappara of  $\mathrm{R_2CuO_4}$ \cite{Jin2003,Berggold2006}. Lower panel: The same analysis for
various data of \lacuod \cite{Nakamura91,Yan2003,Sun2003,Berggold2006}. Inset: The maximum of the calculated \kmag vs.
the N\'{e}el temperature of \lacuod.  It is worth to note, that for high excess oxygen contents a spatial phase separation into hole-rich and hole-poor regions occurs  in \lacuod at $T\lesssim290~\mathrm{K}$ \cite{Yu1996,Zolliker90}. This explains the very small \kmag and the dip at $\sim 275~\mathrm{K}$ for the sample with $T_N=245~\mathrm{K}$ shown in Fig.~\ref{fig:kmag_2D_Berggold2006} \cite{Berggold2006}.
Image adapted from \cite{Berggold2006}.}
\label{fig:kmag_2D_Berggold2006}
\end{figure}

Having qualitatively established that a sizable magnon contribution \kmag is present in the in-plane thermal conductivity of \lacuo, it is interesting to move one step further and to extract the temperature dependence of \kmag.  In order to estimate this temperature dependence, it is of paramount importance to determine its phononic part as accurately as possible. The first step in this estimation is to exploit that \kmag is expected to roughly follow $\kappa_\mathrm{mag}\propto T^2$ (see below) at low temperature  and therefore to become negligible in the temperature range of the phononic low-temperature peak. 
Thus one can estimate $\kappa_\Vert\approx\kappa_\mathrm{ph}$ at $T\lesssim40~\mathrm{K}$  and extrapolate $\kappa_\mathrm{ph}(T)$ towards higher temperature. One possible approach for this extrapolation is to use the Callaway model \cite{Callaway59} for fitting the low-temperature ($T\lesssim40~\mathrm{K}$) thermal conductivity and to use the thereby obtained fit parameters for the high-temperature extrapolation. This procedure has been applied to \srcuocl and $\mathrm{R_2CuO_4}$ ($\mathrm{R = La, Pr, Nd, Sm, Gd}$) \cite{Hofmann2003,Berggold2006}, see Fig.~\ref{fig:SrCuOCl_Hofmann2003} and Fig.~\ref{fig:kmag_2D_Berggold2006} for the fits and the resulting \kmag. 

For \lacuo the application of the Callaway model for extrapolating \kaph is bound to be error-prone due to the dominant importance of soft-phonon scattering which is difficult to incorporate into the model. The step-like change of \kapc of \leucuo at $T_\mathrm{LT}$ (Fig.~\ref{fig:leucuo}) tellingly demonstrates this complication. 
One possibility to overcome this problem is to use a phenomenological approach
the \textit{measured} purely phononic \kapperp perpendicular to the $\mathrm{CuO_2}$ planes and to exploit  the negligibly small \kmag in the temperature range of the phononic low-temperature peak. 
Thus one can estimate $\kappa_\Vert\approx\kappa_\mathrm{ph}$ at $T\lesssim40~\mathrm{K}$  and extrapolate $\kappa_\mathrm{ph}=A\cdot\kappa_\bot$ for higher temperatures, where $A$ is a suitable scaling factor. Fig.~\ref{fig:leucuo} shows examples of in this way estimated \kaph for the in-plane heat conductivity \kappara of \lacuo and \leucuo and the resulting \kmag \cite{Hess03}. 

As we shall see further below, the magnetic heat conductivity is well detectable also in polycrystalline samples. However, due to the polycrystalline nature of such samples, anisotropic information on $\kappa$ is averaged over. None of the above described approaches for estimating \kaph (and thus \kmag) can applied to such data in a simple way because on the one hand the application of the Callaway model to polycrystalline data requires further, non-clarified assumptions, and, on the other hand, for a polycrystal \kapperp can obviously not be obtained independently. 
Hess et al. therefore estimated the phonon contributions $\kappa_{\mathrm {ph}}^{\mathrm {poly}}$ by fitting
$\kappa$ at the high-temperature edge of its maximum (see Fig.~\ref{fig:laznkap}) by $\kappa_{\mathrm {ph}}=\alpha/T+\beta$ and by extrapolating this fit towards high temperature \cite{Hess03}. In turn, $\kappa_{\mathrm{mag}}^{\mathrm
{poly}}$ on the polycrystals is obtained by $\kappa_{\mathrm{mag}}^{\mathrm
{poly}}=\kappa-\kappa_{\mathrm {ph}}$. Note, that the measured $\kappa_{\mathrm{mag}}^{\mathrm {poly}}$ is smaller than the intrinsic $\kappa_{\mathrm{mag}}$ of these compounds by the factor of 2/3 due to averaging over all three
components of the $\kappa$ tensor \cite{Hess03}.

\subsection{Analysis of the magnetic heat conductivity}

\subsubsection*{Kinetic model}
The application \cite{Hess03,Hess2007b} of the kinetic model (Eq.~\ref{eq:kinetic_general}) to the 2D thermal conductivity of a single magnon
dispersion branch (labeled by $i$) yields 

\begin{equation}
 \tilde{\kappa}^i=\frac{1}{2}\frac{1}{(2\pi)^2}\int
v_{\bf k}l_{\bf k} \frac{d}{dT}n_{\bf k}\epsilon_{\bf k}d{\bf k} ,
\end{equation}

with $v_{\bf k}$, $l_{\bf k}$, $n_{\bf k}$ and $\epsilon_{\bf k}$ the velocity, mean free path, Bose-function and energy of a magnon, respectively. Note that $\kappa_{\mathrm{mag}}^i$ of a three-dimensional ensemble of planes, as realized in \lacuo, results from the multiplication of $\tilde{\kappa}^i$ with the number of planes per unit length, i.e., $\kappa_{\mathrm{mag}}^i=\frac{2}{c}\tilde{\kappa}^i$, where $c=13.2~\text{\rm\AA}$ is the lattice constant of \lacuo perpendicular to the planes. Then the total $\kappa_{\mathrm{mag}}$ is given by summing up $\kappa_{\mathrm{mag}}^i$ of each magnon branch.

In order to calculate $\kappa_{\mathrm{mag}}^i$ one can approximate the magnon dispersion relation
$\epsilon_{\bf k}$ of  the two branches $i=1,2$  with the 2D-isotropic expression

\begin{equation}
\epsilon_{\bf k}=\epsilon_k=\sqrt{\Delta^2_i+(\hbar v_0k)^2} , 
\end{equation}

which describes the dispersion observed experimentally \cite{Keimer93,Coldea01} for
small values of $k$. Here, $v_0$ is the spin wave velocity while $\Delta_1$ and $\Delta_2$ denote
the spin gaps of each magnon branch. 
% \begin{widetext}
Assuming a momentum independent mean free path, i.e., $l_{\bf k}\equiv
l_{\mathrm{mag}}$ one finds for each magnon branch \cite{Hess03,Hess2007b},

\begin{equation}\label{eq:fit2d}
\kappa^i_\mathrm{mag}=\frac{{k_B^3T^2l_\mathrm{mag}}}{2\pi\hbar^2v_0c}
\int_\frac{\Delta_i}{k_BT}^\infty x^2\sqrt{x^2-x_{\mathrm{0,i}}^2}\frac{e^x}{(e^x-1)^2} \, dx,
\end{equation}

with the spin wave velocity $v_0\approx1.287\cdot10^5~\mathrm{m/s}$ \cite{Hayden91a}. 
The integral is temperature dependent via its lower boundary $x_{\mathrm{0,i}}=\Delta_i/(k_BT)$, where $\Delta_{1}/k_B\approx26~\mathrm{K}$ and $\Delta_{2}/k_B\approx58~\mathrm{K}$ \cite{Keimer93}. However, the temperature dependence is weak in the $T$-range where the experimental data are discussed and thus one roughly has $\kappa_{\mathrm{mag}}\propto T^2$, if one presumes \lmag to be temperature independent.

\subsubsection*{Low-temperature characteristics}
Eq.~\ref{eq:fit2d} has been used to fit the \kmag data of \lacuo and \leucuo shown in Fig.~\ref{fig:leucuo} assuming a \textit{temperature-independent} magnon mean free path $l_{\mathrm{mag}}$  \cite{Hess03}. In this procedure, an additive shift of the $\kappa_{\mathrm{mag}}$-curve was allowed for which yields a further free parameter apart from $l_{\mathrm{mag}}$
and accounts for the aforementioned uncertainties in the magnitude of $\kappa_{\mathrm{mag}}$.
As can be seen in the figure, for both compounds satisfactory fits (solid lines) were obtained at intermediate temperature ranges (70-158~{\rm K} for \lacuo and 54-131~{\rm K} for \leucuo, \cite{Hess03}). While the slight deviations between the fitted and experimental data towards low $T$ are due to the uncertainties in $\kappa_{\bot,\mathrm{ph}}$ in this range, the deviations at high temperature can be understood in terms of $l_{\mathrm{mag}}$ becoming temperature-dependent due to enhanced magnon scattering off magnons or phonons. 

Notably, the data are consistent with a temperature-independent $l_{\mathrm{mag}}$ for $T$ both, within the fit interval
and below, indicating that in this range the mentioned temperature-dependent scattering processes (magnon-magnon
scattering, magnon-phonon scattering, or effects of a finite spin-correlation length at $T>T_N$) are frozen out and thus may be discarded. Therefore, relevant processes seem
to be sample-boundary scattering or scattering off defects within the $\mathrm{CuO}_2$-planes. The analysis yields
$l_{\mathrm{mag}}\approx1160~\text{\rm\AA}$ and $l_{\mathrm{mag}}\approx560~\text{\rm\AA}$ for \leucuo and \lacuo, respectively. Since these values are far too small to
correspond to the crystal dimensions, which are of the order of millimeters, these values naturally can be interpreted as the size of two-dimensional defect-free grains. Thus, magnon-defect scattering is the most likely scattering process which dominates the low-temperature magnon transport.
This conclusion is consistent with the fact that $\kappa_{\mathrm{mag}}$ and
$l_{\mathrm{mag}}$ are quantitatively different for \leucuo and \lacuo: since the
magnetic properties of both compounds are expected to be identical in essence (i.e., the same spin wave velocity $v_0$), unequal
$\kappa_{\mathrm{mag}}$ can arise only due to a difference in densities of the
magnetic defects that restrict $l_{\mathrm{mag}}$.

\begin{figure}
\centering
\includegraphics[width=0.7\textwidth]{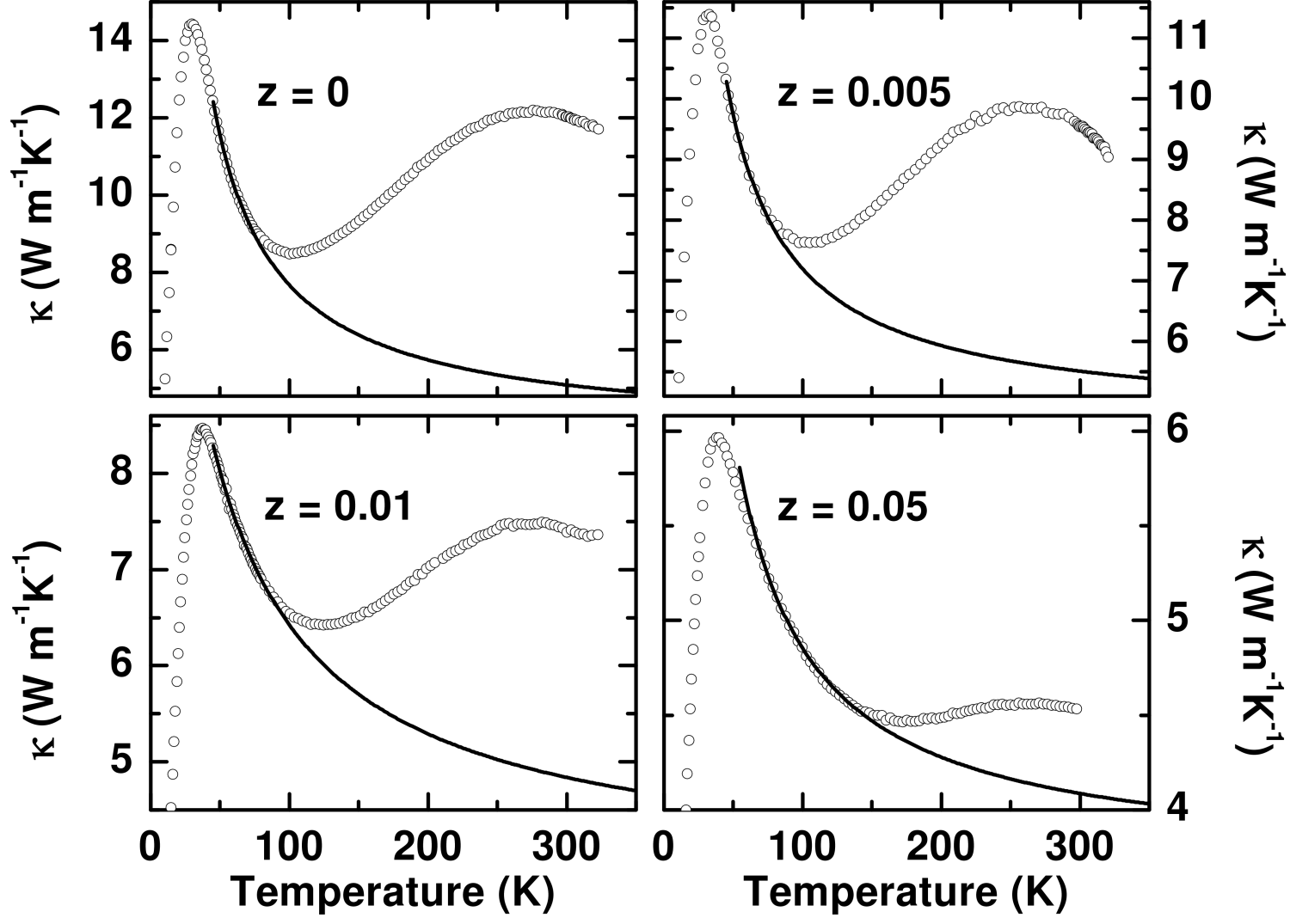}
\caption{Open circles: Thermal conductivity $\kappa$ of \lazn
polycrystals ($z=0$, 0.005, 0.01, 0.05) as a function of $T$. Solid lines: extrapolated
phonon heat conduction \kaph assuming $\kappa_{\mathrm {ph}}=\alpha/T+\beta$. Similar thermal conductivity data have been obtained on single crystalline \lazn by Sun et al. \cite{Sun2003} in qualitative agreement with the shown data. Image reproduced from \cite{Hess03}.}
\label{fig:laznkap}
\end{figure}

\subsubsection*{Magnon-defect scattering}
There is no clear alternative way to measure the defect density of a given crystal directly, in order to allow a quantitative comparison of the magnon mean free path \lmag with the defect density (or the defect distance). Hess et al. have therefore performed measurements of $\kappa$ on samples with a well-defined density of magnetic defects. Such defects can be induced in \lacuo, e.g., by substituting a small amount of non-magnetic Zn$^{2+}$-ions for the magnetic Cu$^{2+}$-ions. Representative results obtained for such samples of \lazn are shown in 
Fig. \ref{fig:laznkap}. The Zn-impurities represent both structural and magnetic impurities and therefore should affect the phononic as well as the magnetic peak in the heat conductivity. In fact, as can be inferred from the figure, the Zn-doping leads indeed to a gradual suppression of both, the phonon as well as the magnon contribution to $\kappa$. The analysis of the data for \lazn using Eq. \ref{eq:fit2d} in an analogous way as in the undoped case\footnote{It is worth mentioning that a slight reduction of the spin wave velocity \cite{Brenig91} and changes of the spin gaps induced by the Zn-ions can be safely ignored. These effects lead to corrections smaller than the experimental error.} yields the interesting result for the low-temperature magnon mean free path that it scales linearly with the reciprocal Zn content \cite{Hess03}: $l_{\mathrm{mag}}\approx 0.74\cdot a/z$ (with the lattice constant $a$).
Hence, these result suggests that \lmag is about equal to the unidirectional distance between the Zn-ions within the $\mathrm{CuO_2}$ planes, i.e., \kmag can be directly used to measure these distances.

\begin{figure}
\centering
\includegraphics[width=\textwidth]{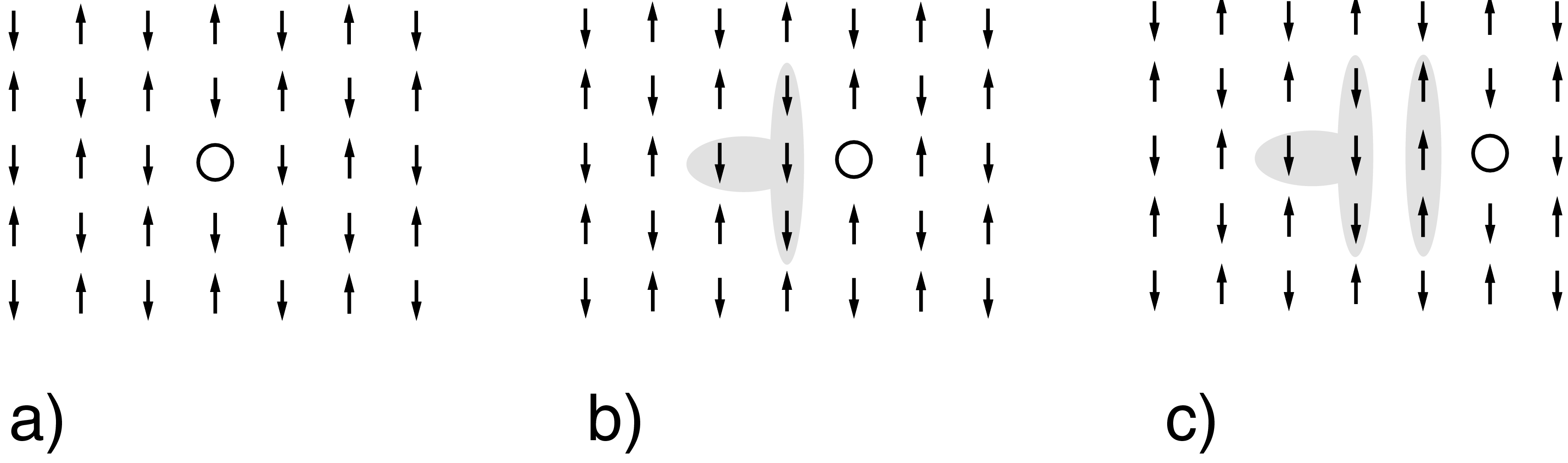}
\caption{Sketch of a hole moving in an 2D-antiferromagnet. a) A hole in an antiferromagnet. b) Antiferromagnet after hopping of the hole to a neighboring site. c) Antiferromagnet after two hopping processes. The gray shaded areas highlight the regions of destroyed antiferromagnetic correlation.}
\label{fig:spinloch}
\end{figure}

It is important to note that the nature of the doped defects plays a decisive role in the effective strength of the magnon-defect scattering. In particular, \textit{mobile} defects can be induced in \lacuo by hole-carrier doping. Such doping can be achieved either by substituting Sr for La  or by enhancing the oxygen content of \lacuo, which at  doping levels beyond about 5\% leads to high-temperature superconductivity. Interestingly, in the case of Sr-doping, one observes a severe suppression of \kmag already at $\sim1\%$ doping level \cite{Sun2003,Hess2007a}, i.e. at much lower doping levels as compared to the afore-discussed Zn-doping. The different effect of these two doping schemes is consistent with the different impact of the defects on the antiferromagnetic correlations in the 2D-HAF. The doped charge strongly couples to the spin excitations resulting in a dressed quasiparticle (the spin polaron) which moves through the antiferromagnetic background. This motion of the spin polaron very effectively destroy the antiferromagnetic correlation of the spins in the plane (see Fig.~\ref{fig:spinloch} for an illustration) \cite{Martinez1991,Chernyshev1999}, whereas static non-magnetic defects (which can be viewed as immobile holes, see Fig.~\ref{fig:spinloch}a) only dilute the antiferromagnet and lead to a very moderate reduction of the spin correlation \cite{Brenig91,Uchinokura95,Hucker99a}. The strong suppression of \kmag already at $\sim1\%$ hole concentration suggests that the magnon mean free path amounts only a few lattice spacings, i.e. the effective 'size' of disturbed antiferromagnet correlation due to the movement of a single hole is very large.

Berggold et al. pointed out that the magnitude of the peak in \kmag of \lacuo significantly varies throughout the literature (see Fig.~\ref{fig:kmag_2D_Berggold2006}, bottom panel) and suggested that the difference arises due to different levels of excess oxygen in the material \cite{Berggold2006}. More specifically, they pointed out that the reported maximum \kmag depends monotonically on the N\'eel temperature $T_N$ (inset of Fig.~\ref{fig:kmag_2D_Berggold2006}). This relation is plausible since at small hole doping levels $T_N$ is known to sensitively depend on the hole content not only for the case of Sr doping \cite{Hucker99a} but also for excess oxygen \cite{Chen1991}. However, one bear in mind that, as has been shown for the case of \lazn, a reduced \kmag can also be caused by chemical impurities through limiting the mean free path, yet having little effect on $T_N$, in particular, in very clean crystals. 
Furthermore, it is natural to expect that structural defects apart from chemical impurities play a crucial role, i.e. the degree of crystal perfection should have an impact on the size of the two-dimensional grains, which limit the mean free path, as well.
This notion is confirmed by recent data for the heat conductivity of ultrapure \lacuo with  $T_N=325~\mathrm{K}$ and \kmag up to $\sim40$~\wkm \cite{Mohanunpublished}, i.e., a much higher value than those collected in Fig.~\ref{fig:kmag_2D_Berggold2006}.

% \textcolor{red}
\section{Two-leg spin ladders}

\subsection{Materials aspects}\label{matpropladder}
\begin{figure}[h]
\begin{center}
\includegraphics[width=\textwidth,clip]{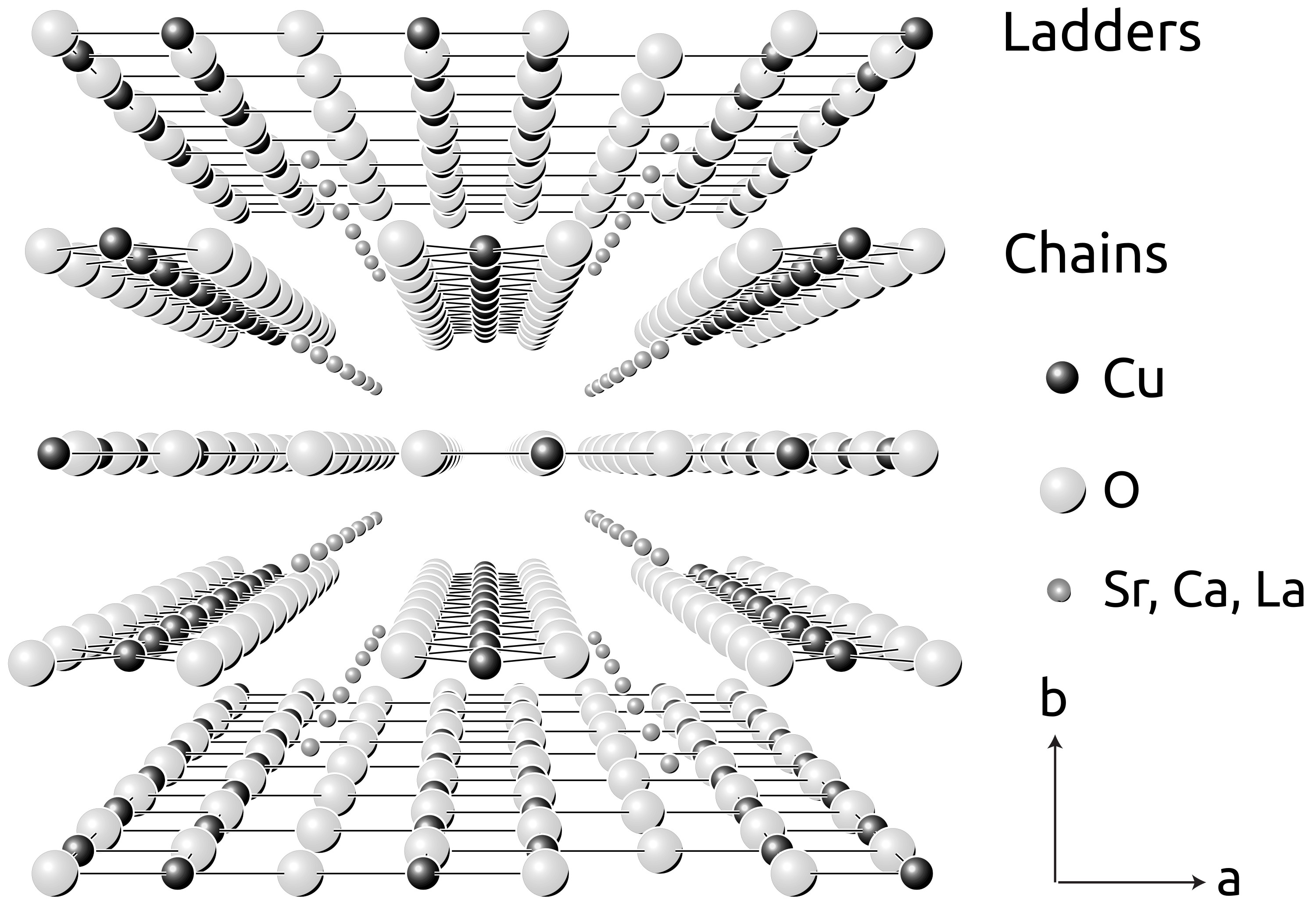}
\end{center} \caption{\label{strukleit1} Three-dimensional representation of the crystal structure of \srcala with a view in the direction of the CuO$_3$-ladders and CuO$_2$-chains ($c$-axis). From
\cite{Ammerahldiss}.}
\end{figure}

\begin{figure}[t]
\begin{center}
\includegraphics[width=\textwidth,clip]{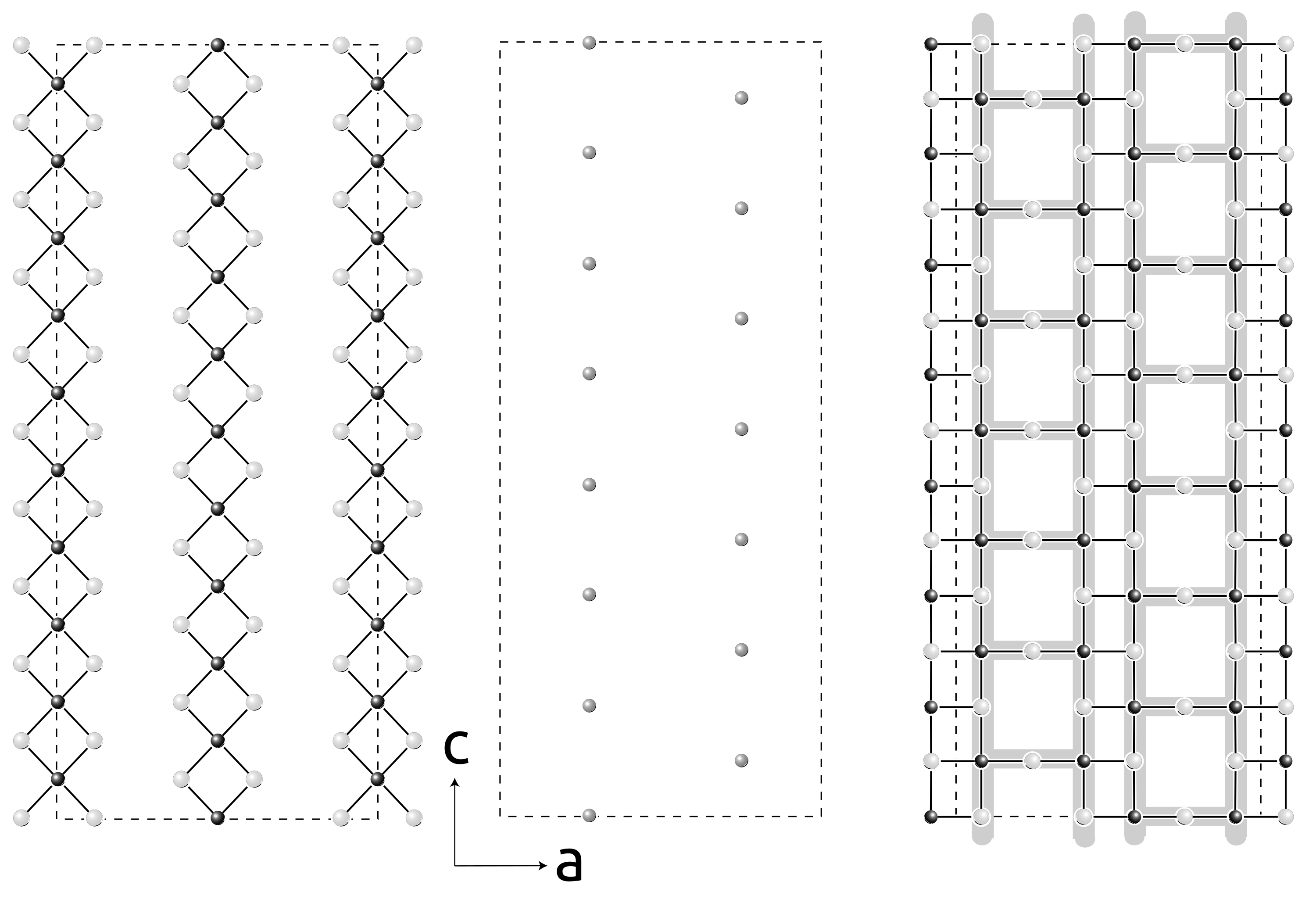}
\end{center} \caption{\label{strukleit2} Structural elements of \srcala. Left: Plane of CuO$_2$ spin chains. Center: layer of (Sr, Ca, La). Right: Plane of Cu$_2$O$_3$ spin ladders with the structure of two separate two-leg ladders indicated in grey. The dashed line indicates the size of the in-plane average unit cell. Dark and light grey bullets represent the Cu and O sites, respectively. From \cite{Ammerahldiss}.}
\end{figure}

Magnetic thermal transport in a cuprate-based two-leg spin ladder system up to present primarily has been observed in the material \srcala. The crystal structure (see Fig.~\ref{strukleit1}) and also the connected physical properties of this material are significantly more complex as compared to all materials discussed in this review and thus deserve a special attention. 
\srcala is composed of two main structural elements which define the physical properties. These are, on the one hand, Cu$_2$O$_3$ planes which realize $S=1/2$ two-leg spin ladders of the type as described in Section~\ref{intromat}. Each of these planes is formed by a parallel network of individual two-leg ladders in the $ac$-plane (see right panel of Fig.~\ref{strukleit2}). Each Cu spin (with $S=1/2$) interacts strongly antiferromagnetically via 180$^\circ$ Cu-O-Cu bonds with its two neighboring Cu spins along the ladder legs (leg interaction $J_\Vert$, parallel to the $c$-axis) and also antiferromagnetically with its one neighboring Cu spin on the same rung within one two-leg ladder (rung interaction $J_\bot$, parallel to the $a$-axis). These interactions result in an effective $S=1/2$ two-leg ladder model as described by Eq.~\ref{Hamleit} with  $J_{i,j}=J_\Vert$ along the legs and $J_{i,j}=J_\bot$ along the rungs for each of the individual ladders. The interaction of the spins of one ladder to its neighboring ones is strongly frustrated and thus causes an effective decoupling of the individual ladders from each other \cite{Gopalan94}. The frustration arises since each Cu spin of one ladder interacts relatively weakly ferromagnetically via 90$^\circ$ Cu-O-Cu bonds with two  strongly antiferromagnetically coupled Cu-spins (interaction $J_\Vert$) of the neighboring ladder.


On the other hand, the structure contains planes (again in the $ac$-direction) of CuO$_2$ chains of edge-shared CuO$_4$ plaquettes with the main exchange path of two neighboring Cu spins within one chain via the approximately 90$^\circ$ Cu-O-Cu bonds, running along the $c$-axis (see left panel of Fig.~\ref{strukleit2}). The resulting magnetic interaction is about one order of magnitude smaller than those of the ladder and have been shown not to directly contribute to the magnetic heat transport \cite{Hess01}. 

The total crystal structure is composed of an alternating stacking of the ladder and chain planes along the $b$-axis where adjacent planes are separated by layers (Sr, Ca, La) ions (see Fig.~\ref{strukleit1} and central panel of Fig.~\ref{strukleit2}).
The chains and the ladders possess along the $c$-axis different translational periods and thus form two incommensurate sublattices. Together with the (Sr, Ca, La) ions, the ladders constitute a sublattice unit cell with $a\approx 11.3\dots11.5$~\AA, $b\approx 12.5\dots13.4$~\AA, and $c_L\approx 3.9$~\AA. 
The sublattice unit cell of the chains has the same $a$ and $b$ lattice constants but is somewhat shorter along the chain direction with $c_C\approx2.75$~\AA. The almost commensurable average common unit cell is $c=7\times c_L\approx10\times c_C$. It should be noted, that the exact values of the lattice constants depend significantly on the relative ratio of the Sr, Ca, and La constituents \cite{Ammerahldiss}.

% \subsubsection{Electronic and magnetic properties of \srcala}
Interestingly,  the formal valencies of the stoichiometric compound are (Sr$^{2+}$)$_{14}$(Cu$^{2.25+})_{24}$(O$^{2-}$)$_{41}$. This means, the magnetically active ladder and chain structures are intrinsically hole-doped. These holes are not equally distributed among both structures. On the one hand, about one hole per formula unit is located in the ladders \cite{Osafune97,Nucker00}, whereas the amount in the chains is estimated to about 6 holes per formula unit. The intrinsically doped holes have been shown to undergo a charge ordered state in both the chains and in the ladders. In the former, the ordered holes are found to yield a state of non-interacting spin dimer singlets with an excitation gap of about 130~K \cite{Ammerahl00,Matsuda99,Regnault99,Klingeler2005,Klingeler2006}, whereas the holes in the ladders form a long-range ordered hole crystal \cite{Abbamonte2004,Rusydi2006}. We shall see further below that this charge ordering has profound impact on the magnetic heat transport properties.

The isovalent substitution of Ca for Sr in \srcuoladder has strong consequences for the physical properties. The charge transport properties gradually develop from a semiconductor-like temperature dependence for \srca at $x=0$ to an increasingly metallic characteristics at maximal $x\approx12$ \cite{Hess04a,Vuletic2006}. Remarkably, at $x\gtrsim5$ the charge ordered state collapses \cite{Matsuda99,Ammerahl00,Kataev01} and at very high Ca doping levels, superconductivity occurs at high pressure \cite{Uehara96,Isobe98}. The increasing metallic nature upon Ca doping commonly is interpreted as a result of a doping induced transfer of holes from the chains into the ladder substructures \cite{Osafune97,Nucker00,Dagotto99}.

Alternatively, the divalent Sr and Ca ions can be substituted by trivalent La, which causes a drastic reduction of the total hole content in (Sr,Ca)$_{14-y}$La$_y$Cu$_{24}$O$_{41}$ with increasing $y$. Thus, at nominally $y=6$ the system reaches the realization of 'hole-free' ladder and chain subsystems. It should be noted, however, that the highest La content at which phase pure single crystals can be achieved is $y\approx5.2$ \cite{Ammerahl99,Ammerahldiss}. Quite importantly, already a moderate La-content of $y\approx2$ yields practically 'hole-free' spin ladders
\cite{Nucker00}. A further consequence of the La-doping is a rapid destruction of the dimer ground state in the chains and the formation of an antiferromagnetic ground state of the chains at $y\gtrsim4$ 
\cite{Ammerahldiss,Matsuda96,Matsuda98,Matsuda00b,Kumagai00,Ammerahl00a}.

The hole-free ladders in (Sr,Ca)$_{14-y}$La$_y$Cu$_{24}$O$_{41}$ at finite $y$ exhibit the theoretically expected properties of an $S=1/2$ two-leg quantum spin ladder in a remarkable way. Inelastic neutron scattering (INS) and optical spectroscopy experiments yield a large $J_\Vert/k_B\approx 1300\dots2200$~K and $J_\bot/k_B\approx1300\dots1440$~K with a gapped ground state and a large triplet excitation gap $\Delta\approx 310\dots410$~K \cite{Matsuda00b,Windt01,Notbohm2007}. For \srcuoladder, despite the finite amount of doped holes, the magnetism is barely different \cite{Eccleston98,Deng2013}, whereas for \srca controversial report exist concerning the spin gap $\Delta$. INS results indicate that $\Delta$ remains practically unchanged up to the highest Ca-content \cite{Deng2013}, whereas findings from nuclear magnetic resonance (NMR) experiments suggest a gradual decrease of $\Delta$ \cite{Kumagai97,Takigawa98,Imai98,Magishi98}. It should be mentioned that the NMR findings suggest a somewhat larger $\Delta/k_B$ for $x=0$ (up to $\sim650$~K) than those from INS.

\subsection{Qualitative proof of magnon heat transport in the spin ladders}

\label{sec:ladders}
\begin{figure}
\centering
\includegraphics[width=0.6\textwidth]{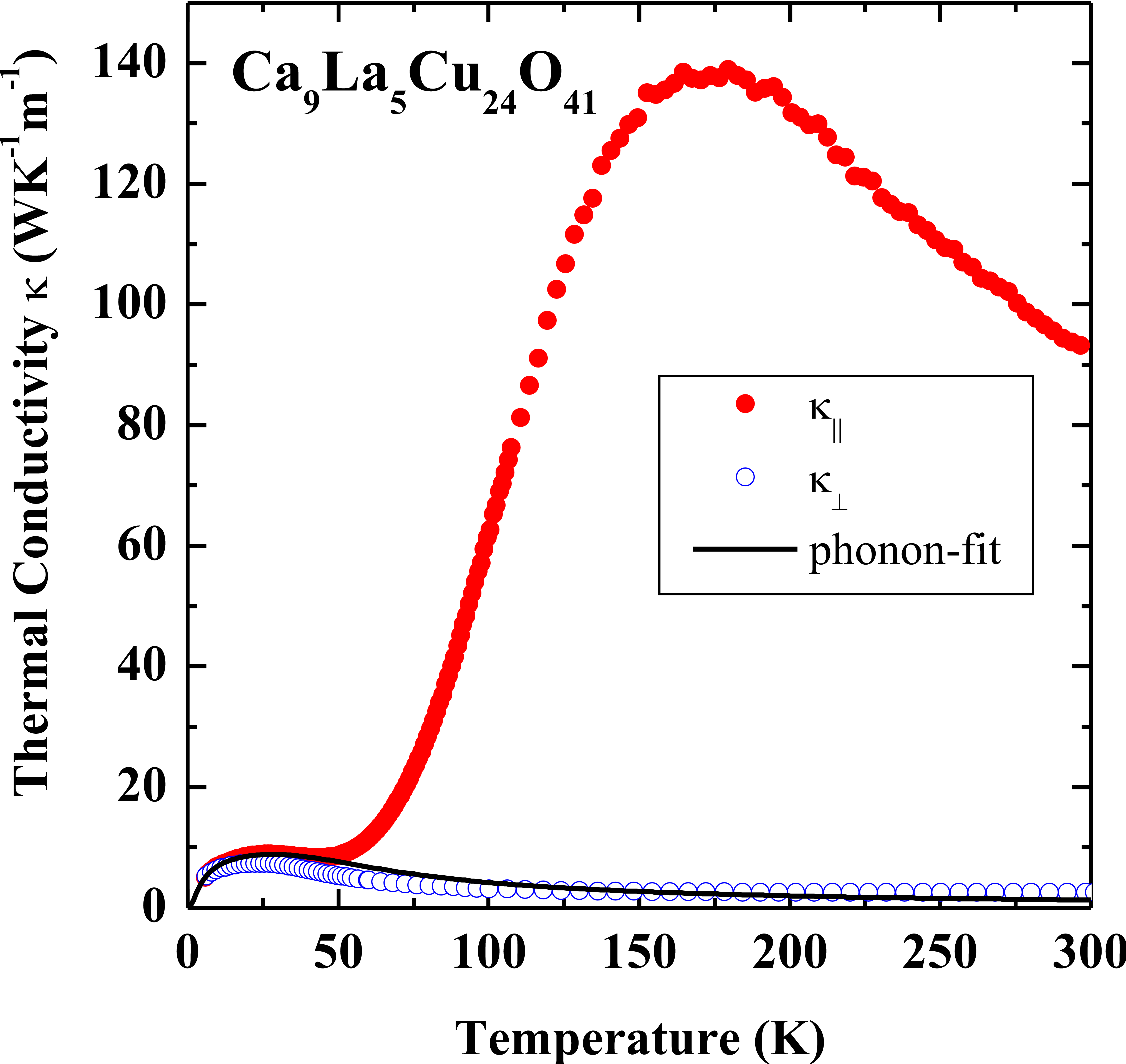}
\caption{Thermal conductivity of \laf as a function of temperature measured along the $a$ and $c$ axes, \kapperp and \kappara, respectively. The solid line represents an estimate of the phonon contribution to \kapc based on the Callaway model. Figure adapted from \cite{Hess01}.}
\label{fig:La5_kap}
\end{figure}

Kudo et al. were the first to report about unconventional magnon heat transport  of the spin ladders in the material \srca with the Ca content ranging from $x=0$ to $x=9$ \cite{Kudo1999}. The data presented in their pioneering study are qualitatively compatible with the data shown in Figures~\ref{fig:signatures}, \ref{fig:La5_kap}, and~\ref{fig:SrCa_kmag} and also with data by Sologubenko et al. \cite{Sologubenko00}. In the following, instead of addressing these early works on first-generation single crystals of this material in detail, we focus on more recent work which allows better to carve out the relevant physics.

The data shown in Fig.~\ref{fig:signatures} for the pristine material \srcuoladder are qualitatively similar to the afore-discussed case of \lacuo with the remarkable difference that the high-temperature anomaly in \kappara, which is measured parallel to the spin ladders along the $c$-axis, is purely one-dimensional, since it is absent in both directions perpendicular to the ladders \cite{Hess01}. The magnetic nature of the anomaly in \kappara is thus evident, the more so as no structural instabilities as is the case for the two-dimensional layered cuprates are known to be present in this material. However, for the sake of meticulous correctness, one should exclude other thinkable one-dimensional contributions to the heat transport parallel to the ladder structures in the material. In particular, one could imagine unusual optical phonon modes related to the quasi-one-dimensional structural elements present in the materials as well magnetic heat transport by the also  present $\mathrm{CuO}_2$-chain structure in \srcala.

Fig.~\ref{fig:La5_kap} shows the thermal conductivity parallel (\kappara) and perpendicular (\kapperp) to the spin ladder structures in the compound \laf \cite{Hess01}. In this material, the $\mathrm{Sr^{2+}}$ site of the pristine compound \srcuoladder is substituted by two different ions, namely $\mathrm{Ca^{2+}}$ and $\mathrm{La^{3+}}$. The resulting structural disorder leads to a strong defect-scattering of the phonons in the system and, correspondingly, to a drastic suppression of its phonon heat conductivity \kaph, which is evident from a direct comparison of \kapperp of \laf and \srcuoladder in Fig.~\ref{fig:La5_kap} and Fig.~\ref{fig:signatures}, respectively. Interestingly, the high-temperature peak in \kappara is not affected by this suppression. In contrast, in the direct comparison, this peak is even enhanced in \laf.
We shall see further below that this enhancement can be attributed to the vanishing hole content in the ladders of \laf.
% Note that the value of \kappara at room temperature is with almost 100~\wkm remarkably large and almost comparable to the heat conductivity of a metal.

\begin{figure}
\centering
\includegraphics[width=\textwidth]{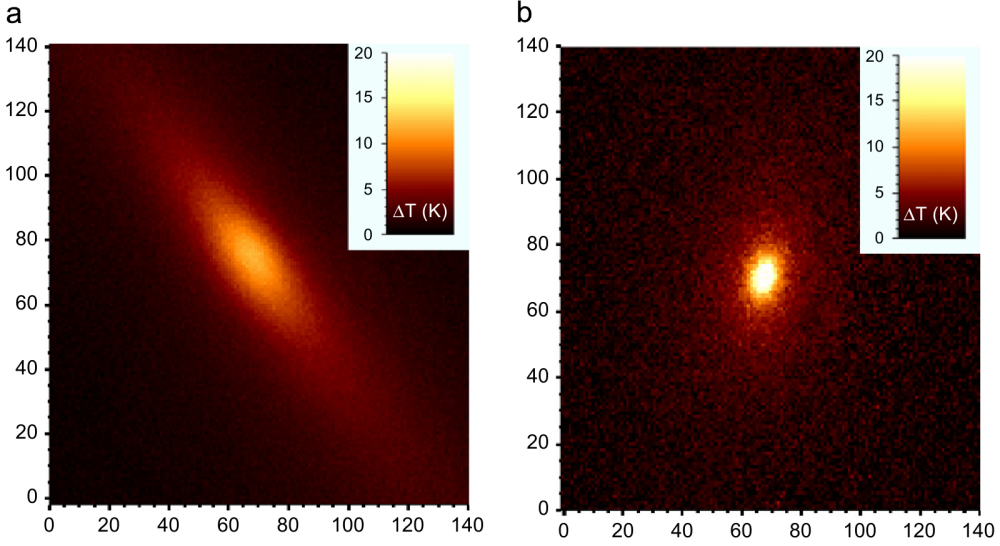}
\caption{a) Fluorescent microthermal image of the $ac$-plane of \laf generated by a localized heat pulse in the center of the image. It shows a highly anisotropic pattern due to the high thermal conductivity in the (diagonal) c-direction (\kappara). b) Fluorescent microthermal image of the $ab$-plane of \laf generated by a localized heat pulse in the center of the image shows an isotropic pattern. The distances on both x- and y-axis are in mm. Figure adapted from \cite{Otter2009}.}
\label{fig:Otter}
\end{figure}

This observation clearly rules out that the high-temperature peak stems from unconventional phononic modes since the suppressed \kaph demonstrates clearly that these modes are heavily disturbed by the disorder in the material. Furthermore, also thinkable magnetic heat transport of the $\mathrm{CuO_2}$ chains can now safely be excluded because the magnetic state of these chains is fundamentally changed upon the substitution of La and Ca, for Sr \cite{Ammerahl00a,Klingeler2006,Klingeler2005,Matsuda98}.
Thus, the high-temperature peak observed in \kappara of \srcala unambiguously represents the magnon heat conductivity \kmag in the material. Furthermore, and remarkably, one finds the unusual materials property that $\kappa_\mathrm{ph}\ll\kappa_\mathrm{mag}$ at $T\gtrsim40~\mathrm{K}$ which is most evident for \laf.
This remarkable anisotropy of the heat conductivity tensor has recently been visualized by fluorescent microthermal imaging at room temperature, see Fig.~\ref{fig:Otter} \cite{Otter2009}.

\subsection{Extraction of the magnetic heat conductivity}

\begin{figure}
\centering
\includegraphics[width=0.6\textwidth]{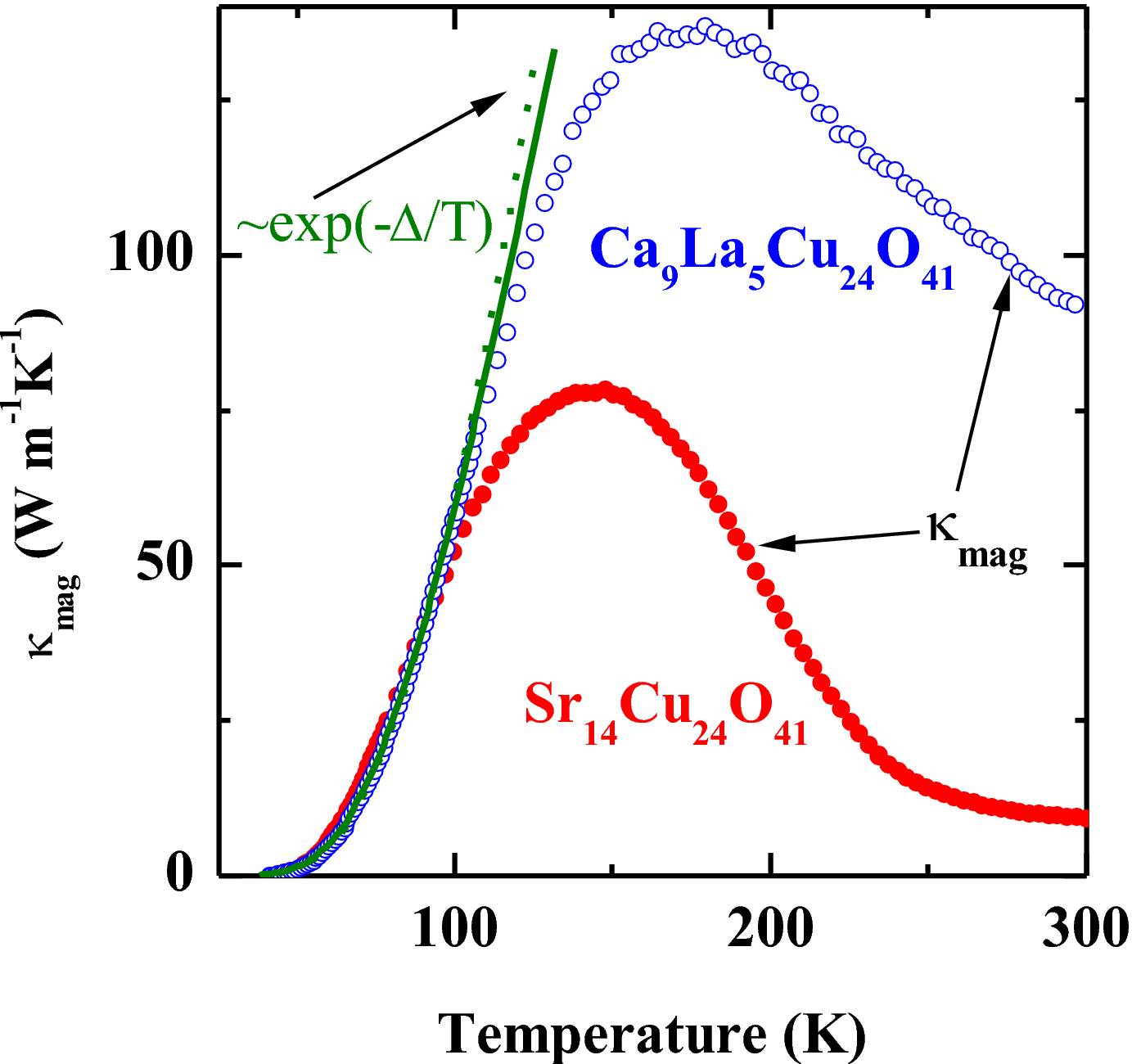}
\caption{Temperature dependence of \kmag of the ladder compounds \srcuoladder and \laf. The solid line represent a fit to the low-temperature data of \kmag of \laf on basis of the kinetic model (Eq.~\ref{eq:kappaladder}). Figure adapted from \cite{Hess01}. Similar data for \kmag of \srcuoladder have been found by Sologubenko et al. \cite{Sologubenko00}.}
\label{fig:kmag_59_sr14}
\end{figure}

Due to the presence of a sizable spin gap in \srcala of the order of $\Delta/k_B\sim300\dots500~\mathrm{K}$ \cite{Kumagai97,Eccleston98,Imai98,Matsuda00b,Notbohm2007} one can safely assume that for low temperatures the magnon heat conductivity falls off as $\kappa_\mathrm{mag}\propto \exp[-\Delta/(k_BT)]$ and becomes negligible at  $T\lesssim40~\mathrm{K}$ in comparison to the phonon heat conductivity \kaph. Thus, one can safely fit \kaph using the Callaway model \cite{Callaway59} in this temperature range and extrapolate this fit towards higher temperature in order to estimate the phononic contribution to \kappara \cite{Sologubenko00,Hess01}. The solid lines in Fig.~\ref{fig:signatures} and Fig.~\ref{fig:La5_kap} represent corresponding results for \srcuoladder and \laf, respectively. Note, that the extreme suppression of \kaph in \laf renders the possible errors in the estimation of \kaph (and thus \kmag) very small, which is particularly important at low temperature where \kmag is still small.
Assuming again $\kappa_\Vert=\kappa_\mathrm{ph}+\kappa_\mathrm{mag}$, the magnon heat conductivity data can now be computed from the measured data. The resulting temperature dependence of \kmag for \srcuoladder and \laf is depicted in Fig.~\ref{fig:kmag_59_sr14} \cite{Hess01}. 

\subsection{Analysis of the magnon heat conductivity}
The temperature dependence of the shown data is for both compounds very similar in the region of the low-temperature increase of \kmag. Differences become apparent only at $T\gtrsim 100~\mathrm{K}$: The increase  of \kmag of \srcuoladder weakens and after reaching a maximum ($\sim80$~\wkm) at $\sim140~\mathrm{K}$, \kmag decreases strongly and saturates at $\sim10$~\wkm for $T\gtrsim240~\mathrm{K}$. In contrast, \kmag of \laf increases much stronger at $T\gtrsim 100~\mathrm{K}$. Also here a maximum value is reached ($\sim140$~\wkm), but at clearly higher temperature $T\approx180~\mathrm{K}$, after which \kmag decreases only moderately and stays large at $\sim90$~\wkm even at room temperature, i.e. it attains a value which is comparable to the heat conductivity of a metal.

\subsubsection*{Kinetic model}

In the one-dimensional case, the kinetic model (Eq.~\ref{eq:kinetic_general}) for the thermal conductivity of a single ladder yields \cite{Sologubenko00,Sologubenko01,Hess01}

\begin{equation}
 \tilde{\kappa}=\frac{1}{2\pi}\int v_{k}l_{k} \frac{d}{dT}n_{k}\epsilon_{k}dk ,
 \label{eq:kinetic_1D}
\end{equation}

again with $v_{k}$, $l_{k}$, and $\epsilon_{k}$ the velocity, mean free path, and energy of a magnon\footnote{It should be noted that the term 'magnon' refers to the specific $\Delta S=1$ triplet excitations of the $S=1/2$ two-leg quantum spin ladder, which sometimes are also dubbed as 'triplon' excitations. They are not be confused with the spin-wave type magnon excitations of the 2D-HAF.}. The mathematical form of the occupation function $n_{k}$ is, however, unclear for the case of a two-leg spin ladder. Based on the fact that the elementary excitations of a two-leg ladder are triplet excitations, i.e. bosons, Sologubenko et al. approximated it by the Bose-function \cite{Sologubenko00}. However, Hess et al. argued that the Bose function leads to unphysically large triplet densities at higher temperatures and suggested an occupation function of the form

\begin{equation}\label{eq:triplets}
n_k=\frac{3}{e^{\frac{\epsilon_k}{k_BT}}+3}
\end{equation}

to account, on average, for the hard-core constraint of no on-site double occupancy for the triplet
excitations \cite{Hess01}. The use of it and assuming $l_\mathrm{mag}\equiv l_{\bf k}$ leads then to the expression \cite{Hess01,Hess2007b}
\begin{equation}\label{eq:kappaladder}
\kappa_\mathrm{mag}=\frac{3 n_s {k_B}^2 }{\pi\hbar}l_\mathrm{mag}T \int_\frac{\Delta}{k_BT}^\frac{\epsilon_\mathrm{max}}{k_BT}
x^2 \frac{e^x}{(e^x+3)^2}~dx,
\end{equation}
where $n_s$ is now the number of ladders per unit area. Note, that Eq.~\ref{eq:kappaladder}) differs from the
expression used by Sologubenko {\em et al.}~\cite{Sologubenko00} for the heat-conductivity of one-dimensional bosons
not only by the distribution function~(\ref{eq:triplets}) but also by an overall factor of three accounting for the
triplet degeneracy. Thus, the following considerations and the discussion of analyzing the \kmag data for the two-leg ladder compounds will be based on Eq.\ref{eq:kappaladder}. Note further, that this result for a one-dimensional system does not depend on the specific form of the dispersion function $\epsilon_k$ and the velocity $v_k$ \cite{Hess01}. The dispersion enters only through its band minimum, i.e., the spin gap $\Delta$ and its maximum $\epsilon_\mathrm{max}$. Experimentally, one finds a lower bound $\epsilon_\mathrm{max}\sim200$~meV in \srcala \cite{Eccleston98,Matsuda00b,Notbohm2007}. At $T\lesssim300~\mathrm{K}$, the integral in Eq.~\ref{eq:kappaladder} therefore does only depend weakly on $\epsilon_\mathrm{max}$. Its exact value thus does not play an important role and one might even set $\epsilon_\mathrm{max}=\infty$ at very low temperature \cite{Hess2007b}. In any case, due the presence of a spin gap \kmag is expected to be exponentially suppressed at low temperature $k_BT\ll \Delta<J$ and one might approximate $\kappa_\mathrm{mag}\propto \exp[-\Delta/(k_BT)]$ presuming a temperature independent \lmag.

\subsubsection*{Low-temperature increase -- thermal occupation of magnons}
Apparently, for both materials \kmag follows such an exponential increase at low temperature (see dotted line in Fig.~\ref{fig:kmag_59_sr14}. An even more accurate description of the data is obtained if the data is described by Eq.~\ref{eq:kappaladder} \cite{Hess01}. The solid line in the figure shows a corresponding fit to the data of \laf with a temperature independent \lmag and the spin gap $\Delta$ as free parameters in the temperature range 54-102~K. The data for \srcuoladder can be fitted similarly well but in the somewhat reduced temperature interval 61-91~K (see \cite{Hess01}, for the details).
In both cases, this analysis yields very similar values for the magnon mean free path \lmag and the spin gap. For the latter one finds $\Delta/k_B=418\pm15~\mathrm{K}$ and $\Delta/k_B=396\pm10~\mathrm{K}$ for \laf and \srcuoladder, respectively, which is in the same order of magnitude than results reported from neutron scattering and NMR measurements 
(see Section~\ref{matpropladder}) \cite{Kumagai97,Eccleston98,Imai98,Matsuda00b,Notbohm2007}.
The results for the low-temperature mean free path $l_0$ are almost identical, specifically $l_0=2980\pm110$~{\rm\AA} and $l_0=2890\pm230\rm$~{\rm\AA} for \laf and \srcuoladder, respectively.\footnote{Note, that $l_0$ specifically denotes the low-temperature value of \lmag used for fitting the low-temperature increase of \kmag. As we shall see later, \lmag becomes temperature dependent at higher temperature. Note further, that the slightly smaller values of $l_0$ and $\Delta$ for $\mathrm{Ca_9La_5Cu_{24}O_{41}}$ as compared to the values reported in the original paper by  \cite{Hess01}, are a consequence of the usage of more accurate lattice parameters and an optimized fit-interval. It is stressed that these small corrections have no further consequences on the conclusions drawn in the original paper.} This corresponds to about 750 lattice spacings along the ladder, which is surprisingly large in view of the rather complicated crystal structure of \srcala \cite{Gopalan94}, and, even more intriguing, 
in view of
the spin-spin correlation length in a two-leg spin ladder which exponentially vanishes already after about 3-4 lattice spacings  
\cite{Dagotto99,Dagotto96}. This means, the mean free path becomes orders of magnitude larger than the size of antiferromagnetically correlated spin configurations. This underpins the fundamentally different quantum nature of the magnons in the two-leg spin ladders as compared to antiferromagnetic spin wave type magnons which emerge from an antiferromagnetic ground state with infinite spin-spin correlation length, as is the case for the 2D-HAF.

\subsubsection*{Ballistic heat transport?}
The report of a large magnon thermal conductivity in \srcuoladder and \laf \cite{Sologubenko00,Hess01} and pertinent large mean free paths soon after triggered a number of controversial theoretical works which addressed the conjecture that the heat transport of a two-leg spin ladder similar to that of a Heisenberg chain was \textit{ballistic}, despite the non-integrability of the model (see also Section~\ref{sec:chains} for further elaboration on the connection between the integrability of the spin model and the thermal Drude weight) \cite{Alvarez02,Gros2004,Heidrich04,Zotos04,Jung06,Boulat2007,Steinigeweg2016}.
One particular aspect of this context has been addressed by Alvarez and Gros, who claimed that the heat transport of a two-leg ladder indeed was ballistic \cite{Alvarez02}. More specifically, based on exact diagonalization results they found a finite thermal Drude weight and evaluated a much shorter magnon mean free path as compared to the above findings through using the kinetic model. Theoretically, these findings have been challenged by several groups: Possible problems due to finite size effects of the exact diagonalization results have been pointed out \cite{Heidrich04}. Furthermore, high-temperature approximations yielded a vanishing Drude weight \cite{Zotos04}. It has been argued, however, that the two-leg ladder model despite being non-integrable is still close enough to integrability to yield finite but large transport coefficients \cite{Boulat2007,Jung06}.

\begin{figure}
\centering
\includegraphics[width=0.6\textwidth]{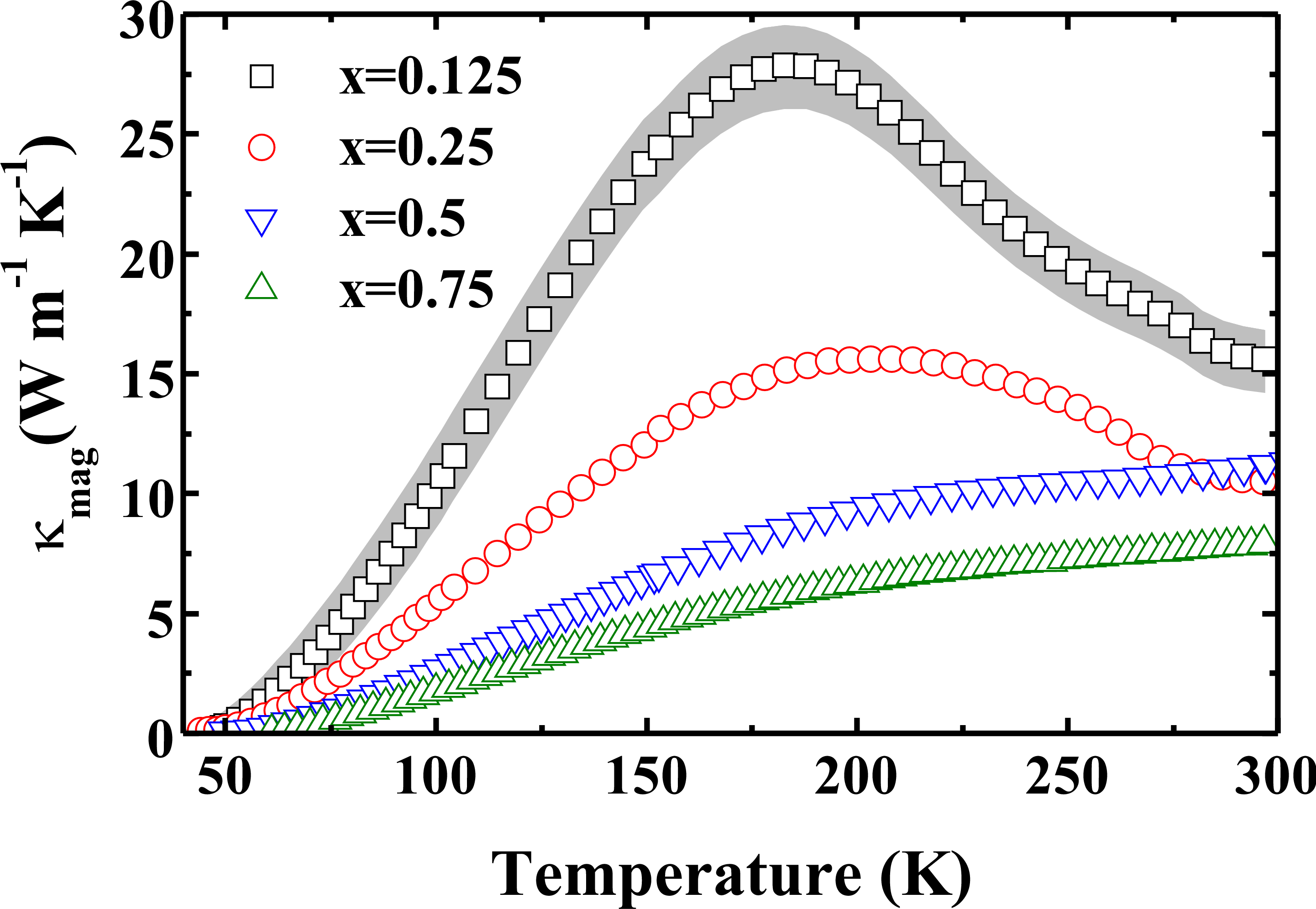}
\caption{$\kappa_{\mathrm{mag}}$ of $\mathrm{Sr_{14}Cu_{24-x}Zn_xO_{41}}$ at $x=0.125, 0.25, 0.5, 0.75$. The
gray shaded areas display the experimental uncertainty of $\kappa_{\mathrm{mag}}$ resulting from the uncertainties of $\kappa_{\mathrm{ph}}$ for the representative case $x=0.125$. Figure adapted from \cite{Hess2006}.}
\label{fig:ladderZn_kmag}
\end{figure}

\begin{figure}
\centering
\includegraphics[width=0.5\textwidth]{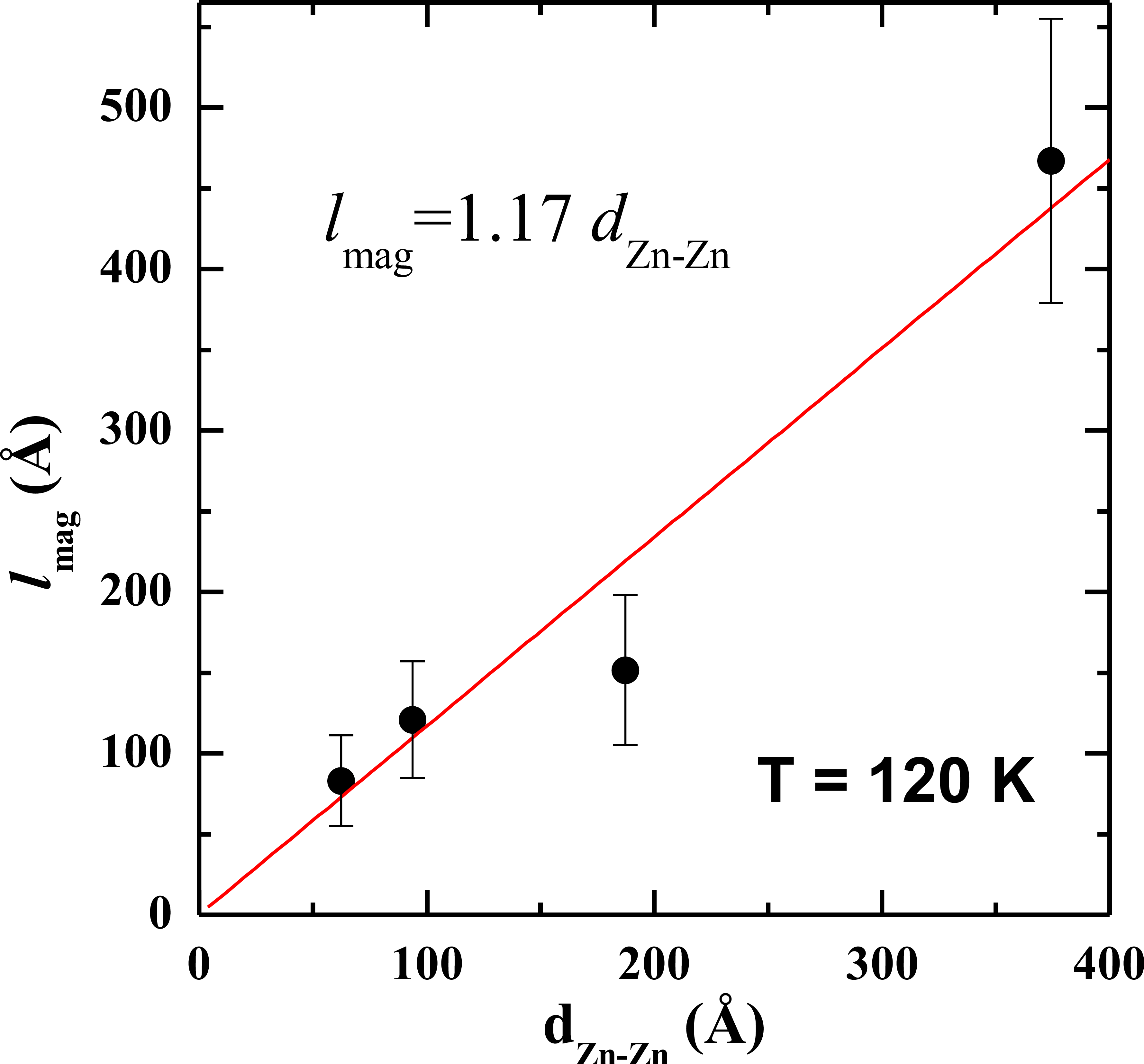}
\caption{$l_{\mathrm{mag}}$ as a function of $d_{\mathrm{Zn-Zn}}$. Solid line: linear fit line through the
origin. The error bars arise due to uncertainties in determining $\kappa_{\mathrm {ph}}$. Figure reproduced from \cite{Hess2006}.}
\label{fig:dZn_lmag}
\end{figure}

Experimentally, the magnitude of the mean free paths extracted by means of the kinetic model have been verified by a specific doping experiment. Similarly as in the 2D-case of \lacuo the heat conductivity of Zn-doped \srcuoladder has been studied, in order to render the determination of the mean free path model-independent \cite{Hess2006}. The non-magnetic Zn-ions act as scatterers for the magnons and cause a gradual suppression of \kmag as can be inferred from Fig.~\ref{fig:ladderZn_kmag} which reproduces the obtained data for \cuzn. Since the mean distance of Zn-ions within a ladder, $d_\mathrm{Zn-Zn}$, can be computed from the Zn content $z$, it is instructive to compare this distance with the magnon mean free path which has been extracted from the \kmag data in Fig.~\ref{fig:ladderZn_kmag} using the kinetic model (Eq.~\ref{eq:kappaladder}). The resulting \lmag vs. $d_\mathrm{Zn-Zn}$ is shown in Fig.~\ref{fig:dZn_lmag}. 
As is evident from the figure, $l_{\mathrm{mag}}$ roughly scales with $d_{\mathrm{Zn-Zn}}$, and a linear fit to the data points yields a slope of $1.17\pm 0.25$, which is close to unity. This confirms unambiguously that the values for
$l_{\mathrm{mag}}$ obtained from the kinetic model are in fair agreement with real scattering lengths in the material  \cite{Hess2006}.

The above findings allow to stress several important points: Despite the strong quantum nature of the two-leg spin ladder (which manifests itself in a large spin gap and a short-range spin correlation) the experimentally observed magnetic heat conductivity can be successfully described by the simple Boltzmann-type kinetic model (Eq.~\ref{eq:kappaladder}). The model is non-integrable and thus ballistic heat transport is not expected from fundamental conservation laws \cite{Zotos2005,Zotos97}, yet the heat conductivity seems anomalously large because the experimental mean free paths are up to several
orders of magnitude larger than the spin correlation lengths in the system.

\subsubsection*{Temperature dependent scattering processes}\label{sec:tdepmagnonscatt}
We will now briefly address the characteristics of \kmag of the spin ladders at higher temperatures, in order to study the influence of temperature-dependent scattering processes on the magnon heat transport, i.e. scattering of magnons off other excitations in a solid such as phonons, charge carriers, and also other magnons. The spin ladder material \srcala is an ideal 'playground' in this regard because it can be doped in very different ways. Definitely, the  possibility of doping the spin ladders with charge carriers is intriguing and unique among one-dimensional cuprate quantum magnets.

\begin{figure}
\centering
\includegraphics[width=0.6\textwidth]{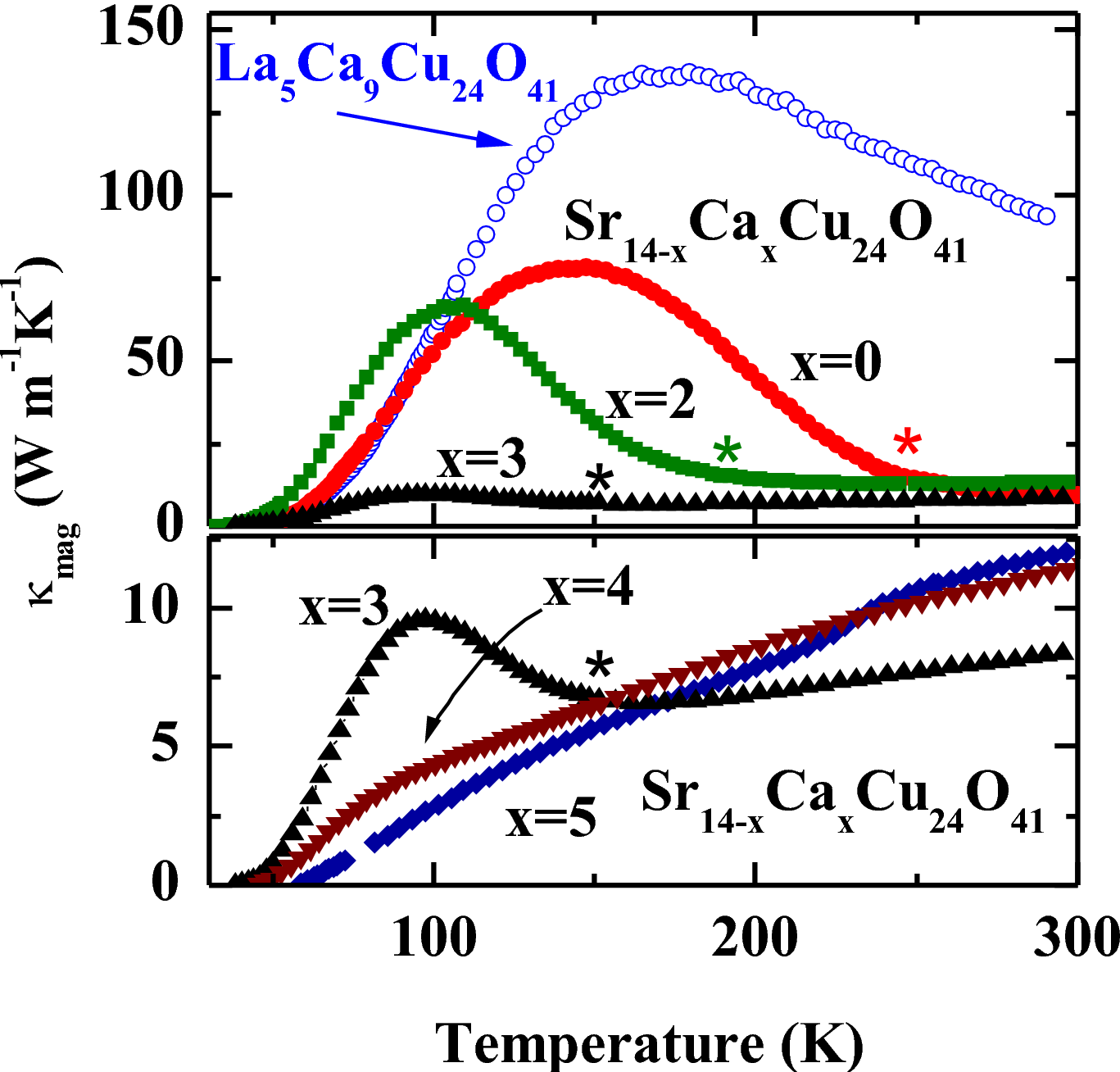}
\caption{$\kappa_{\mathrm{mag}}(T)$ of $\mathrm{Sr_{14-x}Ca_xCu_{24}O_{41}}$ ($x=0, 2, 3, 4,
5$). Top panel: Data for ($x=0, 2, 3$) in comparison with $\kappa_{\mathrm{mag}}$ of \laf. Lower panel: enlarged representation for $x=3, 4, 5$. Stars indicate the approximative $T^*$ for $x\leq3$. Similar results for $x=0, 2$ have been reported by \cite{Sologubenko00}. Figure adapted from \cite{Hess04a,Hess2007b}.}
\label{fig:SrCa_kmag}
\end{figure}

\subsubsection*{Magnon-hole scattering}
In the previous discussion of the low-temperature increase at $T\lesssim100$~K the fact that the hole content in the ladders of \srcuoladder and \laf is very different did not play an important role. Apparently, at this low temperatures the presence of the hole in the ladders of \srcuoladder is unimportant for the magnetic heat transport, as is reflected by the very similar values for \kmag (see Fig.~\ref{fig:kmag_59_sr14}) and thus the mean free path and the spin gap of both compounds.
New light is shed on the relevance of the holes in the ladders from
the direct comparison of the magnon heat conductivity \kmag of \laf with that of \srcuoladder in Fig.~\ref{fig:kmag_59_sr14} at $T\gtrsim100$~K.
The impact of the higher hole content in the ladders of \srcuoladder can directly be read off the figure: while \kmag is practically identical for both compounds at $T \lesssim100~\mathrm{K}$, one observes the already described differences at higher temperature.
It is straightforward to explain the comparatively stronger high-temperature suppression of \kmag of \srcuoladder by a significant magnon-hole scattering because the presence of holes in the ladders is the essential difference to the undoped ladders in \laf. The equality of both \kmag data below the characteristic temperature $T_0\approx100~\mathrm{K}$ implies in addition that this scattering process, on the one hand, is completely unimportant at low temperature $T\lesssim T_0$, but on the other hand unfurls its full strength above another characteristic temperature $T^*\approx 240~\mathrm{K}$ 
where \kmag saturates
\cite{Sologubenko00,Hess01,Hess02,Hess04a}.

This surprising temperature dependence of the magnon-hole scattering strength has been investigated towards its robustness against changes of the hole concentration \cite{Sologubenko00,Hess04a}. \kmag was studied in a series of \srca single crystals. Results for \kmag at low doping levels $x\leq5$ \cite{Hess04a} are shown in Fig.~\ref{fig:SrCa_kmag}. One can infer quite clearly that $T^*$ is gradually shifted towards lower temperature with increasing $x$, i.e., the temperature interval in which \kmag is suppressed due to magnon-hole scattering is extended to lower temperatures. At  $x=4$, 5 the magnon-hole scattering is apparently so dominant, that even the low-temperature peak is suppressed. 

The observed doping dependence of the magnon-hole scattering can clearly be related to the charge-ordered state \cite{Abbamonte2004,Rusydi2006} of the holes in the ladder below $T^*$ \cite{Sologubenko00,Hess01,Hess02,Hess04a}. 
More specifically, the charge order is accompanied by a drastic enhancement of the magnon mean free path \lmag of the magnons where the scattering probability turns out to be practically one (inferred from comparing the \lmag with the estimated distance of holes in the ladders, see below) in the case of completely mobile holes ($T>T^*$), and vanishes in the long-range ordered state ($T< T_0$) \cite{Sologubenko00,Hess04a}. Note, that charge order is also present at $x=4$, 5 \cite{Vuletic2005,Vuletic2003}. However, its onset temperature $T^*$ is already so low that \kmag is significantly suppressed. Note that the magnon heat conductivity in these samples is still well detectable, despite the substantial suppression, see Fig.~\ref{fig:SrCa_kmag}. Sologubenko et al. have shown that a large \kmag exists even at a very high doping level $x=12$ \cite{Sologubenko00}, where no charge order is present \cite{Rusydi2006,Vuletic2005,Vuletic2003}.

The above qualitative analysis of the magnon-hole scattering in \srca demonstrates, that \kmag of a quantum magnet beyond being intriguing in itself can also be used as a \textit{probe} for the charge dynamics and order in the material. A very interesting quantitative result which can be inferred from the above discussion for the undoped \srcuoladder is that through the determination of the low-temperature value of the mean free path $l_0$, i.e. the mean free path deep in the charge ordered phase, the correlation length of the charge order $\xi$ can be estimated. Since $l_0\sim3000$~{\AA} is much larger than the mean hole distance in the ladders and $l_0$ is practically identical for the hole-free \laf, the charge order should be perfect at least on this length scale, i.e. $\xi\gtrsim3000$~{\AA} \cite{Hess04a}.

It should be noted, that the magnon-hole scattering and the impact of the charge order on it can be well analyzed by studying the temperature dependence of the magnon mean free path $l_\mathrm{mag}(T)$ which can be extracted by comparing the measured \kmag data with Eq.~\ref{eq:kappaladder} if the previously extracted values for $\Delta$ are plugged into it. Hess et al. have decomposed $l_\mathrm{mag}(T)$ into a temperature independent magnon-defect part $l_0$ and a temperature-dependent part $l_h(T)$ which represents solely the magnon-hole scattering, using Matthiessen's rule:
\begin{equation}
l_{\mathrm{mag}}(T)^{-1}=l_0^{-1}+l_h(T)^{-1}.
\label{eq:matthiesen_hole}
\end{equation}
It could be shown that on the one hand $l_h$ at high temperatures $T>T^*$ is of the same order of magnitude than the mean distance of holes in the ladders. On the other hand, it was found that $l_h$ has a very similar temperature dependence as the electrical resistivity, which further confirms that \kmag depends sensitively on the charge dynamics in the material  \cite{Hess04a}.

\begin{figure}
\centering
\includegraphics[width=0.6\textwidth]{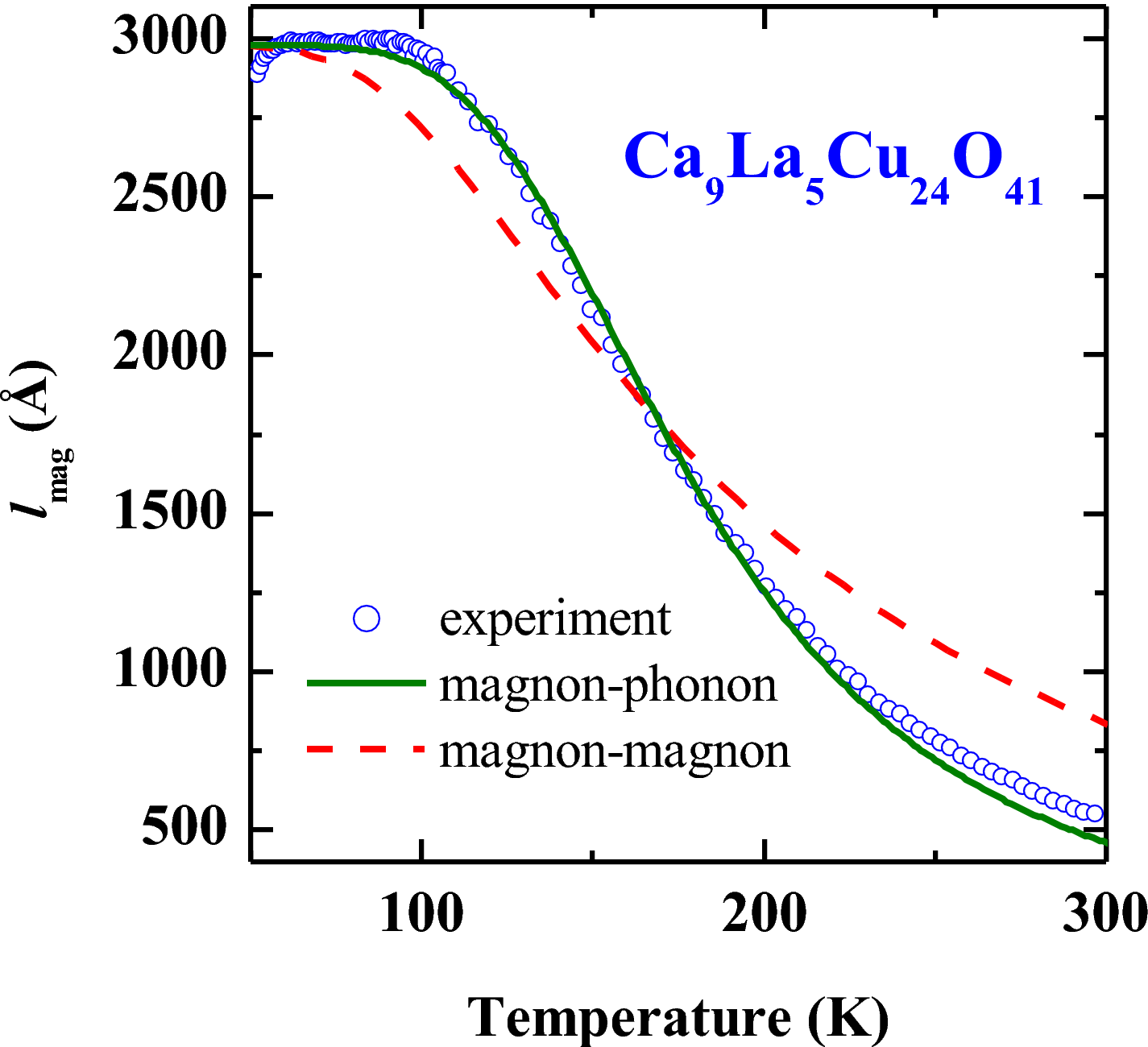}
\caption{$l_{\mathrm{mag}}$ of \laf as a function of temperature $T$. The solid and broken lines represent fits of $l_{\mathrm{mag}}$ accounting for magnon-phonon and magnon-magnon scattering respectively. Adapted from~\cite{Hess01,Hess05,Hess2007b}.}
\label{fig:La5_lmag_fits}
\end{figure}

\subsubsection*{Magnon-phonon scattering}
It is clear from the above discussion of magnon-hole scattering in \srca, that it must be unimportant in \laf. In spite of this, the measured \kmag substantially deviates for $T\gtrsim100~\mathrm{K}$ from the theoretically calculated \kmag based on a constant mean free path. Formally, this deviation can be captured by a temperature dependent mean free path $l_\mathrm{mag}(T)$, where the temperature dependence suggests the presence of a further temperature dependent scattering process. Hess et al. have analyzed this in more detail \cite{Hess01,Hess05}, the results of which have been summarized by \cite{Hess2007b}:

Apart from magnon-hole scattering, which is unimportant in \laf, only magnon-magnon scattering or magnon-phonon scattering are thinkable processes which could cause this temperature dependence of \lmag.
As has been mentioned above, the $T$-dependence of $l_{\mathrm{mag}}$ in \laf can be calculated from the $\kappa_{\mathrm{mag}}$ data using Eq.~\ref{eq:kappaladder} by plugging in the previously extracted $\Delta/k_B=418~\mathrm{K}$ \cite{Hess01,Hess05}. The resulting $l_{\mathrm{mag}}(T)$ as shown in Fig.~\ref{fig:La5_lmag_fits} reflects the different $T$-regimes which govern $\kappa_\mathrm{mag}$. For $T\lesssim110~\mathrm{K}$, $l_\mathrm{mag}$ is $T$-independent with a mean value $l_0=2980$~{\AA} which reflects the scattering of magnons off static defects. In order to describe the $T$-dependent $l_{\mathrm{mag}}$ at higher $T$ it was assumed (as in the case of magnon-hole scattering, see Eq.~\ref{eq:matthiesen_hole}) that all scattering mechanisms were independent of each other and Matthiessen's rule applied 
\begin{equation}
l_{\mathrm{mag}}^{-1} = l_0^{-1}+\gamma_\mathrm{ph}d_\mathrm{ph}^{-1}+\gamma_\mathrm{mag}d_\mathrm{mag}^{-1}.
\end{equation}

Here, $d_\mathrm{ph}$ and $d_\mathrm{mag}$ are the mean 'distances' of phonons and magnons respectively, as calculated from the particle densities with $\gamma_\mathrm{ph}$ and $\gamma_\mathrm{mag}$ the corresponding scattering probabilities. Since it is unclear as to what extent the separate scattering mechanisms contribute to $l_{\mathrm{mag}}$, its behavior was analyzed based on the assumption that only one mechanism is active in addition to magnon-defect scattering.

The case of dominant magnon-phonon scattering was modeled by three energy-degenerate non-dispersive optical branches along the ladder direction, yielding 
\begin{equation}
\frac{1}{d_\mathrm{ph}}=\frac{7.6\cdot10^{9}\rm m^{-1}}{\exp(\Delta_\mathrm{opt}/(k_BT))-1} 
\end{equation}
with $\Delta_\mathrm{opt}$ the optical gap (cf. \cite{Hess05} for details). The experimental $l_{\mathrm{mag}}$ was then fitted with $l_{\mathrm{mag}}^{-1} = l_0^{-1}+\gamma_\mathrm{ph}d_\mathrm{ph}^{-1}$
using $\gamma_\mathrm{ph}$ and $\Delta_\mathrm{opt}$ as free parameters. The fit (solid line in Fig.~\ref{fig:La5_lmag_fits}) describes the data fairly well. Remarkably, the value found for $\Delta_\mathrm{opt}/k_B=795~\mathrm{K}$ is of the same order of magnitude as the energy of the longitudinal {\rm Cu-O} bond stretching mode which is involved in the two-magnon-plus-phonon absorption observed in optical spectroscopy \cite{Gruninger2000,Windt01}.
The scattering probability is obtained as $\gamma_\mathrm{ph}=3.2\cdot10^{-2}$, i.e. significantly smaller than that of magnon-hole scattering for mobile holes.

For the assumption of dominant magnon-magnon scattering, a less satisfactory agreement was obtained with 
$l_{\mathrm{mag}}^{-1} = l_0^{-1}+\gamma_\mathrm{mag}d_\mathrm{mag}^{-1}$, where 

\begin{equation}
\frac{1}{d_\mathrm{mag}}=\frac{1}{\pi c_L}\int_0^\pi\frac{3}{3+\exp(\epsilon_k/k_BT)}dk
\end{equation}
(broken line in Fig.~\ref{fig:La5_lmag_fits}). $c_L$ is the lattice constant along the ladders and $\epsilon_k$ was taken from Johnston et al. for the case of isotropic ladder coupling \cite{Johnston00}, with $\epsilon_{k=\pi}/k_B=\Delta/k_B=418~\mathrm{K}$ employed. 
Note that $\gamma_\mathrm{mag}=0.05$ and thus of similar magnitude as $\gamma_\mathrm{ph}$.

The comparison between both fits suggests that scattering off optical phonons is dominant in this compound, 
since an almost perfect description of \lmag is obtained without the necessity to invoke magnon-magnon scattering as additional scattering mechanism. Nevertheless, a contribution from magnon-magnon scattering cannot be excluded on basis of this analysis.
Scattering off optical phonons is a plausible scattering mechanism for magnons since the longitudinal {\rm Cu-O} bond stretching mode directly affects the {\rm Cu-O} distance and hence the magnetic exchange constant of the ladders $J$. It is worth mentioning that the scattering off acoustic phonons appears unlikely, since the magnon energies with a $\Delta/k_B\sim400~\mathrm{K}$ clearly are higher than that of acoustic phonons, unlike that of optical phonons. One should note, however, that a Debye temperature\footnote{For example, one finds $\Theta_D= 296~\mathrm{K}$ for \srcuoladder  \cite{McElfresh89}.} $\Theta_D~\sim300~\mathrm{K}$ in \srcala is very close to the lower bound of the spread of measured values for the spin gap in two-leg spin ladder materials, see, e.g. \cite{Notbohm2007}. Thus, though optical phonons are a prominent candidate, acoustic phonons and also magnons cannot be excluded to play a role in the dissipation of magnon heat transport in the two-leg spin ladder compounds \srcala.

\subsection{Time-dependent measurements}
All data discussed so far have been obtained with the use of the so-called steady-state method,  see, e.g. \cite{Berman}. It should be noted, however, that concerning results for the two-keg spin ladders significant fluctuations of the absolute magnitude of the magnetic contribution to kappa of the very same sample  have been observed by several groups. Details of this phenomenon are discussed in \cite{Hess_diss2002}.\footnote{
A careful investigation of the impact of these fluctuations yields that all aspects of the aforementioned analysis of \kmag in \srcala remain valid. For example, for \srcuoladder fluctuations of \kmag leave the extracted value of the spin gap practically unaffected while in the same measurement the fluctuations of the extracted mean free path value amount to  not more than about 10\% \cite{Hess_diss2002}.} There are a few attempts to overcome these experimental difficulties by dynamical heat transport studies focusing on a single material, namely \laf. Interestingly, these studies consistently yield a lower value of \kappara, which at room temperature amounts to about 60\% of the data shown in Fig.~\ref{fig:La5_kap}. This finding is also in reasonable agreement with more recent steady-state studies by Naruse et al. \cite{Naruse2013}.

Apart from this quantitative result, these dynamic studies are however conflicting with respect to the extraction of the magnon-phonon relaxation time. Thus, the results of these studies \cite{Otter2012,Montagnese2013,Hohensee2014} are just briefly summarized here, while for an in-depth discussion the reader is referred to the original papers.

\begin{figure}
\centering
\includegraphics[width=0.65\textwidth]{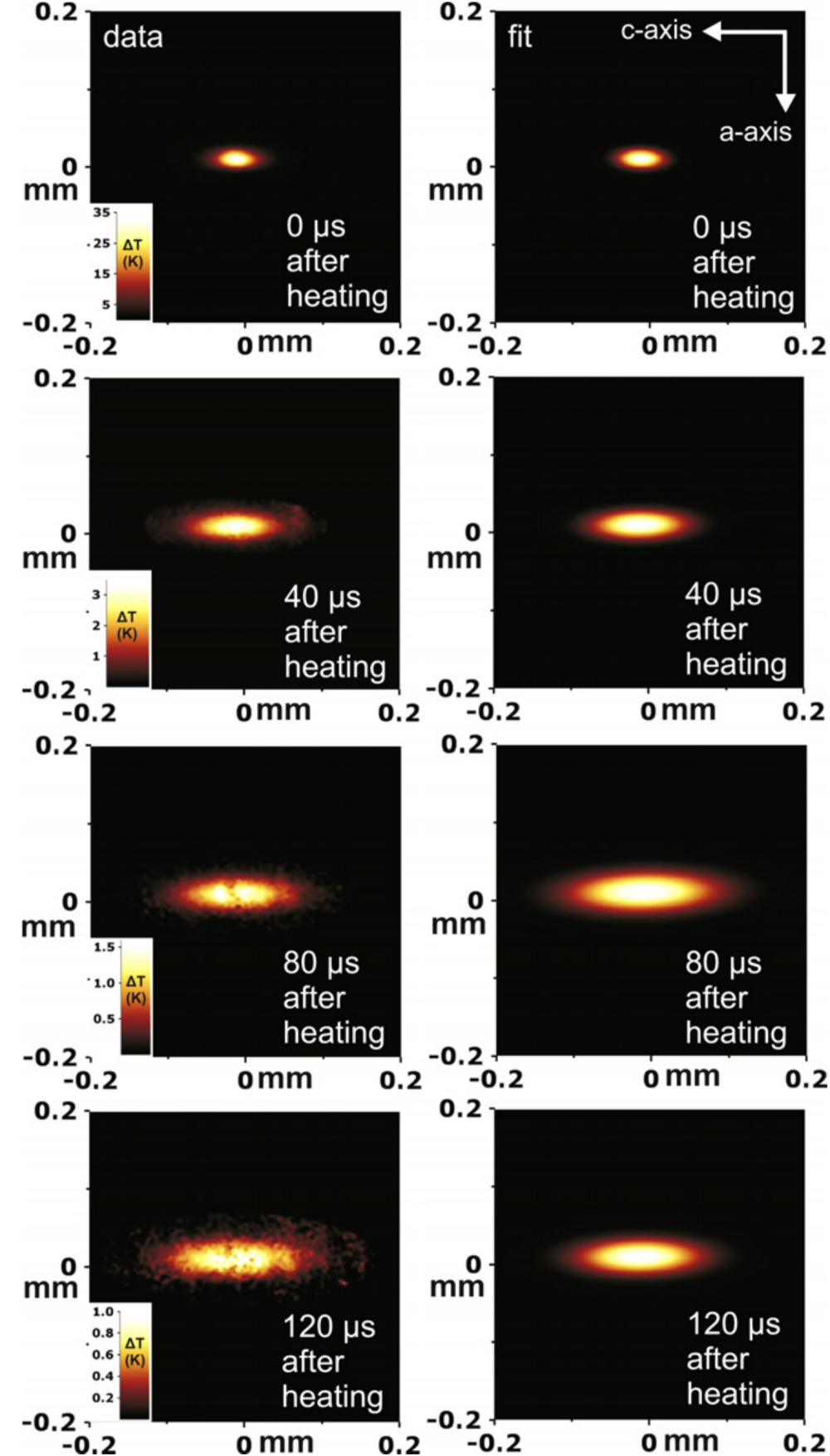}
\caption{Time evolution of the heat diffusion from a hot spot in the spin ladder
compound \laf. The left column shows the data, while the right column
shows the best Gaussian fit to the data. Heating is done by a laser with a pulse
duration of $\mathrm 20~\mu s$. The integration time for the probe UV-pulse is $\mathrm 20~\mu s$. The ladder
 direction is horizontal. The anisotropy of the diffusion process is clearly seen.
 Figure taken from~\cite{Otter2012}.}
\label{fig:Otter2}
\end{figure}

Otter et al. have exploited the  fluorescent microthermal imaging technique which has been mentioned already in the beginning of this section further to extract the time-dependence of the anisotropic heat spread after locally heating the surface of a \laf crystal in the $ac$-plane (see  Fig.~\ref{fig:Otter2}) \cite{Otter2012}. By analyzing the time-dependent anisotropic heat spread with the three-dimensional heat diffusion equation, the obtained data were used to extract the room-temperature heat conductivity along the $a$- and $c$-directions, i.e. \kapperp and \kappara, respectively. Interestingly, while the extracted value for \kapperp was found to be in good agreement with the steady-state results shown in Fig.~\ref{fig:La5_kap}, the value for \kappara amounts, as already mentioned, only about 60\% of the steady-state \kappara. 
% However, more recent steady state studies by Naruse et al. \cite{Naruse2013} are in reasonable agreement with the results from the dynamic method.

Montagnese et al. employed a different but related dynamic method where they measured the time-of-flight of a heat pulse through a \laf sample for various sample thicknesses. This study allowed to extract the  phonon-magnon equilibration time, yielding a very large $\tau_\mathrm{mp}\approx\mathrm 400~\mu s$ as compared to that for the spin chain compound \srcuodouble where $\tau_\mathrm{mp}\approx\mathrm 10^{-12}~s$ \cite{Montagnese2013}, and for which the mentioned fluctuations in \kappara are absent. This finding is, however, in contrast with very recent measurement results using the  time-domain thermoreflectance \cite{Hohensee2014} where a six orders of magnitude faster phonon-magnon equilibration time is found. At present, the origin of these strong discrepancies between both these pioneering studies remains unclear and is not discussed further at this point.

\section{Spin chains}
\label{sec:chains}
\subsection{Theoretical preliminaries}

\begin{figure}
\centering
\includegraphics[width=\textwidth]{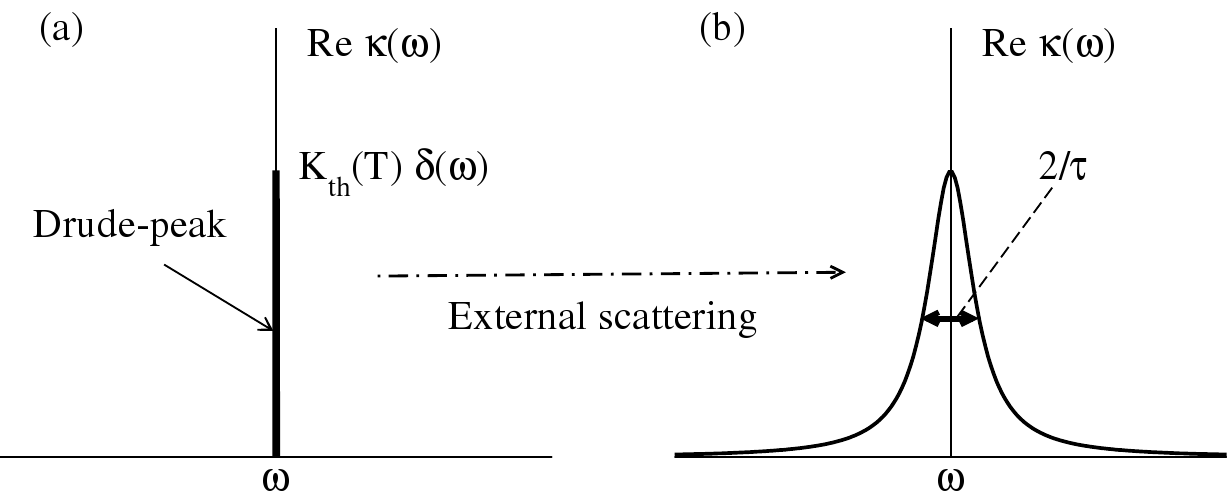}
\caption{a) Sketch of the real part of the thermal conductivity $\kappa(\omega)$ of the Heisenberg chain as a function of frequency $\omega$. The conductivity is given by $ \operatorname{Re}~\kappa(\omega) = K_\mathrm{th}(T)\delta(\omega)$ due to the exact conservation of the energy-current operator \cite{Niemeijer1971,Zotos97}. Thus, the thermal Drude weight \kth is nonzero at any finite temperature, and any contribution at finite frequencies vanishes. b) In a real experiment for spin chain materials, one may expect the Drude weight to be broadened into, e.g., a Lorentzian in frequency space by external scattering. $\tau$ is the inverse width of such a Lorentzian, and it is related to the life time of the current, or the inverse scattering rate, respectively.
Reproduced from \cite{Heidrich_diss}.}
\label{fig:kth_width}
\end{figure}

The thermal conductivity of spin chains appears particularly interesting because unlike the situation in the previously discussed cases of spin planes and spin ladders, rigorous theoretical predictions exist, in particular for the spin-1/2 Heisenberg chain model which is assumed to be relevant for the cuprate spin chains discussed here. The most spectacular one is certainly the prediction of ballistic, i.e. dissipationless, heat transport which would result in a diverging heat conductivity \cite{Zotos97}, which can be expressed in the form of a delta peak at zero frequency in the real part of the frequency-dependent thermal conductivity, i.e., $ \operatorname{Re}~\kappa(\omega) = K_\mathrm{th}(T)\delta(\omega)$ \cite{Heidrich_diss}. Moreover, the temperature dependence of the thermal Drude weight \kth has been calculated \textit{exactly} \cite{Kluemper2002}. Thus, in experimental investigations of the heat conductivity of spin-1/2 Heisenberg chain materials, one should 
expect to find signatures of the extraordinary properties of the model, though external scattering processes should complicate the picture.

At low temperatures $T\lesssim 0.15 J/k_B$, i.e., $T\lesssim 300$~{\rm K} for $J/k_B\sim2000~\mathrm{K}$, the thermal Drude weight depends linearly on temperature, where \cite{Kluemper2002,Heidrich02,Heidrich03,Hess2007}

\begin{equation}
 K_\mathrm{th}=\frac{(\pi k_B)^2}{3\hbar}vT\,, \label{eq:kth}
\end{equation}
with the velocity $v$ of the spinons at long wave lengths.
In a real material, one might qualitatively expect that the external scattering processes cause the delta peak to broaden into, e.g., a Lorentzian in frequency space \cite{Heidrich_diss} with a width $1/\tau$, i.e. the  scattering rate of the heat current (see Fig.~\ref{fig:kth_width}). Hence, the heat conductivity of a single chain, $\tilde{\kappa}_\mathrm{mag}$, is rendered finite and may be approximated by $\tilde{\kappa}_\mathrm{mag}= K_\mathrm{th} \,\tau /\pi$ \cite{Hess2007}. If one naturally relates the experimental spinon mean free path \lmag with the relaxation time $\tau$ via $l_\mathrm{mag}=v\tau$ and considers the number of spin chains per unit area $n_s$, one obtains \cite{Hess2007}

\begin{equation}
l_\mathrm{mag}=\frac{3}{\pi}\frac{\hbar}{k_B^2n_s}\frac{\kappa_\mathrm{mag}}{T}\,,\label{eq:lmag_chain}
\end{equation}
It is interesting and important to note, that the kinetic model in one dimension (Eq.~\ref{eq:kinetic_1D}), if one uses a Fermi distribution in order to account for the fermionic character of the spinons, yields \cite{Hess2007b,Sologubenko01} 
\begin{equation}
 l_\mathrm{mag}=\frac{\pi}{2}\frac{\hbar}{k_B^2n_s}\frac{\kappa_\mathrm{mag}}{T}\left[\int_0^\frac{J\pi}{2k_BT}x^2\frac{\exp(x)}{(\exp(x)+1)^2}dx\right]^{-1}.
\label{eq:lmag_chain_kinetic}
\end{equation}
The integral in (\ref{eq:lmag_chain_kinetic}) is only weakly temperature dependent and approaches $\pi^2/6$ for $T\rightarrow0$. Hence, at low temperatures $k_BT\ll J$, which with $J/k_B\sim2000~\mathrm{K}$ holds even at room temperature one obtains the same result as in the case of the Drude weight approach, i.e. Eq.~\ref{eq:kth} and \ref{eq:lmag_chain} \cite{Hess2007,Hess2007b}.

\subsection{Spinon heat transport in 'dirty' spin chains}
\begin{figure}
\centering
\includegraphics[width=0.6\textwidth]{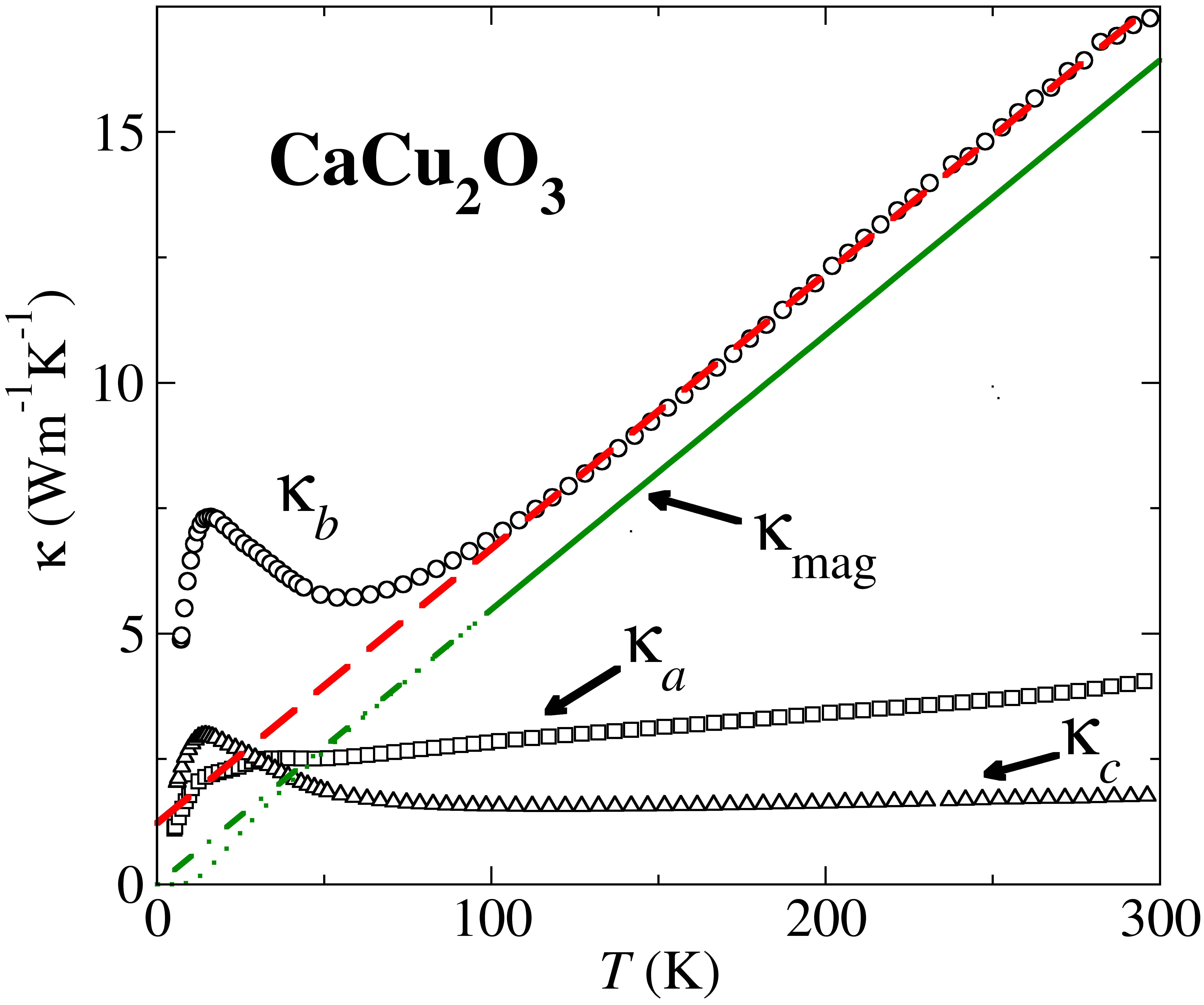}
\caption{$\kappa_a$ ($\square$), $\kappa_b$ ($\bigcirc$) and $\kappa_c$ ($\triangle$) of \cacuoladder as a function of
temperature. The dashed and solid lines represent a linear fit of the experimental data in the range $100\dots300$~{\rm K} and the estimated $\kappa_\mathrm{mag}$ 
in this range. Extrapolations of $\kappa_\mathrm{mag}$ towards low temperatures (assuming a temperature independent
magnetic mean free path as extracted for $T>100~\mathrm{K}$) corresponding to a finite ($\Delta=3$~{\rm meV}) and a vanishing spin gap are represented by  dotted and dashed-dotted lines, respectively. 
% Inset: $\kappa_b$ obtained from two different measurements (open symbols depict the same curve as in the main panel). The different slopes of these data at higher temperature have been used to extract the error of \lmag. 
Figure reproduced from \cite{Hess2007}.}
\label{fig:kappa_123}
\end{figure}

We start our discussion of the spinon heat conductivity in cuprate spin chain materials with the results for \cacuoladder for which we have already briefly discussed its signatures of magnetic heat conductivity in Section~\ref{sec:signatures} (Fig.~\ref{fig:signatures}a, see \cite{Hess2007} for details on this study). The $\mathrm{Cu_2O_3}$ planes in this material actually form, like in \srcala, a two-leg ladder structure. However, due to strongly bent {\rm Cu-O-Cu} bonds along the rungs (bonding angle $\sim 123^\circ$) the rung exchange interaction is believed to be strongly reduced ($J_\bot/k_B\sim 100\dots300~\mathrm{K}$) as compared to the ladder legs ($J_\Vert\sim2000~\mathrm{K}$), and all other exchange interactions present in the compound are expected to be similarly small or even smaller \cite{Goiran06,Kim03,Kiryukhin01}. Thus, it is reasonable to consider this material rather as a chain compound with the chains running along the crystallographic $b$-axis. Indeed, INS confirms the 
absence of a spin gap for energies above $\sim3$~{\rm meV} and excitation spectra which are compatible with weakly coupled spin-1/2 Heisenberg spin chains \cite{Lake2010}. Concerning its crystal structure, the material is rather disordered due to an inherent significant Ca and oxygen deficiency being balanced by excess Cu \cite{Ruck01,Kim03}. We shall see that this disorder has a substantial impact on both the phononic as well the magnetic heat transport properties.

Fig.~\ref{fig:kappa_123} shows once more the heat conductivity of \cacuoladder as a function of temperature, for all principal axes of a crystal (see \cite{Hess2007}, for all details). $\kappa_a$ and \kapc were measured perpendicular to the chains  (thus both are labeled \kapperp in Fig.~\ref{fig:signatures}) and $\kappa_b$ was measured parallel to the chains. Since the material is electronically insulating, electronic heat conduction is negligible and we, therefore, expect these components to originate from phononic heat conduction plus a possible magnetic contribution. 
The thermal conductivity perpendicular to the chain direction is typical for a strongly suppressed phononic thermal conduction with a  high phonon scattering rate \cite{Berman50,Berman}: $\kappa_a$ and $\kappa_c$ exhibits only a weak $T$-dependence and  possess absolute values ($\lesssim 4~\rm Wm^{-1}K^{-1}$) which are much smaller than the phonon heat conductivity \kaph of other chemically undoped chain or ladder cuprates such as \srcuodouble, \srcuosingle (see further down),  or \srcuoladder (Fig.~\ref{fig:signatures}b). The thermal conductivity perpendicular to the chains can, therefore, be considered to be purely phononic and the strong suppression is naturally explained as a direct consequence of the strong off-stoichiometry of \cacuoladder.

The completely different behavior of \kapb, which is measured \textit{parallel} to the chains, in particular, the strong increase at $T\lesssim50~\mathrm{K}$  must arise from magnetic heat conduction in the chains. Such a strong increase of $\kappa$ with rising $T$ cannot be understood in terms of conventional phonon heat conduction by acoustic phonons, and also thinkable contributions from dispersive optical phonons which possibly play a role in the $T$ dependence of $\kappa_a$ can be excluded \cite{Hess2007}. Interestingly, the high-temperature increase of $\kappa_b$ turns into \textit{linear} in $T$ for $T\gtrsim100~\mathrm{K}$.
It is immediately clear that this linearity in temperature not only holds for the total thermal conductivity $\kappa_b=\kappa_\mathrm{ph}+\kappa_\mathrm{mag}$ but in particular also for the magnetic contribution \kmag because \kaph is only weakly temperature dependent as can be inferred from the phononic \kapa and \kapc. Therefore, the magnetic thermal conductivity of \cacuoladder at $T\gtrsim100~\mathrm{K}$ apparently perfectly follows the expected temperature dependence of the thermal Drude weight \kth as given in Eq.~\ref{eq:kth}.

\begin{figure}
\centering
\includegraphics[width=0.6\textwidth]{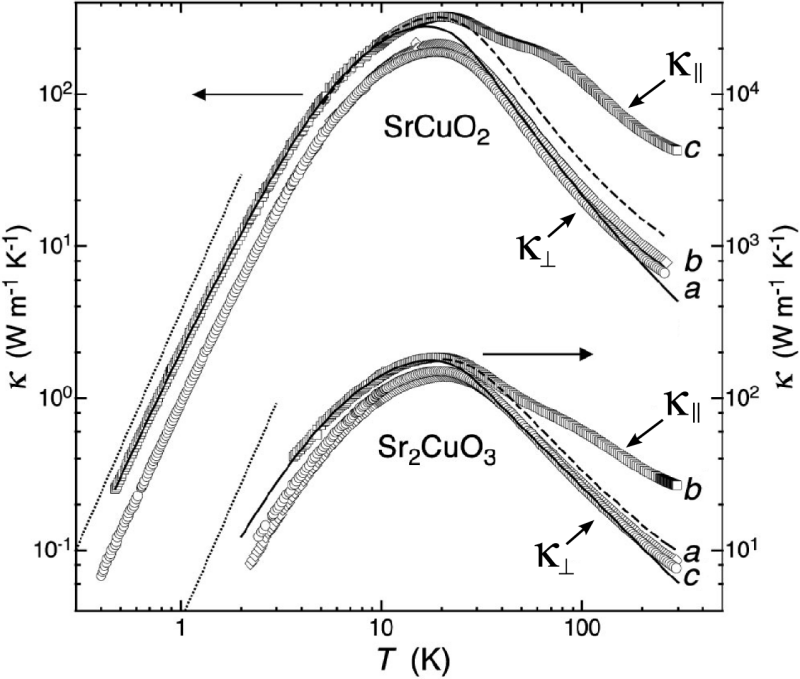}
\caption{Temperature dependences of the thermal conductivities 
of \srcuodouble and \srcuosingle along the $a$, $b$, and $c$ axes. The dotted
lines represent estimated limits of the thermal conductivity due to
the finite size of the samples. The solid and dashed lines represent
different evaluations of the phonon contribution, see \cite{Sologubenko01}, where the figure has been adapted from, for details.}
\label{fig:kappa_SCO112_213_Solo}
\end{figure}

Hess et al. have investigated this result further and extracted in a simple procedure the magnetic heat conductivity by presuming that the observed slope in the data is that of \kmag \cite{Hess2007}.
As can be seen in Fig.~\ref{fig:kappa_123}, a linear fit in the range $100\dots300$~{\rm K} (dashed line) describes the data almost perfectly and thus yields $\kappa_{\mathrm{ph},b}=1.2~\mathrm{Wm^{-1}K^{-1}}$ and $\kappa_\mathrm{mag}=0.055{\rm ~Wm^{-1}K^{-2}}\times T$ in this temperature regime. The  extracted $\kappa_\mathrm{mag}$ for $T>100~\mathrm{K}$  is plotted in Fig.~\ref{fig:kappa_123} as a solid line. $\kappa_\mathrm{mag}$ at lower $T$ cannot be inferred from the data.

Obviously, the mean free path of such a linear \kmag should be temperature independent in the framework of the model described above. Comparison with Eq.~\ref{eq:lmag_chain} yields $l_\mathrm{mag}=22\pm 3$~{\rm\AA} for the entire range $100\dots300$~{\rm K} corresponding to about 6 lattice spacings \cite{Hess2007}. Such a mean free path appears to be is extremely short, in particular in view of the much larger mean free paths which were obtained on \lacuo and \srcala (see the previous sections). The large impurity density which is implied by it is however qualitatively consistent with the fact that the material is intrinsically disordered. Furthermore, it is even quantitatively consistent with an independent measurement of the impurity density of the material, which in this special case is possible via the magnetic susceptibility (see \cite{Hess2007}, for details).

A temperature-independent \lmag as is found here seems at first glance surprising. It may indicate, however, that spinon-impurity scattering arising from effective chain cuts through the impurities is dominating over all other scattering processes, in particular, spinon-phonon scattering.
In the Boltzmann-type view that has been introduced above one may understand this, provided Matthiessen's rule holds, by assuming

\begin{equation}
 l_\mathrm{mag}^{-1}=l_0^{-1}+l_\mathrm{sp}(T)^{-1}+l_\mathrm{ss}(T)^{-1}\, ,
 \label{eq:matthiesen_allg}
\end{equation}
where $l_{sp}$ and $l_\mathrm{ss}$ stand for \textit{a priori} thinkable temperature-dependent spinon-phonon and spinon-spinon scattering lengths, respectively. A constant \lmag may then naturally arise if $l_0 \ll l_{sp},\, l_\mathrm{ss}$, a situation which seems to be plausible in the situation of a 'dirty' spin chain as obviously is realized in \cacuoladder. In order to experimentally obtain access to these apparently masked scattering processes it seems straightforward to study \kmag of spin chain materials which are chemically cleaner than \cacuoladder and where $l_0 \gg l_{sp},\, l_\mathrm{ss}$ can expected to be realized. In this situation, which shall be discussed in the next section, it should be possible to study the relevance of  spinon-phonon and spinon-spinon scattering directly.

\subsection{'Ballistic' spinon heat transport in 'clean' spin chains}

The spin chain compounds \srcuodouble and \srcuosingle  are both considered as being excellent realizations of the spin-1/2 Heisenberg model and are not prone to the stoichiometric problems that arise in \cacuoladder. They are therefore ideal candidates to further studying relevant spinon scattering processes. 
Sologubenko et al. were the first to study the heat conductivity of these materials already in 2000 and 2001 \cite{Sologubenko00a,Sologubenko01}. Fig.~\ref{fig:kappa_SCO112_213_Solo} presents their data for the heat conductivity of these materials measured along all crystal directions.
As can be seen in the figure, unlike the previously discussed cases, there is no clear double-peak structure or high-temperature increase visible in \kappara. Yet, there obviously is an enhancement which develops in \kappara at elevated temperature and causes a significant anisotropy between \kappara and \kapperp. Sologubenko already at that time concluded the presence of substantial spinon heat conduction in these materials based on these data. The solid and dashed lines in the Figure represent two different approaches to evaluate the phononic contribution to \kappara. The detailed evaluation and analysis of the resulting spinon heat conductivity (see \cite{Sologubenko01}) is omitted here because more recent data allow a more precise view on the matter.

These new data emerged recently by two groups who independently from each other discovered that the heat conductivity in \srcuodouble and \srcuosingle depends crucially on the chemical purity of sample, where in particular the spinon heat conductivity \kmag turned out to be extremely sensitive to chemical impurities \cite{Hlubek2010,Hlubek2012,Kawamata2010,Kawamata2008}.

\begin{figure}
\centering
\includegraphics[width=0.6\textwidth]{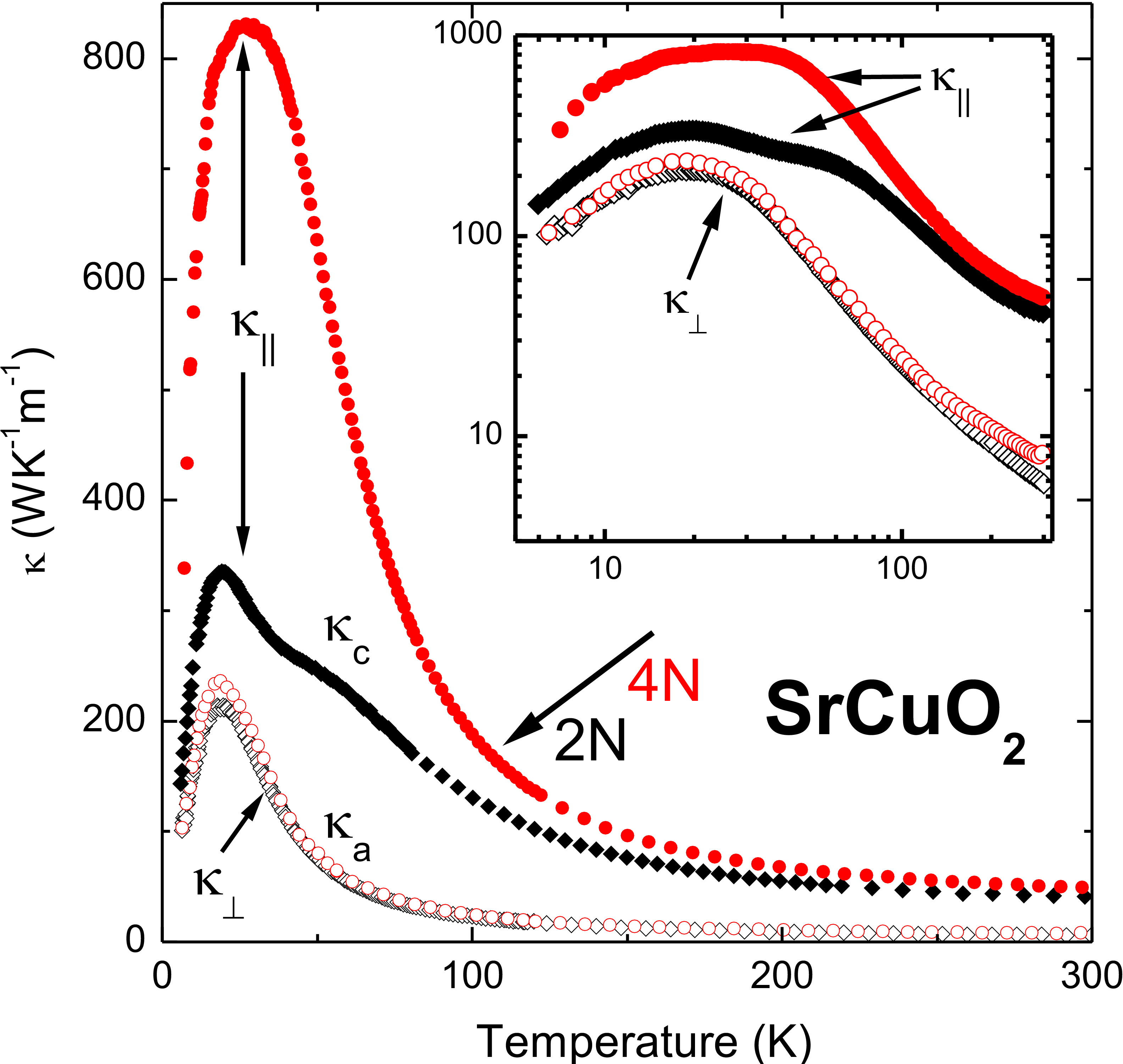}
\caption{\kapa and \kapc of \srcuodouble for different purity values. Closed (open) symbols represent $c$ axis ($a$ axis) data, red circles
(black diamonds) correspond to '$4N$' ('$2N$') purity. Inset: The same data as in the main panel in double-logarithmic representation. Adapted from \cite{Hlubek2010}.}
\label{fig:srcuo2_purity}
\end{figure}

\subsubsection*{Ballistic spinon heat transport in SrCuO$\bf _2$}

Hlubek et al. \cite{Hlubek2010} and Kawamata et al. \cite{Kawamata2010} studied independently from each other the effect of chemical purity on the magnetic heat conductivity of double-chain compound \srcuodouble, with very similar results.
Fig.~\ref{fig:srcuo2_purity} presents the thermal conductivities \kappara and \kapperp of \srcuodouble which were obtained on single crystals with 99\% ('$2N$') and 99.99\% ('$4N$') chemical purity as obtained by Hlubek et al. \cite{Hlubek2010}.\footnote{The chemical purity refers to that of the primary chemicals {\rm CuO} and $\mathrm{SrCO_3}$ used in for the crystal growth by the traveling solvent floating zone method \cite{Hlubek2010}. }
Hlubek et al. report for both \kapperp and \kappara of the '$2N$' sample very similar results as previously found by Sologubenko et al. (see Fig.~\ref{fig:kappa_SCO112_213_Solo}). However, they observed a drastically enhanced spinon heat conductivity of the samples with the higher purity ('$4N$') \cite{Hlubek2010}: 
The thermal conductivity parallel to the spin chains of the '$2N$' sample, $\kappa_{\Vert,2N}$, exhibits a low-temperature peak at $\sim20~\mathrm{K}$ and a shoulder at $T\gtrsim40~\mathrm{K}$. This shoulder and the fact that $\kappa_{\Vert,2N}$ remains at all temperatures much larger 
than the thermal conductivity perpendicular to the chains $\kappa_{\bot,2N}$ (the figure shows data for the $a$-direction $\kappa_{a,2N}$ only, because data along the $b$ direction are practically identical \cite{Hlubek2010}, see also Fig.~\ref{fig:srcuo2_purity}) have been interpreted as the primary signatures of magnetic heat transport in the compound \cite{Sologubenko01}. 
Interestingly, for the '$4N$' sample, the heat transport perpendicular to the chains, $\kappa_{\bot,4N}$, is only slightly enhanced as compared to $\kappa_{\bot,2N}$ which is indicative of a weakly reduced phonon-defect scattering. In contrast, the heat transport parallel to the chains, $\kappa_{\Vert,4N}$ is drastically enhanced exhibiting a broad peak centered at $\sim28~\mathrm{K}$. At all temperatures $\kappa_{\Vert,4N}>\kappa_{\Vert,2N}$ up to room temperature. At low temperature, the enhancement is largest (exceeding a factor of 2 at $T\lesssim 70~\mathrm{K}$) where both curves approach each other at further increased temperature.

\begin{figure}
\centering
\includegraphics[width=0.6\textwidth]{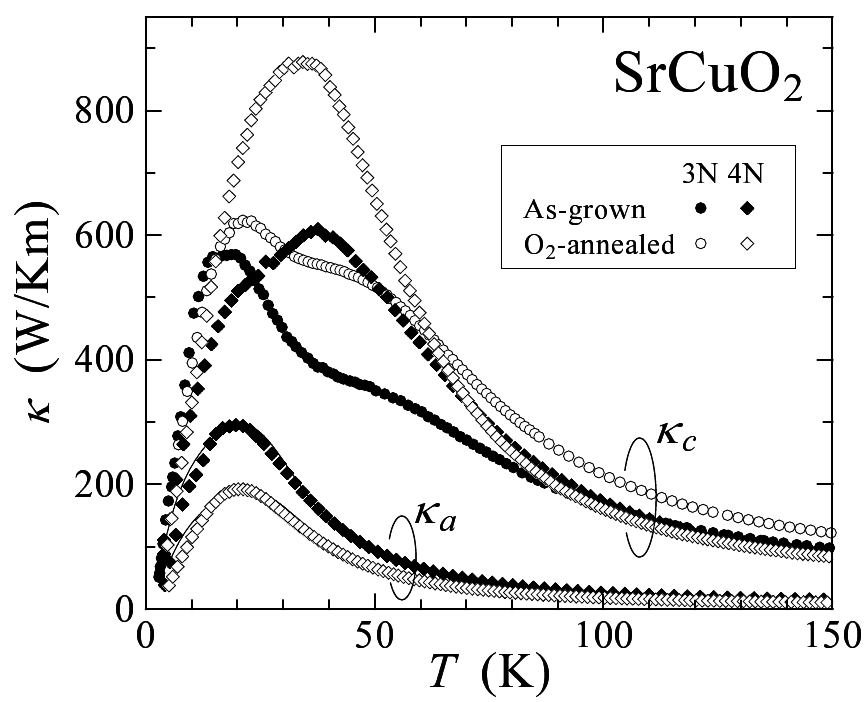}
\caption{Temperature dependence of the thermal conductivity along the $c$-axis (parallel
to spin-chains, \kapc) and along the $a$-axis (perpendicular to spin-chains, \kapa) for as-grown and $\mathrm O_2$-annealed single-crystals of \srcuoladder grown from raw materials with '$3N$' and '$4N$' purity. Solid lines are fitting results for \kapa using the Callaway model. Reproduced from \cite{Kawamata2010}}
\label{fig:SrCuO2_Kawamata}
\end{figure}

Kawamata et al. pointed out that not only the chemical purity of the growth materials but also the oxygen stoichiometry, which can be influenced by annealing the samples, has a strong influence on the heat conductivity of \srcuodouble parallel to the chains \cite{Kawamata2010}. Fig.~\ref{fig:SrCuO2_Kawamata} shows corresponding data for samples with '$3N$' and '$4N$' chemical purity. As can be inferred from the figure, the heat conductivity parallel to the chains of as-grown samples is strongly enhanced by annealing the samples in $\mathrm O_2$ atmosphere. In contrast to this, the annealing procedure leads to a decrease of \kapa of the '$4N$' sample (see Figure). Note that the data for \kapc of the annealed '$4N$' sample are very similar to those of Hlubek et al. while only a very rough agreement is found for \kapa (see Fig.~\ref{fig:srcuo2_purity}). The data by Hlubek et al. have been obtained on as-grown samples for which the annealing process lead to slight enhancement of the phonon heat conductivity \cite{Hlubekdiss}.

Hlubek et al. drew several qualitative conclusions from their data \cite{Hlubek2010}: i) the increase of \kappara upon enhancing the chemical purity in contrast to the negligible increase found in \kapperp implies that the increase of \kappara primarily concerns the magnetic heat conductivity \kmag. ii) the extreme low-temperature sensitivity to impurities of \kmag suggests that spinon-defect scattering is the dominating process for dissipating the heat current in this regime. iii) upon rising the temperature the spinon-defect scattering is increasingly masked by a further scattering process which leads to 
$\kappa_{\Vert,4N}$ and $\kappa_{\Vert,2N}$ approaching each other.
Hlubek et al. suggested spinon-phonon scattering as the most reasonable candidate for the latter process, since phonons are inevitably present, and the
only thinkable alternative,  spinon-spinon scattering, is expected to be unimportant for the heat transport of the $S=1/2$ Heisenberg chain model \cite{Zotos97,Kluemper2002,Heidrich03,Heidrich05}.

Again assuming $ \kappa_\Vert=\kappa_\mathrm{ph}+\kappa_\mathrm{mag}$, the phonon heat conductivity parallel to the spin chains \kaph may be reasonably approximated from the measured \kapperp, i.e., $\kappa_\mathrm{ph}\approx\kappa_\bot$ where a certain amount of anisotropy should be taken into account \cite{Sologubenko01,Hlubek2010}. Thus, \kmag can again be estimated from the measured data by $\kappa_\mathrm{mag}=\kappa_\Vert-\kappa_\bot$.
The relatively large anisotropy of the heat conductivity of the '$4N$' sample implies that any error in \kaph has only very little effect on \kmag, as long as $\kappa_\mathrm{mag}\gg\kappa_\bot$.

\begin{figure}
\centering
\includegraphics[width=0.6\textwidth]{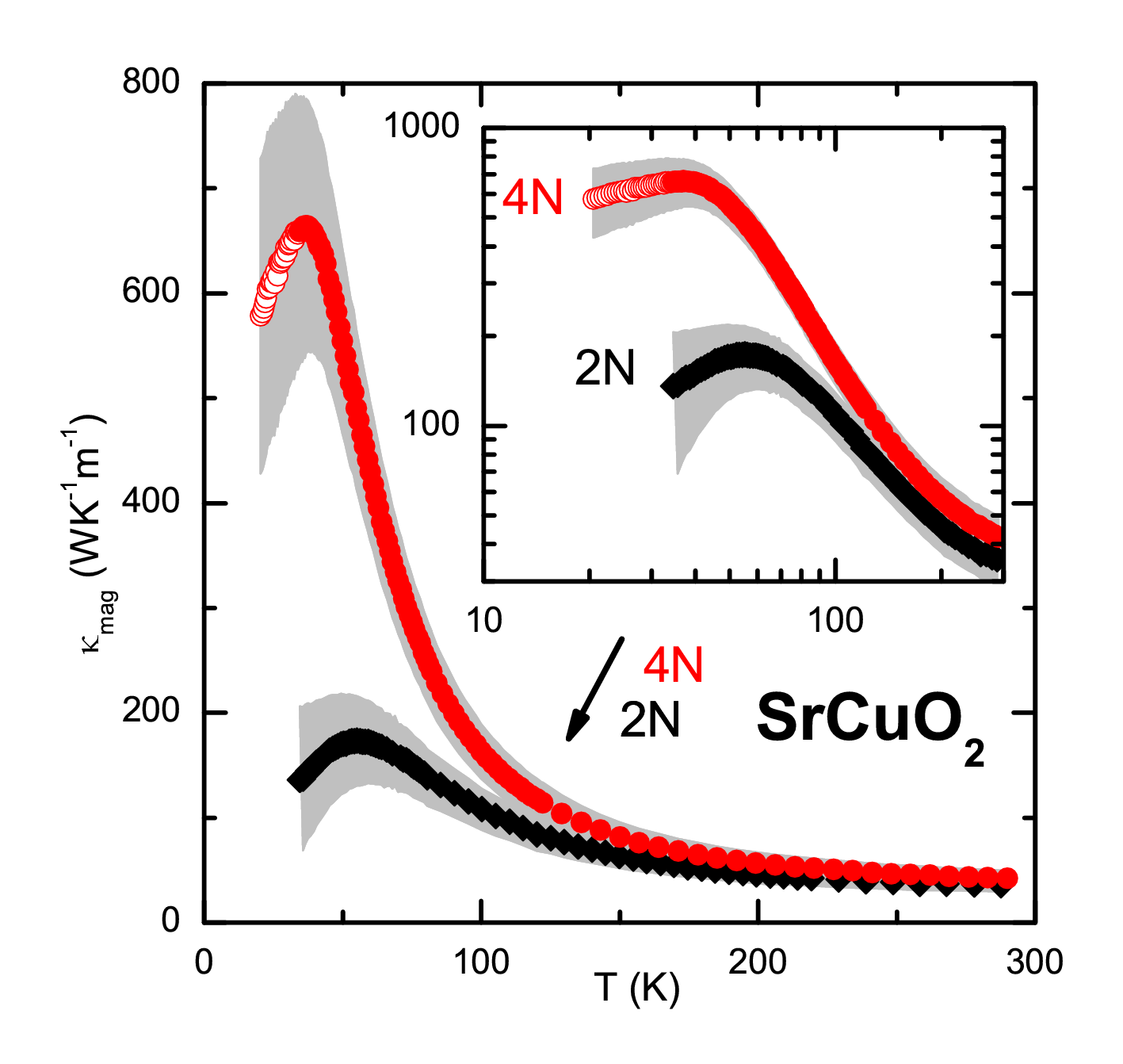}
\caption{\kmag of \srcuodouble for different purities. Open symbols represent low-$T$ \kmag which is disregarded in the
further analysis. The shaded areas show the uncertainty of the estimation of \kmag due to an assumed relative error of 30\% for the phononic background. Inset: The same data as in the main panel in double-logarithmic representation. Adapted from \cite{Hlubek2010}.}
\label{fig:srcuo2_kmag}
\end{figure}

Fig.~\ref{fig:srcuo2_kmag} shows the thus extracted spinon heat conductivity \kmag of both the '$2N$' and the '$4N$'  samples as a function of temperature $T$ \cite{Hlubek2010}. Both curves exhibit the characteristic peak structure which qualitatively reflects the competition between the increasing occupation of quasiparticle states and the growing importance of temperature dependent scattering processes with rising temperature \cite{Berman}. In the regime of the low-temperature increase, the latter are unimportant and only temperature independent boundary scattering  dominates. The increase then is the result of the growing number of thermally excited spinons. The most reasonable candidate for the temperature dependent scattering is of course, as mentioned above, the spinon-phonon scattering. Within this scenario, the difference between both curves arises naturally from the different purity of both compounds which yields a corresponding different importance of boundary scattering. This corresponds very well with the concept that a single impurity within a chain serves as a boundary in one dimension.
Note, that the peak value of $\sim660$~\wkm of the '$4N$' is to the best of our knowledge a `record` as it exceeds the largest reported magnetic heat conductivities \cite{Sologubenko01,Hess01} by more than a factor of  3.

\begin{figure}
\centering
\includegraphics[width=0.6\textwidth]{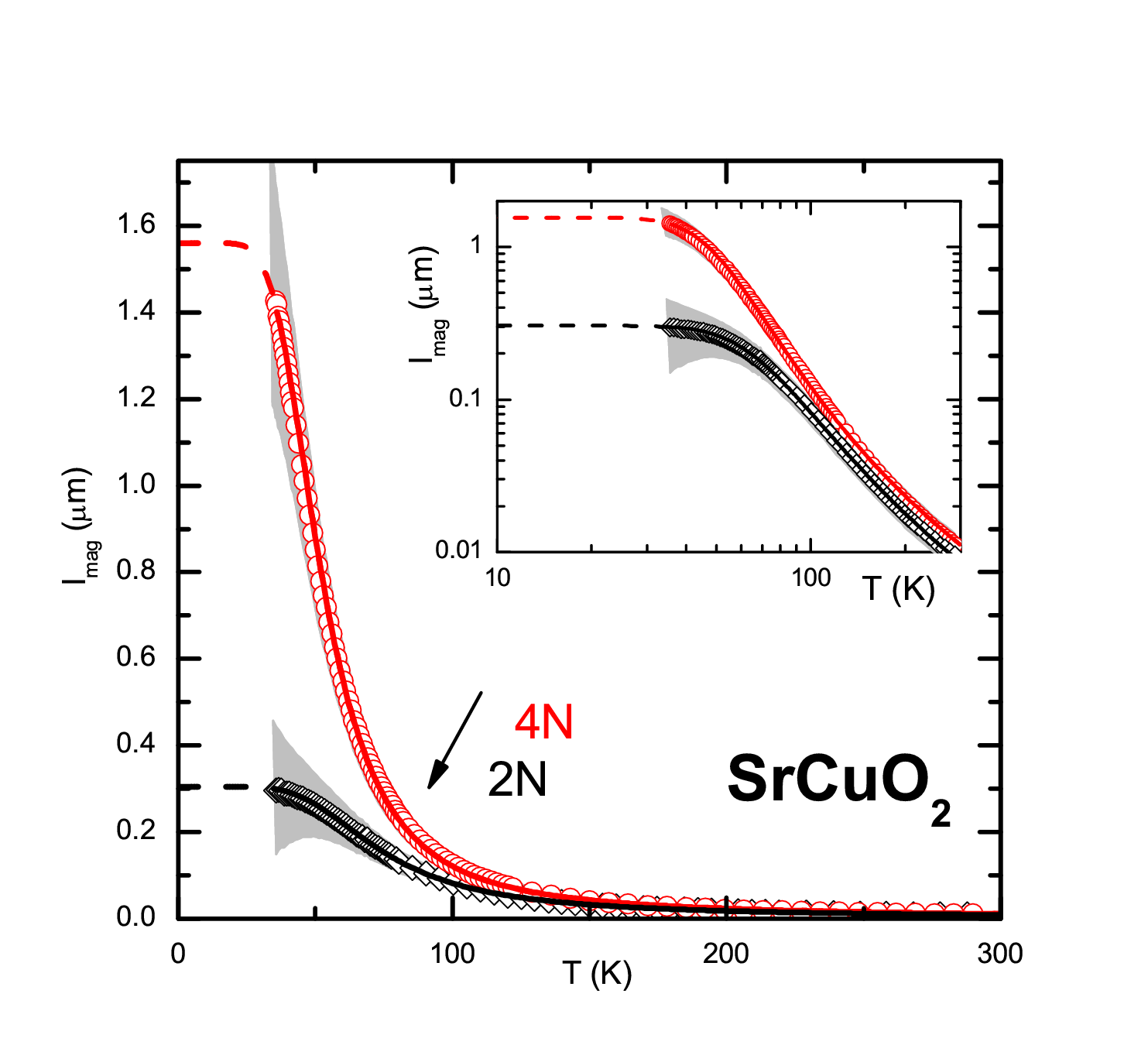}
\caption{Magnetic mean-free paths of \srcuodouble for different purities. The solid lines were calculated according to Eq.~\ref{eq:matthiesen_spez}. The shaded area illustrates the uncertainty from the estimation of the phononic background.
Adapted from \cite{Hlubek2010}.}
\label{fig:srcuo2_lmag}
\end{figure}

A more direct view on the temperature dependence of spinon scattering processes is provided by the spinon mean free path that can computed according to Eq.~\ref{eq:lmag_chain}. Resulting data are shown in Fig.~\ref{fig:srcuo2_lmag}. Both data sets show a strong decrease with increasing temperature, consistent with the above expectation of a growing importance of spinon scattering with increasing temperature. At the same time, the different magnitude of the curves which is most pronounced at lowest temperature signals the different importance of impurity scattering.

Following a conjecture of Sologubenko et al. \cite{Sologubenko01}, the data in Fig.~\ref{fig:srcuo2_lmag} have been modeled by Hlubek et al. using Matthiessen's rule for the \textit{extrinsic} spinon-defect and spinon-phonon scattering, i.e., employing a reduced version of Eq.~\ref{eq:matthiesen_allg}, where the \textit{intrinsic} spinon-spinon scattering as expressed by $l_\mathrm{ss}$ has been discarded:

\begin{equation}
 l_\mathrm{mag}^{-1}=l_0^{-1}+l_\mathrm{sp}(T)^{-1}\, .
 \label{eq:matthiesen_spez}
\end{equation}
An empirical expression has been used to model the spinon-phonon scattering, which represents a general umklapp process \cite{Sologubenko01,Hlubek2010}:

\begin{equation}
 l_\mathrm{sp}(T)=\frac{\exp(T^*_u/T)}{A_sT}\, .
 \label{eq:spin-phonon_emp}
\end{equation}
Here,  $T^*_u$ represents the energy scale of the scattering phonons, and $A_s$ a scattering cross section.
The solid lines in Fig.~\ref{fig:srcuo2_lmag} represent fits to the data employing Eqs.~\ref{eq:matthiesen_spez} and \ref{eq:spin-phonon_emp}, where for both purity levels the same  $T^*_u$ has been used \cite{Hlubek2010}. This reasonable assumption accounts for the expectation that the relevant energy scale for the scattering is the same for both purity levels.
The other parameters $l_{0,2N}$, $l_{0,4N}$, $A_{s,2N}$, $A_{s,4N}$  were employed as free fit parameters. Ideally, the purity should be solely captured by the spinon-defect scattering lengths $l_{0,2N}$, $l_{0,4N}$, since one would expect also  $A_{s,2N}\approx A_{s,4N}$, where deviations between these two values (up to 30\%) should be allowed for in order to compensate geometrical errors of the individual measurements of the heat conductivity. As can be inferred from the figure, these constraints allow excellent fits. 
Hlubek et al. report the energy scale $T^*_u\sim200~\mathrm{K}$ which is of the same order of magnitude as the Debye temperature of the material, corroborating spinon-phonon scattering as being the dominant temperature dependent process. The analysis yields further the spinon-defect scattering lengths $l_{0,2N}\approx300$~{\rm nm} and $l_{0,4N}\approx 1.6~\mu${\rm{m}}, which correspond to more than 750 and 4100 lattice spacings, respectively \cite{Hlubek2010}. 

This result is remarkable in two aspects: i) it suggests that indeed only the extrinsic spinon-defect and spinon-phonon scattering are relevant for relaxing the heat current. Thus the findings provide an experimental corroboration of the theoretical prediction of \textit{ballistic} heat transport in the $S=1/2$ antiferromagnetic Heisenberg chain. ii) in the high-purity sample, an extraordinary scattering length of more than  one micrometer is apparently present at low temperature, which apparently is only limited by the  impurities in the chains. This suggests that much larger spinon mean free paths could be achievable in perfectly clean crystals of \srcuodouble.

\subsubsection*{Ballistic spinon heat transport in Sr$\bf _2$CuO$\bf _3$}
After the pioneering experiments of Sologubenko et al. on the single chain material \srcuosingle \cite{Sologubenko00a,Sologubenko01} with 99\% purity (labeled '$2N$' hereafter), the  spinon heat transport in this compound has been under scrutiny by two further studies. Kawamata et al. investigated the heat transport of the pure material on a crystal where chemicals with 99.9\% purity (labeled '$3N$' hereafter) had been used for the crystal growth \cite{Kawamata2008}. Hlubek et al. investigated a crystal with even higher purity (99.99\% purity of the starting chemicals, labeled  '$4N$' hereafter), and performed a comparative study of the effect of the different purity levels \cite{Hlubek2012}.

\begin{figure}
\centering
\includegraphics[width=0.6\textwidth]{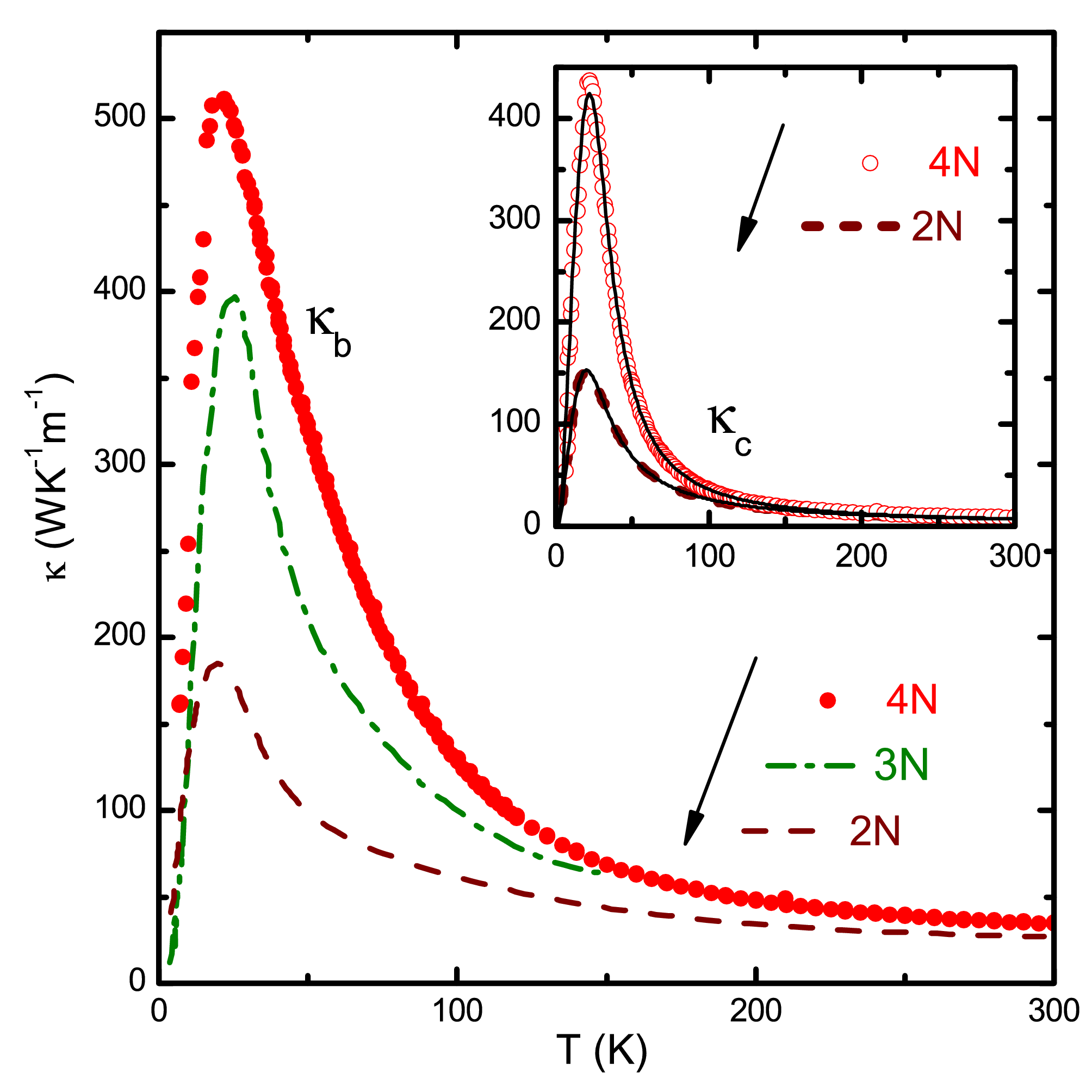}
\caption{Thermal conductivity of \srcuosingle parallel to the spin chains ($\kappa_b$)
for various purities. The dashed lines represent results from Sologubenko et al.
with '$2N$' purity, reproduced from \cite{Sologubenko00a}. The dash-dotted line has been obtained
by Kawamata et al for '$3N$' purity and is reproduced from \cite{Kawamata2008}. Inset: thermal
conductivity of \srcuosingle perpendicular to the spin chains ($\kappa_c$) for '$2N$' (also
reproduced from \cite{Sologubenko00a}) and '$4N$' purity. The solid lines are fits to the Callaway
model \cite{Callaway59}.
Adapted from \cite{Hlubek2012}.}
\label{fig:kappa_213}
\end{figure}

Fig.~\ref{fig:kappa_213} depicts the corresponding data for the thermal conductivity parallel to the spin chains ($\kappa_\Vert$, main panel), and except for the $3N$ purity, perpendicular to them ($\kappa_\bot$, inset).
$\kappa_\bot$, like in \srcuodouble exhibits a temperature dependence that is characteristic for phononic heat conductivity. The height of the characteristic phononic peak at $T \approx 22~\mathrm{K}$ sensitively depends
on the density of impurities in the system, which generate phonon-defect scattering. In fact, the
data for both purities can be described well in the framework of a model by Callaway \cite{Callaway59}, where the difference between both curves is largely captured by different point defect
scattering strength (see Reference \cite{Hlubek2012} for details).

\begin{figure}
\centering
\includegraphics[width=0.6\textwidth]{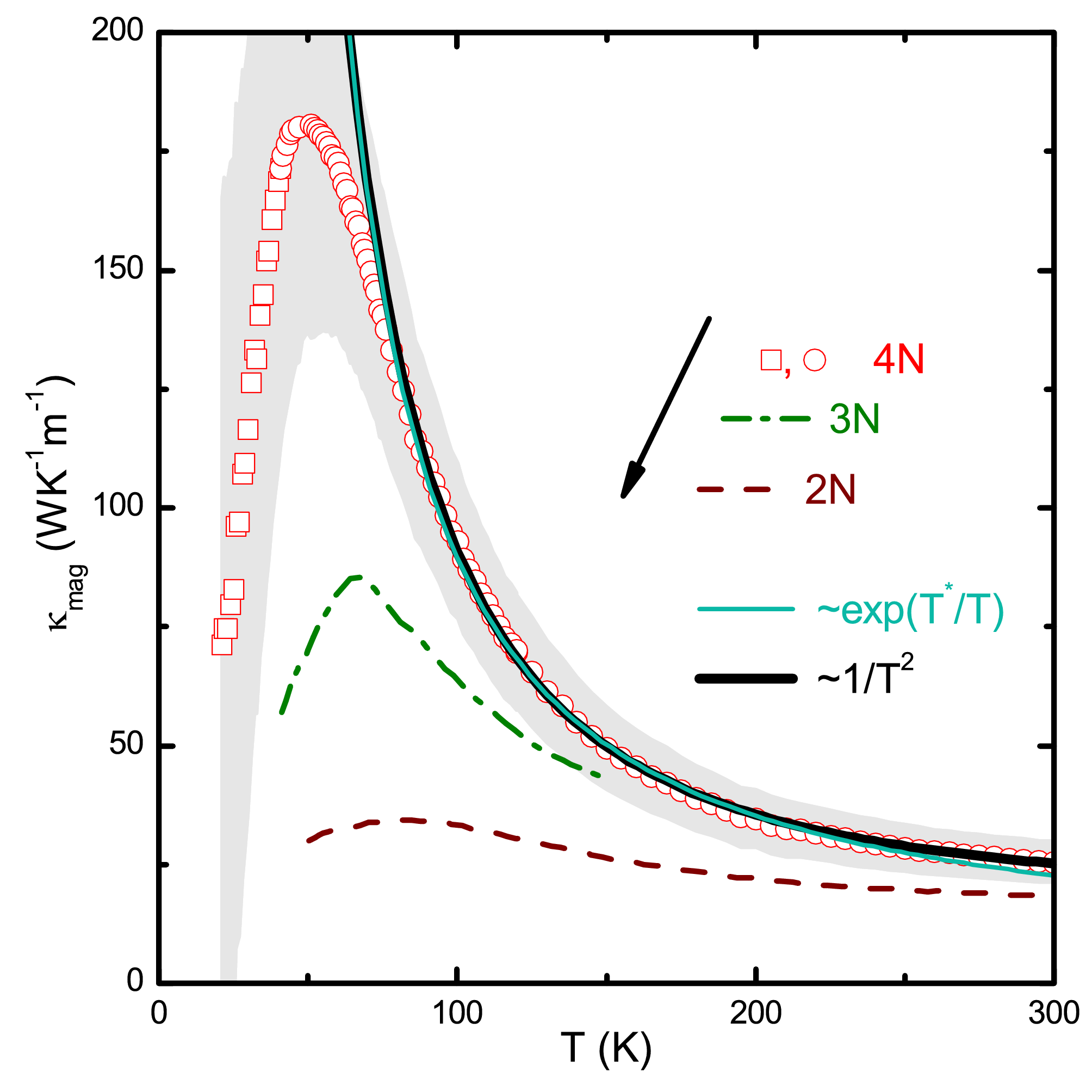}
\caption{Estimated magnetic thermal conductivity of \srcuosingle for '$4N$' (circles,
squares), '$3N$' (dash-dotted line) \cite{Kawamata2008}, and '$2N$' (dashed line) \cite{Sologubenko00a} purity. The shaded
area illustrates the uncertainty from the estimation of the phononic background.
The '$4N$' results shown in squares instead of circles have a large uncertainty. The
thick solid line is a fit for $T > 80~\mathrm{K}$ with $\kappa_\mathrm{mag,fit1}\sim1/T^{2}$. The thin solid line is
a fit with $\kappa_\mathrm{mag,fit2}\sim\exp(T^*/T)$. See Ref. \cite{Hlubek2012} where the figure has been reproduced from for a further discussion.}
\label{fig:kappamag_213}
\end{figure}

Very similarly as described above for \srcuodouble, the spinon heat
conductivity has been obtained via $\kappa_\mathrm{mag} = \kappa_\Vert - \kappa_\bot$ for the '$4N$' sample, which is shown together with similarly obtained results from References \cite{Sologubenko00a,Sologubenko01,Kawamata2008} for
'$3N$' and '$2N$' in Fig.~\ref{fig:kappamag_213}. The very similar temperature dependence of \kmag as compared to that of \srcuodouble, though with somewhat smaller values is evident. Hlubek et al. pointed out that the high-temperature decay of \kmag can be fit 
well with  $\kappa_\mathrm{mag}\sim\exp(T^*/T)$, with $T^*$ a characteristic energy scale, which can be related to prevailing umklapp processes \cite{Hlubek2012}.

\begin{figure}
\centering
\includegraphics[width=0.6\textwidth]{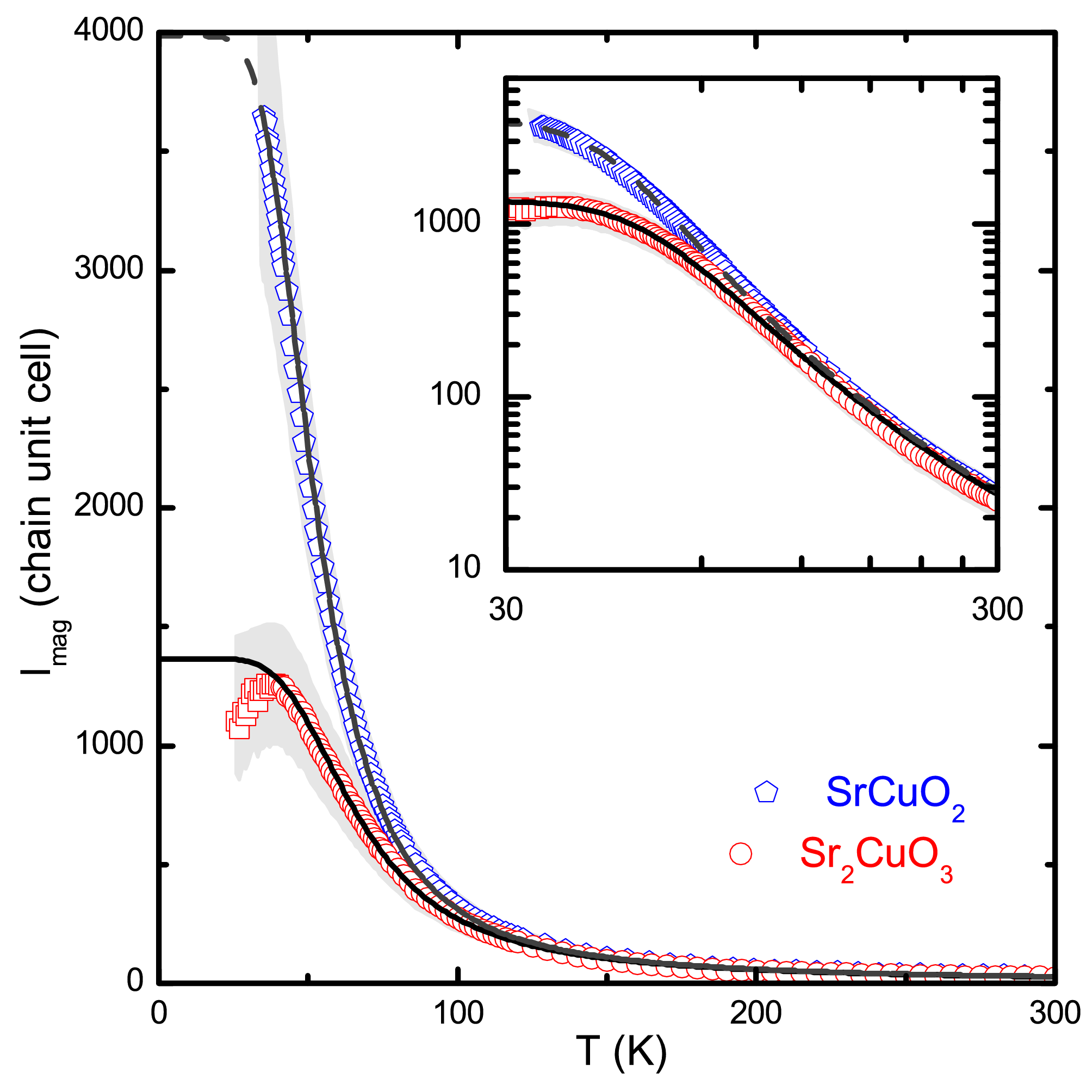}
\caption{Magnetic mean free paths of \srcuodouble (pentagon shape) and \srcuosingle
(circles) for '$4N$' purity. The shaded area illustrates the uncertainty from the
estimation of the phononic background. The lines are fits to the mean free paths.
Figure taken from \cite{Hlubek2012}.}
\label{fig:lmag_213}
\end{figure}

Hlubek et al. extracted from the data in Fig.~\ref{fig:kappamag_213} the spinon mean free path following Eq.~\ref{eq:lmag_chain}, and analyzed the temperature dependence according to
Eqs.~\ref{eq:matthiesen_spez} and \ref{eq:spin-phonon_emp}. Remarkably, the analysis work equally well (see Fig.~\ref{fig:lmag_213}) as before for \srcuodouble with $l_0 = 0.54 \pm 0.05~\mu\mathrm{m}$ (corresponding to approximately 1370 lattice spacings), $T^*_u= 210 \pm 11~\mathrm{K}$ \cite{Hlubek2012}. It is important to point out that the energy scale for the phonon scattering $T^*_u$ is practically the same for both \srcuodouble and \srcuosingle. This is consistent with the fact that in both materials the spin chains are formed by practically the same $\mathrm{CuO_2}$ plaquettes, suggestive of a very similar local spin-phonon coupling.
The low-temperature mean free path limit, $l_0$, is significantly smaller for \srcuosingle than that of \srcuodouble. For a direct quantitative comparison, Fig.~\ref{fig:lmag_213} shows \lmag of both compounds in relation to the number of lattice spacings. Despite the same formal chemical purity, the double chain compound's mean free path is about a factor of three larger, consistent with a relatively higher chemical stability of the compound \cite{Hlubek2012}.

Hlubek et al. attempted for the first time a deeper insight into the spin-phonon interaction by calculating $l_\mathrm{sp}$ directly from a spin-phonon scattering theory for the XY-limit of the Heisenberg model \cite{Louis2006} with promising results at high temperature. However at lower temperature, the agreement between the experimental results and the theoretical $l_\mathrm{sp}$ becomes less satisfactory (see \cite{Hlubek2012} for details).
Very recently, Chernyshev and Rozhkov reanalyzed the experimental data of Hlubek et al. for both \srcuodouble and \srcuosingle with an alternative theoretical model which specifies $l_\mathrm{sp}$ including realistic numerical values for the spin-phonon interaction \cite{Chernyshev2016}. Also in this case, the analysis relies for both compounds on just two types of scattering processes, viz. spinon-defect and spinon-phonon scattering. Thus the earlier statement for a strong experimental confirmation of ballistic heat transport holds for both materials.

\subsection{Spin chains with doping induced disorder}
The ballistic transport in the isotropic Heisenberg chain implies the natural question about the impact of subtle disorder on the transport properties. 
Such disorder should break the integrability of the spin model and thus a very substantial impact on the spinon heat conductivity is expected.
One possibility to approach this problem in \srcuodouble or \srcuosingle is to substitute Ca for Sr in small amounts, thus creating bond disorder due to the different ionic radii of Sr$^{2+}$ and Ca$^{2+}$. It turns out that this disorder type has a strong impact not only on the spinon heat transport but also on the ground state of the spin chains as is evidenced by Nuclear Magnetic Resonance (NMR) measurements \cite{Ribeiro2005,Hammerath2011,Hammerath2014,Hlubek2011,Mohan2014}.
% 
% % \subsubsection*{Nuclear magnetic resonance}
% \begin{figure}
% \centering
% \includegraphics[width=0.6\textwidth]{fig_NMR112}
% \caption{The $^{63}\mathrm{Cu}$ NMR spin lattice relaxation rate $1/T_1$ of
% \srcuodouble (a) and $\mathrm{Sr_{0.9}Ca_{0.1}CuO_2}$ (b) for different field directions.
% Inset: zigzag structure and main couplings. Figure taken from \cite{Hammerath2011}.}
% \label{fig:NMR112}
% \end{figure}
% 
% \subsubsection*{Bond disorder}
Hammerath et al. performed NMR measurements at the $^{63}\mathrm{Cu}$ nucleus for \srcuodoubleca at $x=0, 0.1$ and $\mathrm{Sr_{1.9}Ca_{0.1}CuO_3}$ determined the spin-lattice relaxation rate $1/T_1$ as a function of temperature \cite{Hammerath2011,Hammerath2014}.
% The spin-lattice relaxation rate is related to the dynamic susceptibility $\chi^{''}(\vec{q}, \omega)$ of the electronic spin system:
% \begin{equation}
%  T_1^{-1}\propto T\sum_{\vec{q}}A_\bot^2({\vec{q}}, \omega)\frac{\chi^{''}({\vec{q}}, \omega)}{\omega}.
% \end{equation}
The data very clearly reveal a doping induced spin gap which according to Hammerath et al. is solely induced by the bond disorder on the intrachain coupling \cite{Hammerath2014}. It should be noted, however, that a priori, it cannot be excluded that a finite amount of the doped Ca, instead of replacing the Sr in the structure, replaces Cu inside the $\mathrm{CuO_2}$-chain structures. In this case, a more severe site disorder would be the result which cuts the spin chains (which in the clean compounds possess a large average length of several thousand unit cells, according to the spinon mean free path, see Fig.~\ref{fig:lmag_213}). Indeed, a specific investigation of such a site disorder has been addressed by substituting Ni for Cu \textit{inside} the $\mathrm{CuO_2}$-chain structures of \srcuosingle as well as \srcuodouble \cite{Utz2015,Simutis2013} in NMR measurements and INS measurements. Both approaches led to the conclusion that finite size effects play an important role, i.e. the Ni effectively cuts the chains into finite segments which exhibit a gapped ground state. 

% 
% \begin{figure}
% \centering
% \includegraphics[width=0.7\textwidth]{fig_neutrons112}
% \caption{(a) Background-subtracted time-of-flight (TOF) neutron spectrum measured in $\mathrm{SrCu_{0.99}Ni_{0.01}O_2}$, as a function of momentum transfer along the spin chains and energy transfer. Quasi-vertical streaks at $q_\Vert/(2\pi)\approx\pm0.5$ correspond to the low-energy part of the spinon dispersion. Arrows indicate the spin pseudogap. (b) Momentum-integrated dynamic structure factor measured in $\mathrm{SrCu_{0.99}Ni_{0.01}O_2}$ at several temperatures, using TOF (solid symbols) and three-axis (open symbols) spectroscopy. The lines are calculated using a model which takes finite chain segments into account, where an effective doping level of $x=0.017$ is assumed. See \cite{Simutis2013} for details. Figure taken from \cite{Simutis2013}.}
% \label{fig:neutrons112}
% \end{figure}

% \subsubsection*{Bond disorder and heat conductivity}

\begin{figure}
\centering
\includegraphics[width=\textwidth]{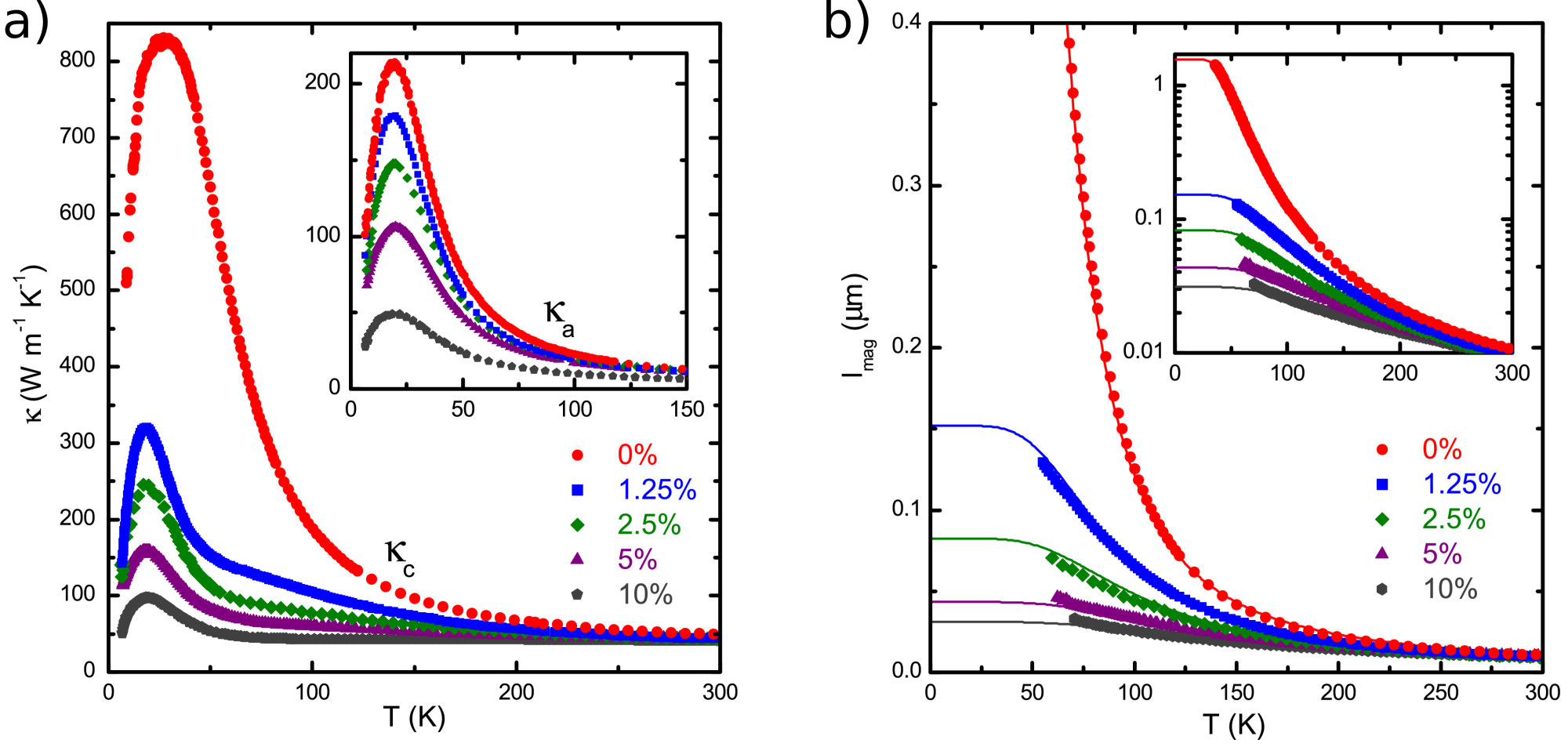}
\caption{a) Thermal conductivity $\kappa_c$ parallel to the spin
chain ($\kappa_\Vert$) for \srcuodoubleca at different doping levels. Inset:
Thermal conductivity $\kappa_a$ for the same doping levels perpendicular to
the spin chain ($\kappa_\bot$). b) \lmag of \srcuodoubleca for different levels of
doping. Toward low temperatures, the error in the estimation of \kmag
increases due to an increase in the phononic heat conductivity. Thus, the values of \lmag are shown
only at temperatures above which the error in \lmag is reasonably small.
The solid lines were calculated according to Eqs.~\ref{eq:matthiesen_spez},~\ref{eq:spin-phonon_emp}.
Figures taken from \cite{Hlubek2011}.}
\label{fig:kappa_112_Ca}
\end{figure}

Remarkably, the Ca-induced disorder has a dramatic impact on the heat conductivity parallel to the spin chains \kappara, while the phononic heat conductivity \kapperp is only moderately affected. This implies a very strong suppression of the spinon heat conductivity \kmag upon increasing the disorder. This is illustrated in Fig.~\ref{fig:kappa_112_Ca}a for the Ca-doped double-chain compound, i.e.  \srcuodoubleca \cite{Hlubek2011}. A very similar result is obtained for Ca-doped \srcuosingle (not shown) \cite{Mohan2014}.

The spinon heat conductivity can be calculated from the heat conductivity data in exactly the same way as it was already explained for the pure compounds above. Similarly, the temperature dependence of the spinon mean free path for various doping levels can be extracted via Eq.~\ref{eq:lmag_chain}, resulting in the data shown in Fig.~\ref{fig:kappa_112_Ca}b. Note, that a possible impact of the spin pseudogap can be excluded to have a significant impact on the spinon heat conductivity and the mean free path because at the size of the gap is too small ($\Delta/k_B\sim50~\mathrm{K}$ \cite{Hammerath2011,Hammerath2014}) to play a significant role for the spinon heat conductivity which is studied at $T\gtrsim50$~{\rm K} only \cite{Hlubek2011,Mohan2014}.

The data in Fig.~\ref{fig:kappa_112_Ca}b, and corresponding data for \srcuosingleca can be analyzed remarkably well as in the case of the clean compounds following Eqs.~\ref{eq:matthiesen_spez},~\ref{eq:spin-phonon_emp}. Thereby, essentially the spinon-defect scattering length $l_0$ turns out as the crucial parameter which is determined by the doping level whereas the phonon scattering term can be set identical for all doping levels (solid lines in Fig.~\ref{fig:kappa_112_Ca}b). 

\begin{figure}
\centering
\includegraphics[width=0.5\textwidth]{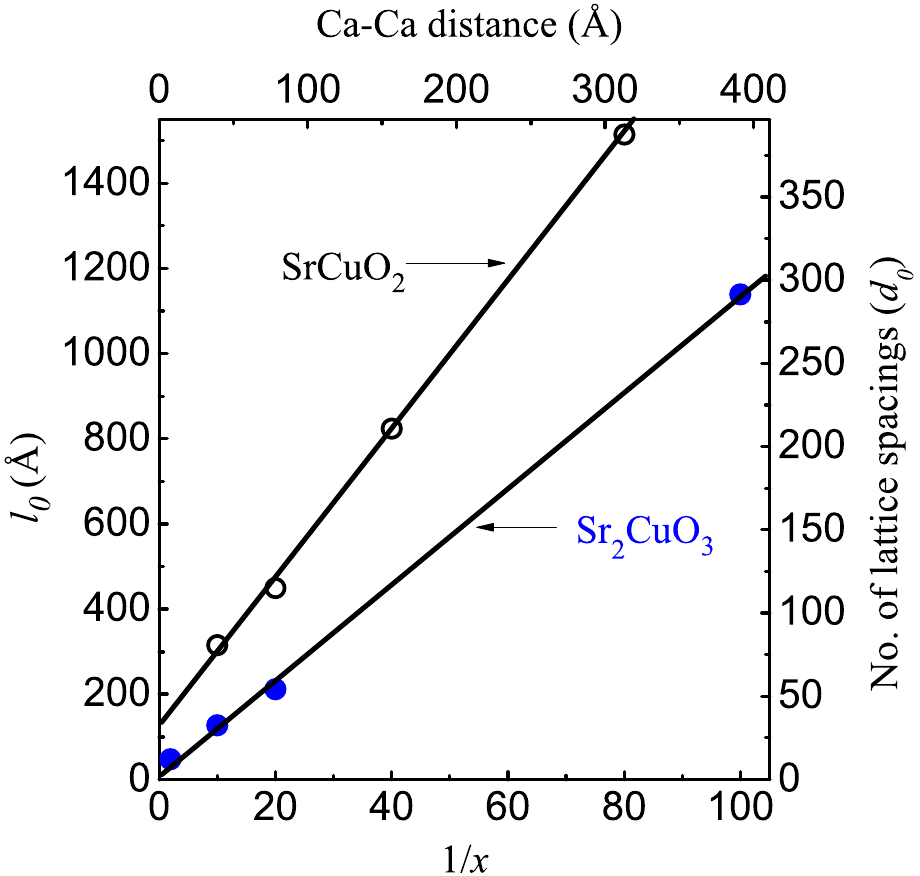}
\caption{Spinon-defect scattering length $l_0$ plotted
against the inverse of Ca concentration ($x$) (lower abscissa) and the
mean distance between two Ca atoms (upper abscissa) for \srcuodouble
(open symbols) and \srcuosingle (filled symbols); the solid lines are
linear fits to the data. The ordinate on the right expresses the
mean free path in terms of the number of lattice spacings between
two Cu sites ($d_0$).
Figure taken from \cite{Mohan2014}.}
\label{fig:lmag_Ca}
\end{figure}

It is very instructive to plot the obtained values of the spinon-defect scattering length $l_0$ as a function of the mean distance
between two Ca atoms and the inverse of Ca concentration as is shown in Fig.~\ref{fig:lmag_Ca}. For the single-chain material \srcuosingleca, $l_0$ apparently scales perfectly with the
inverse of Ca concentration as $l_0 =\frac{2.91\cdot d_0}{x} $, where $d_0 = 3.91~\text{\rm\AA}$ is the lattice spacing between two Cu sites along
the chain \cite{Mohan2014}. This implies that the Ca-induced disorder can perfectly be captured in terms of effective defects
in the chain, where the scattering probability per defect is
equally strong at all concentrations. 

Interestingly, the situation is somewhat different for \srcuodoubleca (see Fig.~\ref{fig:lmag_Ca}) \cite{Hlubek2011}.
There, a linear scaling is present, too, however, with an offset: $l_0 =\frac{4.3\cdot d_0}{x} + l_\mathrm{lim}$,
with $l_\mathrm{lim}\approx30\cdot d_0$ with $d_0=3.92~\text{\rm\AA}$. This indicates that already at intermediate
concentrations ($x = 0.1$) the effect of Ca saturates and the
mean free path of spinons is not reduced any further upon
increasing the Ca concentration. The observed offset was originally interpreted to be
due to a limit set by the disorder-induced long-distance decay
of the spin-spin correlation \cite{Hlubek2011}. However, since the double-chain nature did not enter the latter interpretation, the fact, that the offset is absent in the single-chain material \srcuosingleca implies a different origin of the offset, which however remains open.

\section{Conclusion}
We have seen that in the here discussed cuprate materials with large antiferromagnetic exchange interaction which realize low-dimensional $S=1/2$ Heisenberg systems in the form of chains, two-leg ladders, and planes, a sizeable magnetic heat conductivity arises. The possibility to extract the magnetic heat conductivity in a very clean manner in principle opens up a new approach for sensitively and comprehensively probing magnetic excitations in these systems.
Indeed, for each these quantum systems, using the kinetic model, it is possible to formulate a relatively simple way towards rationalizing the observed temperature dependences of \kmag and to draw basic conclusions about the involved scattering processes of the heat-carrying magnetic excitations.
So far, these considerations have been done individually for the various systems which are very different in terms of their elementary excitations, namely gapless spinons for the one-dimensional chains, triplon excitations with a large excitation gap for the ladders, and spin wave-like excitations in two-dimensions (with only small anisotropy gaps). Important conclusions could be drawn: Quite importantly, the kinetic model apparently yields realistic length scales of the magnetic mean free path \lmag, despite the simplicity of the model. Based on this finding, the discussed data confirmed the ballistic nature of the heat transport of integrable spin chain systems which truly is a fundamental finding. On the other hand, we have seen that in the non-integrable two-leg spin ladders the magnetic heat transport is nevertheless substantial with \lmag exceeding the spin-spin correlation length by about three orders of magnitude.

\begin{figure}
\centering
\includegraphics[width=\textwidth]{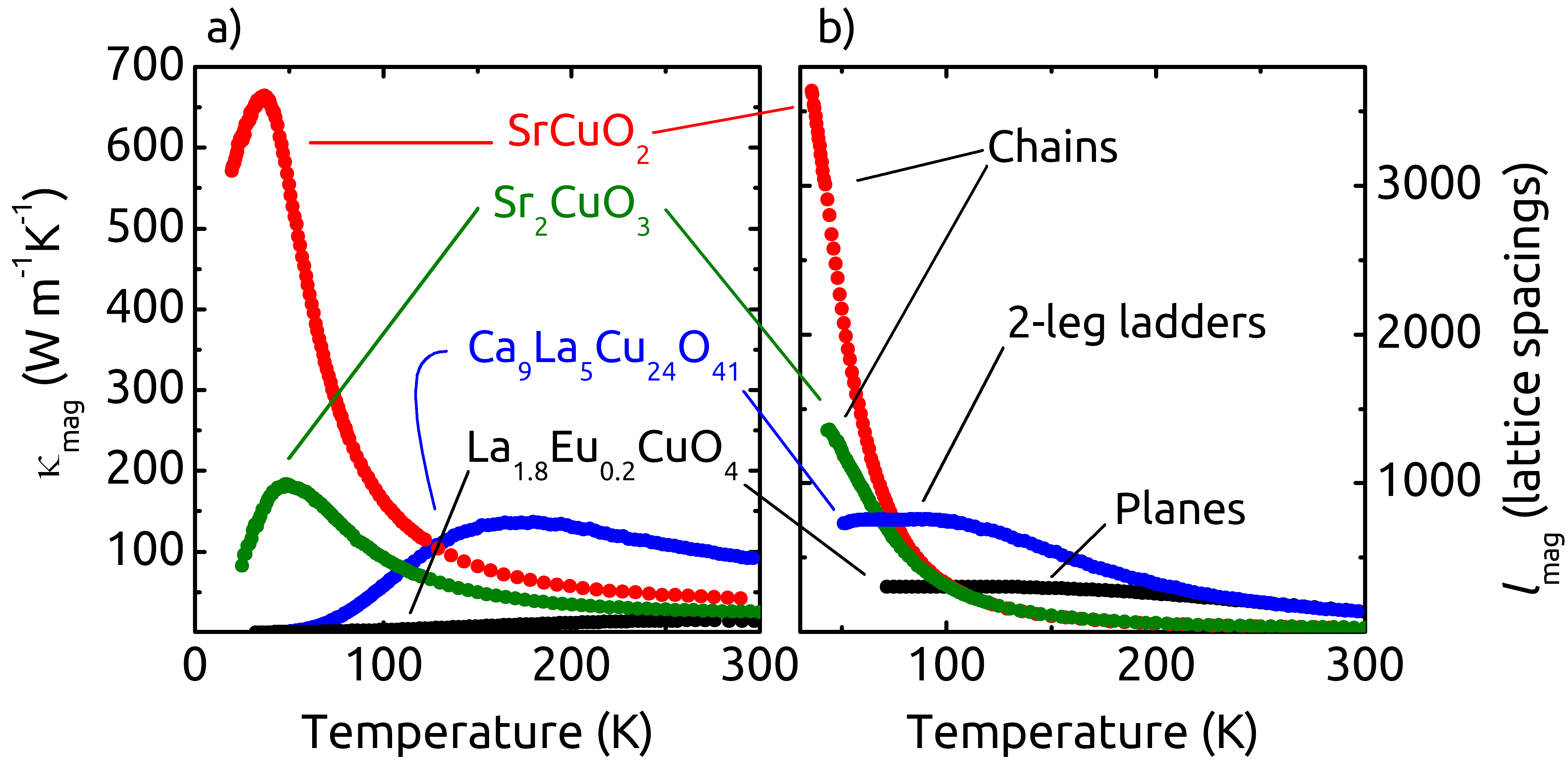}
\caption{Comparison of the magnetic heat transport of representative low-dimensional spin systems, namely for the spin chain compounds \srcuodouble and \srcuosingle \cite{Hlubek2010,Hlubek2012}, the two-leg spin ladder compound \laf \cite{Hess01}, and the 2D-HAF material \leucuo \cite{Hess03}. a) Magnetic heat conductivity as a function of temperature, $\kappa_\mathrm{mag}(T)$. b) Magnetic mean free path as a function of temperature, $l_\mathrm{mag}(T)$, in units of lattice spacings.}
\label{fig:lmag_all}
\end{figure}

Beyond these findings for the individual systems, it is interesting to directly compare the magnetic heat transport of these systems and to investigate whether the differences in the nature of the spin system also yield different transport and whether, nevertheless, similarities emerge.
Fig.~\ref{fig:lmag_all}a shows a collection of data  of \kmag of the chain systems \srcuodouble and \srcuosingle, of the two-leg ladder material \laf, and of the 2D-HAF \leucuo. These data can be considered in so far as being representative as they correspond to clean compounds where the up to present largest mean free paths have been extracted \cite{Hlubek2010,Hlubek2012,Hess01,Hess03}.
The data are the same as discussed in the previous sections, but this figure shows them for the first time together on the same scale and indeed reveals remarkable differences. Quite clearly, the chain compound \srcuodouble exhibits a 'record' value for \kmag of almost 700~\wkm which is many times larger than that of (i) the single chain compound \srcuosingle and, furthermore, (ii) that of  the 2-leg ladder and the 2D-HAF. 

The first observation (i) has been recognized earlier \cite{Hlubek2012} and already discussed in the previous section. 
A closer investigation of the mean free path of these two Heisenberg chain materials as shown in panel (b) of Fig.~\ref{fig:lmag_all} reveals that \lmag apparently can be separated in two regimes. On the one hand, this is at $T\gtrsim100$~K where there is virtually no difference between both curves. Here, according to the analysis, spinon-phonon scattering is dominant, where the data suggest that this type of scattering has no difference in both compounds. On the other hand, at $T\lesssim100$~K \lmag of \srcuosingle approaches a much lower low-temperature limit than that of \srcuodouble. Hence, the difference seems to imply a different inherent perfectness of the chains in the two compounds.

Upon trying to obtain some further understanding of the more interesting observation (ii), it is important to note that the very large value of \kmag of \srcuodouble occurs at relatively low temperature $T\lesssim100$~K where \kmag of both the ladder compound \laf and of the 2D-HAF material \leucuo is very small. In both cases, this comparably small low-temperature value of \kmag can be straightforwardly be understood: for the two-leg ladder material \laf, the significant spin excitation gap $\Delta~\sim400$~K prohibits a significant population of triplons below about 100~K. Similarly, for the 2D-HAF, the low-temperature thermal population of magnetic excitations is much 'slower' than that of the one-dimensional Heisenberg chain. One can convince oneself about this fact from considering the temperature dependence of the magnetic specific heat $c_\mathrm{mag}$ of both systems which is linear in temperature for the chain system whereas the two-dimensional spin plane possesses a quadratic low-temperature increase. Indeed, an estimation of the magnetic specific heat at $T=100$~K (i.e. $k_BT\ll J$) using the estimate $\kappa_\mathrm{mag}\sim c_\mathrm{mag} v_0 l_\mathrm{mag}$ and the respective kinetic expressions for \kmag in Eqs.~\ref{eq:fit2d} and \ref{eq:lmag_chain} yields an almost two orders of magnitude smaller $c_\mathrm{mag}$ for the 2D-HAF than that of the Heisenberg chain. Here, the magnetic velocity $v_0\sim Jd/\hbar$ with $d$ the distance between neighboring spins has been used. Thus, taking further into account the mean free path \lmag at about 100~K being of a similar order of magnitude for all compounds (see further below), it is not surprising that the magnetic heat conductivity of the Heisenberg chain is by far the largest at low temperatures.

Fig.~\ref{fig:lmag_all}b compares the temperature dependence of the mean free paths \lmag as extracted from the data in panel (a) upon using the kinetic model. Here one finds that \lmag of the spin chain compounds additionally exceeds clearly that of the ladder and the plane compounds at $T\lesssim100$~K. These large low-temperature values of \lmag additionally promote the extraordinarily large \kmag of the chains in this regime. 

Another interesting finding as revealed by Fig.~\ref{fig:lmag_all}b is the fact that at $T\gtrsim100$~K the mean free path becomes significantly smaller than that of both the 2-leg ladder and of the plane compounds. Since the temperature-induced reduction of \lmag of the spin chain materials has clearly been assigned to spinon-phonon scattering \cite{Hlubek2010,Hlubek2012,Chernyshev2016}, this observation seems to imply that the spinon heat transport of the $S=1/2$ Heisenberg chain is much more prone to scattering of phonons than the two-leg Heisenberg ladder and the 2D-HAF counterparts. It remains to be clarified to what extent this notion is related to the integrability of the Heisenberg chain model, whose ballistic transport properties potentially could exhibit a particular sensitivity to distortions such as those induced by the phonons. Alternatively, differences in the magnetic and phononic spectrum could play an important role, too. Indeed such seems to be a crucial difference in the spin-phonon scattering of the Heisenberg chains where the gapless spinons can be expected to interact with acoustic phonons and that of the 2-leg spin ladders, where the energy of the acoustic phonons is too small to cause an important interaction with the gapped triplon excitations and instead scattering of optical phonons seems to dominate (see Section~\ref{sec:tdepmagnonscatt}).

Finally, it is to be remarked that little is known from experiments about the temperature evolution of \lmag of the 2D-HAF at elevated temperatures since apparently the data shown in Fig.~\ref{fig:lmag_all}b as extracted from Ref.~\cite{Hess03} barely touch this regime. An obvious aspect of future research is thus to extend the achievable temperature regime of reliable \kmag-data to much higher temperatures and to verify specific predictions about magnon scattering processes involving phonons and the correlation length \cite{Chernyshev2015}.

As it has already been mentioned above, all the here discussed analysis is based on the kinetic model (see Section \ref{kinmod}), the successful applicability of which is surprising in view of the strong quantum nature of the here studied spin models. Another possible direction of future research therefore could address ways to model the magnetic heat transport in the studied systems on a more microscopic level.

It is worth to mention that the large magnetic heat conductivity values of the spin ladder and the chain systems may also open up new technological developments. The exploitation of the resulting anisotropic heat conductivity tensor for thermal management applications already has been investigated (see e.g. \cite{Otter2012}).  Another, yet unexplored direction would be spin information transport experiments,  where electrical pumping and detection of spin currents with the help of the spin-Hall and the inverse spin-Hall effect, as recently successfully used on classical spin systems \cite{Cornelissen2015}.

Upon concluding, it is important to mention further active research directions which could not be touched in this review. One concerns research on $\kappa_{\mathrm{mag}}$ for systems where the magnetic exchange energy becomes comparable or much smaller than the Debye energy, and how it about evolves when the magnetic systems become less quantum in nature, i.e. when $S>1/2$. Initial work has already addressed these aspects. Intriguing heat conductivity results  for one-dimensional spin systems with relatively small exchange interactions have been obtained for both organic and inorganic materials, see e.g. \cite{Sologubenko2007a,Lue2008,Sologubenko2009,Jeon2016,Steckel2016} for  $S=1/2$ systems and \cite{Sologubenko03,Sologubenko2008,Kohama2011,Sun2013,Karadamoglou04,Savin2005,Kordonis06} for systems with large spin, see also the reviews \cite{Sologubenko2007,Zhao2016}.

Another dynamic field where magnetic heat transport is being investigated concerns frustrated spin systems, where it promises to become a good probe for accessing possible topological excitations. The class of spin-ice compounds of the type $R_2$Ti$_2$O$_7$ ($R$ a rare earth element) constitutes a subject on its own where the focus is on magnetic monopole-like excitations \cite{Klemke2011,Kolland2012,Toews2013,Fan2013,Kolland2013,Scharffe2015,Li2015,Tokiwa2016,Hirschberger2015,Toews2018}.
Another class concerns highly frustrated layered compounds \cite{Yamashita2008,Yamashita2010} where very recently $J_\mathrm{eff}=1/2$-materials with Kitaev interactions came into focus \cite{Kitaev2006,Hirobe2017,Hentrich2018,Hentrich2018a,Kasahara2018,Kasahara2018a} since such systems are conjectured as fascinating avenues for exploring the 'magnetic' heat transport of topological fractionalized quasiparticles.

\section*{Acknowledgments}
I am indebted to Neela Sekhar Beesetty, Wolfram Brenig, Bernd B\"uchner, David Cahill, Sang-Wook Cheong, Alexander Chernyshev, Stefan-Ludwig Drechsler, Hanan ElHaes, Jochen Geck, Ioannis Giapintzakis, Hans-Joachim Grafe, Franziska Hammerath, Fabian Heidrich-Meisner, Nikolai Hlubek, Vladislav Kataev, R\"udiger Klingeler, Andreas Kl\"umper, Gernot Krabbes,  Paul van Loosdrecht, Thomas Lorenz, Oleg Mityashkin, Ashwin Mohan, Satoshi Nishimoto, Peter Prelov\v{s}ek, Pascal Reutler, Alexandre Revcolevschi, Patrick Ribeiro, Georg Roth, Romuald Saint-Martin, Chinnathambi Sekhar, Robin Steinigeweg, Yannic Utz, Anja Waske,  Anja Wolter, Babak Zeini,  Andrey Zheludev, and Xenophon Zotos for fruitful discussion or collaborations.

Funding: This work was supported by the Deutsche Forschungsgemeinschaft [grant numbers HE3439/7, HE3439/8, HE3439/9, HE3439/12, HE3439/13, SFB 1143 (project C07)]; and the European Commission [grant numbers FP6-032980, PITN-GA-2009-238475].

\section*{References}

% \bibliography{bibtex/Cuprates,bibtex/LowDTransport,bibtex/TiOX,bibtex/kmag_organic,/home/ch5/Dokumente/Literatur/RuCl/rucl,/home/ch5/Dokumente/Literatur/Spin_liquids/spinliquids}

\end{document}